\documentclass[preprint,12pt]{elsarticle}

\usepackage{graphicx}
\usepackage{amssymb, amsthm}
\biboptions{compress}

\journal{\hspace{-3.5cm} \raisebox{-1mm}{\begin{tikzpicture} \draw [fill = white, white] (0, 0) rectangle (3.5, 0.4); \end{tikzpicture}}}

\usepackage{amsmath}
\usepackage{url}
\usepackage{color}
\usepackage{tensor}
\usepackage{mathrsfs}
\usepackage{nicefrac}
\usepackage{tikz-cd}
\usepackage{tikz}
\usetikzlibrary{decorations.markings}
\usetikzlibrary{matrix,arrows}
\usepackage{array}
\usepackage{ulem}
\usepackage{placeins}
\usepackage{enumitem}

\usepackage{pdflscape}

\usepackage[latin1]{inputenc}      
\usepackage[T1]{fontenc}    
\usepackage{ae,aecompl}                          
\usepackage{dsfont}
\usepackage{appendix}
\usepackage{amsfonts,amsmath,latexsym,bbm}
\usepackage{amssymb}
\usepackage{accents}
\usepackage{graphics}
\usepackage{graphicx}
\usepackage{booktabs,longtable}
\usepackage{bm}
\usepackage{dcolumn}
\usepackage{enumitem}
\usepackage{upgreek}
\usepackage{tensor}

\usepackage{tikz}
\usepackage{tikz-cd}
\usepackage{tikz-3dplot}
\usetikzlibrary{decorations.markings}
\usetikzlibrary{matrix,arrows}
\usetikzlibrary{calc}
\usetikzlibrary{shapes}
\usetikzlibrary{positioning}
\usetikzlibrary{fit}

\usetikzlibrary{decorations.pathreplacing}
\makeatletter
\def\underbrace#1{%
  \@ifnextchar_{\tikz@@underbrace{#1}}{\tikz@@underbrace{#1}_{}}}
\def\tikz@@underbrace#1_#2{%
  \tikz[baseline=(a.base)] {\node[inner sep=2] (a) {\(#1\)};
  \draw[thick,line cap=round,decorate,decoration={brace,amplitude=4pt}]
    (a.south east) -- node[pos=0.5,below,inner sep=7pt] {\(\scriptstyle #2\)} (a.south west);}}
\makeatother

\usepackage[linktocpage=true]{hyperref}

\makeatletter
\renewcommand*{\eqref}[1]{%
  \hyperref[{#1}]{\textup{\tagform@{\ref*{#1}}}}%
}
\makeatother

\setlist{font=\normalfont\itshape} 

\renewcommand{\vec}[1]{\boldsymbol{#1}}
\newcommand{\tsr}[1]{\overset\leftrightarrow{#1}}
\newcommand{\ext}{_{\rm ext}}
\newcommand{\tot}{_{\rm tot}}
\newcommand{\ind}{_{\rm ind}}

\newcommand{\h}{\hspace{1pt}}
\newcommand{\hh}{\hspace{0.5pt}}
\newcommand{\mh}{\hspace{-1pt}}

\newcommand{\de}{\mathrm d}
\newcommand{\e}{\mathrm e}

\newcommand{\lar}[1]{\textnormal{\mbox{\large $#1$}}}
\renewcommand{\i}{\mathrm i}

\definecolor{oldgray}{gray}{0.4}

\newcommand{\ket}[1]{|#1\rangle}
\newcommand{\bra}[1]{\langle #1|}
\renewcommand{\mid}{\h|\h}

\newcommand{\raisemath}[1]{\mathpalette{\raisemith{#1}}}
\newcommand{\raisemith}[3]{\raisebox{#1}{$#2#3$}}

\newcommand{\EE}{_{\raisemath{-2pt}{EE}}}

\newcommand{\BB}{_{\raisemath{-2pt}{BB}}}

\newcommand{\T}{_{\mathrm T}}
\renewcommand{\L}{_{\mathrm L}}

\DeclareMathAlphabet{\mathbbmsl}{U}{bbm}{m}{sl}

\DeclareMathOperator{\Tr}{Tr}

\numberwithin{equation}{section}

\begin{document}

\begin{frontmatter}

%% Title, authors and addresses

%% use the tnoteref command within \title for footnotes;
%% use the tnotetext command for the associated footnote;
%% use the fnref command within \author or \address for footnotes;
%% use the fntext command for the associated footnote;
%% use the corref command within \author for corresponding author footnotes;
%% use the cortext command for the associated footnote;
%% use the ead command for the email address,
%% and the form \ead[url] for the home page:
%%
%% \title{Title\tnoteref{label1}}
%% \tnotetext[label1]{}
%% \author{Name\corref{cor1}\fnref{label2}}
%% \ead{email address}
%% \ead[url]{home page}
%% \fntext[label2]{}
%% \cortext[cor1]{}
%% \address{Address\fnref{label3}}
%% \fntext[label3]{}

\title{Ab initio materials physics and microscopic electrodynamics of media}

%% use optional labels to link authors explicitly to addresses:
%% \author[label1,label2]{<author name>}
%% \address[label1]{<address>}
%% \address[label2]{<address>}

\author[freiberg]{R.~Starke}
\ead{starke.ronald@gmail.com}

\author[heidelberg]{G.A.H.~Schober\corref{cor1}}
\ead{giulio.schober@gmail.com}

\cortext[cor1]{Corresponding author.}

%\address[vienna]{Department of Computational Materials Physics, University of Vienna, \\ Sensengasse 8/12, 1090 Vienna, Austria}
\address[freiberg]{Institute for Theoretical Physics, TU Bergakademie Freiberg, Leipziger Stra\ss e 23, \\ 09596 Freiberg, Germany}
\address[heidelberg]{Institute for Theoretical Physics, Heidelberg University, Philosophenweg 19, \\ 69120 Heidelberg, Germany \vspace{-0.6cm}}

\begin{abstract}
We argue that the amazing progress of first-principles materials physics necessitates a revision of the Standard Approach to electrodynamics of media. 
We hence subject this Standard Approach to a thorough critique, which shows both its inherent
conceptual problems and its practical inapplicability to modern ab initio calculations. We then go on to show that the common \linebreak practice
in ab initio materials physics has overcome these difficulties by taking a different, {\it microscopic} approach to electrodynamics of media, only superficially resembling the Standard Approach whose validity is restricted to the  {\it macroscopic} domain. 
As this paradigm shift went largely unnoticed outside and partly even within the ab initio community, the present article aims at a systematic development 
and paradigmatic discussion of this new, microscopic first-principles approach to electrodynamics of media, for which we propose the name {\it Functional Approach}.
\end{abstract}

\begin{keyword}
%% keywords here, in the form: keyword \sep keyword
ab initio materials physics \sep electrodynamics of media \\[3pt]
{\it Cite as:~} ResearchGate pub.~{\bfseries 303755035}; arXiv:1606.00445

%% MSC codes here, in the form: \MSC code \sep code
%% or \MSC[2008] code \sep code (2000 is the default)

\end{keyword}

\end{frontmatter}

%%
%% Start line numbering here if you want
%%
% \linenumbers

%% main text

\newpage

\ \vspace{-1.3cm}
\tableofcontents

\section{Introduction}

Ever since J.\,C.~Maxwell's introduction of electromagnetic field theory in 1865 \cite{Maxwell65}, the learned world has been struck time and again
by scientific upheavals in electrodynamics, ranging from H.~Hertz' detection of electromagnetic waves in 1887 \cite{Hertz} (cf.~also \cite{Buchwald,Bryant} for historical accounts),
over A.~Einstein's development of special relativity in 1905 \cite{Einstein1905}, up to the mastery of quantum electrodynamics 
in the second half of the twentieth century (see \cite{SchweberQED} for a historical overview).
All these extremely spectacular discoveries notwithstanding, in the wake of many-body quantum mechanics a more silent revolution has made its way,
which---although being extremely relevant for the interpretation of electrodynamics in materials---almost completely escaped attention
outside the reclusive circles of highly specialized experts: the development of {\it ab initio materials physics}
(see e.g.~\cite{Giuliani,Martin,Kohanoff} for modern textbooks, \cite{Liebing, BoeriPRL, Linscheid1, Linscheid2, Draxl16, Klett16, Kresse16, Pavarini16, Scheffler16, Wissgott2016, Lejaeghere, Hahn} for 
recent research articles, in particular in connection with functional renormalization group (fRG) applications see \cite{Scherer1, Maier, LichtensteinRG, Classen, FRGSTM, Evers}).

In fact, although ab initio materials physics enjoys the reputation of being a rather applied field of research,
it has initiated a theoretical paradigm shift \cite{Kuhn} in the strict sense of the word: While it had traditionally been assumed in many research areas
that there are essentially two different types of physical laws---a fundamental one applying in vacuo and a phenomenological one applying in matter---ab initio materials physics
has brought it to light that the fundamental laws (which are those allegedly applying only in vacuo) actually always hold true.
In fact, they can even be used to {\itshape predict} the behavior of materials {\it from first principles}.
By contrast, in the traditional view the fundamental laws would somehow lose their validity
or at least their applicability in materials, where they would have to be replaced
by their phenomenological counterparts. These phenomenological laws then usually involve free parameters which 
are supposed to characterize the material under consideration.

The prime example of this philosophy is, of course, the classical theory of electromagnetism, where
the physical laws most incisively come in two different guises: as Maxwell's equations {\itshape in vacuo}
and {\itshape in media.} In a way, however, the preliminary and tentative character of this traditional distinction
has been clear at least since the works of H.\,A.~Lorentz \cite{Lorentz1902,Lorentz1916}, who identified this dualism with an apparently
similar dichotomy: the one of {\it microscopic} and {\it macroscopic} Maxwell equations. Surprisingly, it turned out that
the Maxwell equations ``in vacuo'' actually hold true ``in medio'' as well, but only microscopically. Macroscopically though,
they had to be replaced by the old Maxwell equations in media. Later, these then promptly went under the name of ``macroscopic Maxwell equations'',
while the old vacuum Maxwell equation were rechristened ``microscopic Maxwell equations''. Correspondingly, the ``macroscopic'' Maxwell equations
were not accepted as fundamental anymore, but instead had to be derived from the ``microscopic'' Maxwell equations by means of suitable averaging procedures.
Henceforth, we will call this the {\it Standard Approach} to electrodynamics in media. It is exposed in the 
traditional textbook literature (\cite{Jackson,Griffiths,Landau}; see also \cite{VanKranendonk} for a systematic theoretical development).

In ab initio materials physics, however, electromagnetic material constants are not interpreted as 
empirical parameters to be determined {\it ex datis}. Instead, these material properties are themselves 
the target of sophisticated computer simulations (such as \cite{VASP, Wien2k, QE-2009}), and hence they have to be determined {\it ex principiis}. 
Correspondingly, ab initio calculations are not only independent of, but in fact even preclude
a priori assumptions about the electromagnetic response. 
In other words, material properties are nowadays not predetermined concepts, which would be outside the scope of physical theories. 
Instead, the material properties are themselves to be determined from fundamental physical laws. This, of course, requires these fundamental 
laws to be free of adjustable material parameters. Hence, from the modern point of view there are not different laws of nature 
depending on the material, but the fundamental laws themselves---i.e., the microscopic Maxwell equations combined 
with a (quantum) field theory for the microscopic charges and currents---determine the electromagnetic properties of {\it any} material.
To distinguish Maxwell's equations ``in media'' from those ``in vacuo'' would be as misleading in the ab initio context as, for example, distinguishing
a many-body Schr\"{o}dinger equation ``in media'' from an alleged counterpart ``in vacuo''.

We conclude that {\it macroscopic electrodynamics is not suitable for first-principles calculations}, which are inherently microscopic and parameter free. 
As such, this insight is far from being new. For example, in the important work of L.\,V.~Keldysh \cite[p.~7]{Keldysh} we find
under the headline ``Averaging Maxwell's equations. Do we need
physically infinitesimal volumes and magnetic permeability?'' the very explicit statement:
``Excluding from the very outset the field structure at microscopic distances around each particle, this approach [i.e.~the Standard Approach in our terminology] does not allow one
to relate the dielectric constant to the microscopic structure of the medium and is therefore inevitably found to be purely phenomenological.''
Even more directly, the textbook \cite[p.~13]{IlinskiiKeldysh} by L.\,V.~Keldysh and Yu.\,A.\,Il'inski asserts that
``[t]he averaging over a physically infinitely small volume immediately
excludes from analysis the field acting on a particle and forces one to forego
the response of the medium to the field; after this, one has to be satisfied
with taking only a phenomenological account of this response.
Therefore, averaging over a physically infinitely small volume has to
be dropped from a microscopic theory of matter's response to 
electromagnetic field [\ldots]''.
Independently, the restriction of electrodynamics in media to the macroscopic domain has been criticized by L.\,L.~Hirst \cite{Hirst}.
Furthermore, the unappropriateness of the Standard Approach for the ab initio computation of electromagnetic materials properties 
has been discussed within the context of the so-called Modern Theory of Polarization (see in particular Refs.~\cite{Vanderbilt, Resta07, Resta10}).
Then, a systematic critique of the Standard Approach has also been presented by K.~Cho \cite{ChoEDBH, Cho08, Cho03, Cho10}.
Finally, critical remarks concerning the Standard Approach can even be found in the modern textbook \cite[Chap.~VI,  Sec.~27]{Fliessbach} on classical electrodynamics.

Superficially, one might now conclude that this situation does not pose any problems because ab initio materials physics
is a fundamental theory, whereas the Standard Approach was meant to be phenomenological anyway. In other words, while one should not ask
about the microscopic mechanisms in the phenomenological theory, the phenomenological concepts on their side do not play any r\^{o}le in the fundamental theory.
On closer inspection, however, we see that this is not true. Although the {\it starting point} of ab initio materials physics is always given by
the many-body Schr\"{o}dinger equation and the microscopic Maxwell equations, the {\it observable quantities} one ultimately wants to predict are in most cases still given by the old
phenomenological concepts such as the electric permittivity, the magnetic permeability, the conductivity or even the refractive index \cite{Baroni,Keith1992,Lichtenstein,Wang06,Lee,Schober12,Laszlo,Quandt,Abeyrathne,Nath,LiLi,Pan}.
Of course, their calculation from first princples would not be possible if they did not have any meaning on the microscopic level at all. 
Quite on the contrary, these concepts are as well defined microscopically as they are macroscopically: they are given in terms of response functions which, at least in principle,
can be calculated via the Kubo formalism from the microscopic states of matter \cite{Giuliani,Bruus,Kubo,Mahan}. The so-called ``macroscopic'' quantities 
are then obtained ex post by taking suitable limites where the wavevectors and frequencies go to zero~\cite{Fliessbach, Adler, Hanke, Wiser, Bechstedt, Dolgov}.

Thus, given that the allegedly ``macroscopic'', phenomenological quantities are actually obtained from their microscopic
counterparts as appropriate limiting values, these microscopic counterparts require an equally microscopic definition in the first place. Needless to say that such microscopic
definitions are standard in ab initio materials physics for a long time: they are given by linear expansions of {\it induced} fields 
as {\it functionals} of {\it external} perturbations \cite{Fliessbach,SchafWegener,Kittel,Ashcroft}, which is in principle both microscopically and macroscopically valid.
This, however, directly implies that within ab initio materials physics a new paradigm for 
electrodynamics in matter has emerged. Correspondingly, we now propose for this new paradigm the name {\it Functional Approach},
thereby underlining the importance of functional dependencies between fields.
This Functional Approach, which systematizes and axiomatizes the common practice implicit in modern ab initio materials physics, 
has been developed explicitly by the authors of this article \cite{ED1,EDOhm,Refr,EffWW} 
with the focus on practical applications and conceptual matters. In the present work, we will compare the Functional Approach to the Standard Approach paradigmatically
in order to show that within the ab initio context, this new approach is imperative and, in fact, already common practice.

This article is organized as follows: In Sec.~\ref{Sec_StandardApp}, we systematically elaborate on our criticism of the Standard
Approach to electrodynamics in media. Next, Sec.~\ref{sec_functional} introduces the Functional Approach to electrodynamics of materials, with Subsec.~\ref{subsec_derivations} being dedicated to its systematic derivation from first principles and Subsec.~\ref{subsec:generalConclusions}  concerned with the most important conclusions
to be drawn from this new paradigm. Finally, Sec.~\ref{sec_thermodynamics} shows that the Functional Approach is imperative also from the thermodynamical
point of view. Apart from this, the appendices deal with the problem of homogeneously polarized bodies (App.~\ref{app_const}), with the notorious
dipole densities (App.~\ref{app_dip}), and with the Kubo formula for the current response tensor (App.~\ref{app_elmKubo}).

\section{Standard Approach to electrodynamics in media}\label{Sec_StandardApp}
\subsection{Consensus on field equations}

In this introductory subsection, we give a short account of the Standard Approach as a classical field theory regardless of its macroscopic character.
It will later turn out that the field equations as such can be upheld if with an adapted interpretation. We use the conventions of Ref.~\cite{ED1}.

Maxwell's equations in the presence of material media (see e.g.~\cite[Eqs. (6.6)]{Jackson} and \cite[Eqs.~(7.55)]{Griffiths}) 
usually come as a pair of homogeneous equations, 
\begin{align}
\nabla \cdot \vec B(\vec x,t) & = 0\,, \label{eq_maxwell_3} \\[5pt]
\nabla \times \vec E(\vec x,t)+ \partial_t\vec B(\vec x,t)& = 0\,, \label{eq_maxwell_4}
\end{align}
and a pair of inhomogeneous equations,
\begin{align}
\nabla \cdot \vec D(\vec x,t) & = \rho_{\rm f}(\vec x,t)\,, \label{eq_maxwell_1} \\[5pt]
\nabla \times \vec H(\vec x,t) - \partial_t\vec D(\vec x,t) & = \vec j_{\rm f}(\vec x,t)\,. \label{eq_maxwell_2}
\end{align}
In these equations, $\vec D,\vec E,\vec H,\vec B$ respectively denote the {\it displacement field},
the {\it electric field}, the {\it magnetic field} and the {\it magnetic induction}, while $\rho_{\rm f}, \h \vec j_{\rm f}$ denote the so-called {\it free charge and current densities}.
In particular, there are two electric field quantities, $\{\vec D,\vec E\}$, and two magnetic field quantities, $\{\vec H,\vec B\}$.
In order to account for their respective differences, one introduces yet another pair of fields, the electric and magnetic polarizations $\vec P$ and $\vec M$,
which are respectively given by
\begin{align}
\vec P(\vec x,t)&=-\varepsilon_0 \h \vec E(\vec x,t)+\vec D(\vec x,t)\,,\label{eq_PED}\\[5pt]
\vec M(\vec x,t)&=\mu_0^{-1}\vec B(\vec x,t)-\vec H(\vec x,t)\,,\label{eq_MBH}
\end{align}
with the so-called vacuum permittivity $\varepsilon_0$ and vacuum permeability $\mu_0$\hh.
For these polarization fields, one introduces further inhomogeneous field equations given by 
(see e.g.~\cite[Sec.~2.2.1]{BornWolf}, \cite[p.~190]{Wachter}, \cite[p.~76]{Kovetz2000} and \cite[p.~11]{Melrose})
\begin{align}
-\nabla\cdot\vec P(\vec x,t)&=\rho_{\rm b}(\vec x,t)\,,\label{eq_inhomo1}\\[5pt]
\nabla\times\vec M(\vec x,t)+\partial_t\vec P(\vec x,t)&=\vec j_{\rm b}(\vec x,t)\,,\label{eq_inhomo2}
\end{align}
where $\rho_{\rm b}$ and $\vec j_{\rm b}$ are mostly called the {\it bound} charge and current densities 
(as for alternative designations and other variants of these equations, see the discussion below).

\subsection{Problems of the Standard Approach}\label{subsec_problems}

Before we come to the development of the Functional Approach to electrodynamics
of materials, we systematically elaborate in this subsection on the shortcomings of the Standard Approach. We thereby restrict ourselves to the most important {\itshape conceptual} problems. 
As we have shown before in Refs.~\cite{ED1, EDOhm, Refr}, the Standard Approach also leads to {\itshape practical}
problems, which regard in particular the description of bianisotropic materials \cite{ED1} and the relativistic covariance \cite{EDOhm}. 
Furthermore, the Standard Approach implies a wrong formula for the refractive index~\cite{Refr}.
On the conceptual side, the most important problems of the standard approach will turn out to be: (i)
the incompleteness of its fundamental field equations, and (ii) the ambiguity of the source splitting prescription. These two problems will be discussed in Subsecs.~\ref{prob_1} and \ref{sec_amb}, respectively.

\subsubsection{Incomplete field equations} \label{prob_1}

From a field theoretical point of view, the fundamental field equations of the Standard Approach are incomplete. 
Generally, field equations have the purpose of determining all the fundamental fields, 
in terms of which a field theory is defined, from some given initial data (i.e., the initial and boundary conditions on the fields). 
In the case of electrodynamics, this means that the field equations should determine the electromagnetic fields as 
functionals of their initial data as well as the charge and current densities. We will now explain this problem in some detail. 
For this purpose, we introduce the total charge and current densities~by
\begin{align}
\rho\tot(\vec x,t)&:=\rho_{\rm f}(\vec x,t)+\rho_{\rm b}(\vec x,t)\,,\label{eq_totalold1}\\[5pt]
\vec j\tot(\vec x,t)&:=\vec j_{\rm f}(\vec x,t)+\vec j_{\rm b}(\vec x,t)\,.\label{eq_totalold2}
\end{align}
By the fundamental field equations of the Standard Approach, Eqs.~\eqref{eq_maxwell_3}--\eqref{eq_inhomo2}, it then follows immediately that the fields \{$\vec E, \vec B$\} obey
\begin{align}
\nabla\cdot\vec E(\vec x,t)&=\rho\tot(\vec x,t)/\varepsilon_0\label{eq_MaxTot1}\\[3pt]
\nabla\times\vec E(\vec x,t)&=-\partial_t\vec B(\vec x,t)\,,\label{eq_MaxTot2}\\[3pt]
\nabla\cdot\vec B(\vec x,t)&=0\,,\label{eq_MaxTot3}\\[3pt]
\nabla\times\vec B(\vec x,t)&=\mu_0 \h \vec j\tot(\vec x,t)+\varepsilon_0 \h \mu_0 \h \partial_t\vec E(\vec x,t)\,.\label{eq_MaxTot4}
\end{align}
Given appropriate boundary and initial conditions, these
equations determine the fields \{$\vec E, \vec B$\} uniquely in terms of the sources \{$\rho\tot, \h \vec j\tot$\}. 
They coincide with the electric and magnetic fields generated by the total charge and current densities, and hence we may write
\begin{align}
 \vec E & \equiv \vec E\tot \,, \\[4pt]
 \vec B & \equiv \vec B\tot \,.
\end{align}
However, the situation is completely different for the remaining fields $\{\vec D,\vec H, \linebreak \vec P,\vec M\}$.
Concretely, by Eqs.~\eqref{eq_inhomo1}--\eqref{eq_inhomo2} we are only given relations for
the longitudinal part of $\vec P$ and the transverse part of $\vec M$, while we lack field equations
for the transverse part of $\vec P$ and the longitudinal part of $\vec M$. Similarly, we lack
field equations for the transverse part of $\vec D$ and the longitudinal part of $\vec H$ (see \cite[Sec.~2.1]{ED1} for the definition of longitudinal and transverse parts of a three-dimensional vector field). 

Furthermore, let us introduce the fields $\vec E_{\rm b}$ and $\vec B_{\rm b}$ analogously to Eqs. \eqref{eq_MaxTot1}--\eqref{eq_MaxTot4} by
\begin{align}
\nabla\cdot\vec E_{\rm b}(\vec x,t)&=\rho_{\rm b}(\vec x,t)/\varepsilon_0\label{eq_MaxBound1}\\[3pt]
\nabla\times\vec E_{\rm b}(\vec x,t)&=-\partial_t\vec B_{\rm b}(\vec x,t)\,,\label{eq_MaxBound2}\\[3pt]
\nabla\cdot\vec B_{\rm b}(\vec x,t)&=0\,,\label{eq_MaxBound3}\\[3pt]
\nabla\times\vec B_{\rm b}(\vec x,t)&=\mu_0 \h \vec j_{\rm b}(\vec x,t)+\varepsilon_0 \h \mu_0\h \partial_t\vec E_{\rm b}(\vec x,t)\,.\label{eq_MaxBound4}
\end{align}
These are the electric and magnetic fields generated by the bound charges and currents. Correspondingly, Eqs.~\eqref{eq_inhomo1}--\eqref{eq_inhomo2} 
then suggest that
\begin{align}
 \vec P_{\rm L} & = -\varepsilon_0 \h (\vec E_{\rm b})_{\rm L} \,, \label{eq_Id1}\\[5pt]
 \vec M_{\rm T} & = (\vec B_{\rm b})_{\rm T} / \mu_0 \,, \label{eq_Id2}
\end{align}
where the subscripts $\mathrm L$ and $\mathrm T$ denote the longitudinal and transverse parts, respectively (see \cite[Sec.~2.1]{ED1}). In fact, the relation \eqref{eq_Id1} between the polarization and the electric field is a strict identity following from Eqs.~\eqref{eq_inhomo1}, \eqref{eq_MaxBound1} and the
Helmholtz vector theorem (see Eqs.~\eqref{eq_HVT_1}--\eqref{eq_HVT}). The corresponding relation \eqref{eq_Id2} between the magnetization and the magnetic field follows similarly from Eqs.~\eqref{eq_inhomo2} and \eqref{eq_MaxBound4}, but only in the static case
where the time derivatives of the fields vanish.
On the other hand, there is apparently no reason why this identification should break down in the dynamical case (which would imply that---even worse---both
the longitudinal and the transverse part of the magnetization are undetermined).
Thus, we conclude that the longitudinal part of the polarization $\vec P$ coincides with the longitudinal part of the electric field $\vec E_{\rm b}$\h, and the transverse part of the magnetization $\vec M$ coincides with the transverse part of the magnetic field $\vec B_{\rm b}$ (up to the conversion factors).  However, while the fields $\{\vec E_{\rm b}, \vec B_{\rm b}\}$ are completely determined by the
Maxwell equations \eqref{eq_MaxBound1}--\eqref{eq_MaxBound4}, the quantities
$\vec P_{\rm T}$ and $\vec M_{\rm L}$ are undefined in the \mbox{Standard Approach.}

This, of course, immediately raises the question of why one should characterize the properties of a material by the underdetermined quantities $\{\vec P,  \vec M\}$, 
while the real and in principle measurable (once the distinction between free and bound charges is specified, see Sec.~\ref{sec_amb}) 
electric and magnetic fields $\{\vec E_{\rm b}, \vec B_{\rm b}\}$ do not play any r\^{o}le in the Standard Approach.
In this already alarming situation, the critique of the Standard Approach is even more exacerbated once we remember
that the ``dipole densities'' $\vec P$ and $\vec M$ are measurable precisely in as far as they themselves are sources
of electric and magnetic fields. In particular, thence the notorious problem arises of finding the electric and magnetic fields of (e.g.~homogeneously)
polarized bodies (see App.~\ref{app_const}). Thus, within the Standard Approach the physical picture amounts to the following: electromagnetic fields induce
in the material underdetermined dipole densities, which in turn are sources of electromagnetic fields.
However, from this vantage point it is completely incomprehensible why the theory should not be based directly on the electromagnetic fields as generated by the material under the action of externally applied \mbox{perturbations.}

Going one step further, we may introduce also the fields $\vec E_{\rm f}$ and $\vec B_{\rm f}$ 
as the electric and magnetic fields generated by the free sources. 
These satisfy analogous equations to \eqref{eq_MaxBound1}--\eqref{eq_MaxBound4}, but with the free instead of the bound charge and current densities. 
By the linearity of Maxwell's equations, we further have the relations
\begin{align}
 \vec E_{\rm f}(\vec x, t) + \vec E_{\rm b}(\vec x, t) & = \vec E_{\rm tot}(\vec x, t) \,, \\[5pt]
 \vec B_{\rm f}(\vec x, t) + \vec B_{\rm b}(\vec x, t) & = \vec B_{\rm tot}(\vec x, t) \,.
\end{align}
Now, by the same logic as above, Eqs.~\eqref{eq_maxwell_1}--\eqref{eq_maxwell_2} also imply the identities
\begin{align}
 \vec D_{\rm L} & = \varepsilon_0 \h (\vec E_{\rm f})_{\rm L} \,, \\[5pt]
 \vec H_{\rm T} & = (\vec B_{\rm f})_{\rm T} / \mu_0 \,.
\end{align}
In Tables \ref{tab_standard_1} and \ref{tab_standard_2}, all the field equations and the ensuing field identities of the Standard Approach are summarized. Already at first glance, 
these tables suggest that the field equations of the Standard Approach can be completed by identifying also the transverse parts of $\vec P$ and $\vec D$ with the transverse parts of the respective electric fields, and the longitudinal parts of $\vec M$ and $\vec H$ with the longitudinal part of the magnetic field (which is zero by Eq.~\eqref{eq_maxwell_3}). In fact, such field identifications---but with  {\itshape induced} and {\itshape external} instead of bound and free quantities---will form the basis of the Functional Approach to electrodynamics of media (see Tables \ref{tab_functional_1} and \ref{tab_functional_2} in Sec.~\ref{sec_functional}).

Before introducing this approach in the next section, let us discuss how one usually deals in the Standard Approach with the problem of the incomplete field equations. In fact, there are at least three competing approaches in the literature which deal with this noteworthy situation: 
(i) the stipulation of a ``gauge freedom'' in the definition of $\{\vec P,\vec D\}$ and $\{\vec M,\vec H\}$, 
(ii) the use of constitutive relations for providing a unique definition of these quantities, and (iii) 
the definition of $\{\vec P, \vec M\}$ in terms of ``dipole densities''. We now briefly describe these 
three approaches and show that none of them provides a satisfactory solution to this problem.

\begin{enumerate}[listparindent=\parindent,parsep=0pt,itemsep=1em]
\item[(i)]{\itshape Gauge freedom of the polarizations?} References \cite{Hirst,Dolgov,Smith,Chatel,Toptygin,Kirzhnitz} stipulate a ``gauge freedom'' in the definition of $\{\vec P,\vec D\}$ and $\{\vec M,\vec H\}$,
meaning that the transverse and respectively longitudinal parts of these pairs can be arbitrarily altered, while only the respective sums, 
\begin{align}
\nabla\times(\vec D(\vec x,t)-\vec P(\vec x,t))&=\varepsilon_0 \h \nabla\times\vec E(\vec x,t)\,,\\[5pt]
\nabla\cdot(\vec M(\vec x,t)+\vec H(\vec x,t))&=\mu_0^{-1} \, \nabla\cdot\vec B(\vec x, t) \,,
\end{align}
are uniquely defined by the homogeneous Maxwell equations \eqref{eq_maxwell_3}--\eqref{eq_maxwell_4}. Correspondingly, 
some textbooks (see e.g.~\cite[Eq.~(6.23)]{Griffiths}, \cite[p.~70]{SlaterFrank}, 
\cite[Eq.~(8-9)]{Panofsky}, \cite[Eqs.~(13.40), (13.55)]{Zangwill} and \cite[Eq.~(B.25)]{Blundell}) explicitly state that the divergences of $\vec M$ and $\vec H$
are in general nonzero, but restricted to obey
\begin{equation}
\nabla\cdot\vec H(\vec x,t)=-\nabla\cdot\vec M(\vec x,t)\,.
\end{equation}
Similar statements can be found for the rotations of $\vec P$ and $\vec D$ (see e.g. \cite[p.~47]{Kirzhnitz} or \cite[p.~228]{Dolgov}).

This state of affairs on the {\itshape theoretical side} is of course not satisfactory, because in the {\itshape experiment} \h $\vec P$ and $\vec M$ 
are identified with observable quantities (see e.g.~\cite{Hur, Murakawa, Ishiwata, Chiba, Tesarova}).
Furthermore, this would necessitate the stipulation of a gauge freedom of the physical response functions as well: Consider, for example, the dielectric tensor defined through
\begin{equation} \label{diel}
 \vec D = \varepsilon_0 \hspace{0.5pt} \tsr \varepsilon_{\rm r} \h \vec E \,.
\end{equation}
If $\vec P$ and $\vec D$ were defined only up to gauge transformations, then also the dielectric tensor would not be defined uniquely by Eq.~\eqref{diel}. In particular, 
in the frequently evoked case of a homogeneous and isotropic medium, the (wavevector- and frequency-dependent) 
dielectric tensor can be written in terms of the longitudinal and transverse projection operators (see e.g.~\cite{Giuliani} or \cite[Sec.~2.1]{ED1}) as
\begin{equation}
 \tsr \varepsilon_{\rm r}(\vec k, \omega) = \varepsilon_{\rm r, \hh L}(\vec k, \omega) \h \tsr P_{\rm L}(\vec k) + \varepsilon_{\rm r, \hh T}(\vec k, \omega) \h \tsr P_{\rm T}(\vec k) \,.
\end{equation}
In this case, Eq.~\eqref{diel} implies that the longitudinal part of $\vec D$ is related to the longitudinal part of $\vec E$ by the {\itshape longitudinal dielectric function}~$\varepsilon_{\rm r, \hh L}$, while the {\itshape transverse dielectric function} $\varepsilon_{\rm r, \hh T}$ mediates between the 
corresponding transverse parts:
\begin{align}
 \vec D_{\rm L}(\vec k, \omega) & = \varepsilon_0 \, \varepsilon_{\rm r, \hh L}(\vec k, \omega) \, \vec E_{\rm L}(\vec k, \omega) \,, \\[5pt]
 \vec D_{\rm T}(\vec k, \omega) & = \varepsilon_0 \, \varepsilon_{\rm r, \hh T}(\vec k, \omega) \, \vec E_{\rm T}(\vec k, \omega) \,.
\end{align}
Now, if the transverse part of the displacement field was arbitrary, so would be the transverse dielectric function. 
This, however, does not make sense, because the latter is actually a significant material property 
which is routinely determined at optical frequencies \cite{Kliewer69, Rajagopal, Bagchi, Kirk, Cockayne}. In fact, 
due to the transverse nature of electromagnetic waves, it is sometimes claimed that the transverse dielectric function is even
the {\itshape only} response function which is directly measurable in {\itshape optical} experiments \cite[p.~274]{Marel}. With a gauge freedom of the transverse electric polarization, however, this quantity would completely lose its meaning.

\item[(ii)] {\itshape Polarizations defined by constitutive relations?} Often, it is argued that the ambiguity of the polarization and 
magnetization can be overcome by postulating specific {\itshape constitutive relations} or {\itshape material relations} 
(see e.g.~\cite[Eqs.~(1.51)--(1.52)]{Melia} or \cite[Secs.~4.8.4 and 5.7.1]{RebhanEdyn})
\begin{align}
\vec P&=\vec P[\vec E,\vec H]\,,\\[5pt]
\vec M&=\vec M[\vec E,\vec H]\,.
\end{align}
In the simplest case, these relations read
\begin{align}
\vec P&=\varepsilon_0 \h \tsr\chi_{\rm e} \h \vec E\,,\label{eq_MatRel10}\\[2pt]
\vec M&=\tsr\chi_{\rm m} \h \vec H\,,\label{eq_MatRel20}
\end{align}
with the electric and magnetic susceptibility tensors $\chi_{\rm e}$ and $\chi_{\rm m}$.

This argument is not consistent either, because it would require the constitutive relations to serve a double purpose: On the one hand, they would
relate the electric and magnetic polarizations to the electric and magnetic fields and therefore constitute
a statement about the actual behavior of the material. On the other hand, they would partly {\itshape define} what
the respective polarizations even mean, and in this respect they would {\itshape not} constitute a statement about
the behavior of the material. In particular, we would run into the same problems as in the case of the alleged ``gauge freedom'' of the polarizations: 
If Eq.~\eqref{eq_MatRel10} is part of the definition of $\vec P$, then the question arises of how the tensor $\chi_{\rm e}$ (in particular its transverse part in a homogeneous and isotropic medium) is defined. 
We conclude that Eq.~\eqref{eq_MatRel10} is not sufficient for defining both $\vec P$ and $\chi_{\rm e}$ at the same time, and hence in any case the susceptibility tensor remains undefined in the Standard Approach.

\item[(iii)]{\itshape Polarizations defined as dipole densities?}
The majority of the traditional textbooks introduces the polarizations $\vec P$ and $\vec M$ as ``dipole moment[s] per unit volume'' \cite[p.~166 and p.~262]{Griffiths}
or ``mean dipole moments per volume'' \cite[p.~190]{Wachter} (see also \cite[p.~192]{Ibach}, \cite[pp.~215, 238]{FitzpatrickED} and \cite[p.~67]{SlaterFrank}). As to the lacking field equations, the textbook by L.\,D. Landau and E.\,M.~Lifshitz states: 
``The exact form of $\vec P$ can be completely determined only by establishing its connection with the dipole moment'' \cite[footnote on p.~35]{Landau}.

However, the na\"{i}ve picture of elementary dipoles which orient themselves in the electromagnetic fields as described, for example, by the Clausius-Mossotti model \cite{Mossotti}, 
has already been proven to be unrealistic for the vast majority of materials by the Modern Theory of Polarization \cite{Vanderbilt, Resta07, Resta10}. In particular,
 the conception of the, say, electric polarization $\vec P$ as a ``dipole density'' appears extremely dubious in the case of the oftentimes
highly delocalized ``liquid'' \cite{Giuliani} of electrons in the solid state. This is all the more problematic since,  more often than not, the dielectric properties of a solid
are dominated precisely by the electronic contributions (as opposed to the nuclear contributions).

Even worse, it is in general not clear how a ``dipole density'' can actually be defined as a field $\vec P(\vec x, t)$ on space-time in terms of the microscopic charge and current densities $\rho(\vec x, t)$ and $\vec j(\vec x, t)$. In fact, in App.~\ref{app_dip} we show that the attempt to define the polarization and magnetization as continuous densities of electric and magnetic {\itshape point dipoles} leads precisely back to the original problem of insufficiently defined fields. Moreover, it is unclear why such dipole densities would contribute to the total electric and magnetic fields as suggested by Eqs.~\eqref{eq_PED}--\eqref{eq_MBH}.
For this reason, some textbooks even reject Eqs.~\eqref{eq_PED}--\eqref{eq_MBH} as fundamental equations and instead
regard them only as ``dipolar approximations'' (see \cite[Eq.~(I.9)]{Jackson} and in particular the textbook by R.\,E.~Raab and O.\,L.~de Lange \cite{Raab}).

Finally, as there are no fundamental, microscopic ``dipole fields'', one has to restrict in this approach the validity of the Maxwell equations in media to the macroscopic domain.
The terms `macroscopic electrodynamics' and `electrodynamics in media' are therefore often used interchangeably (see remarks in the introduction).
As a macroscopic theory though, Maxwell's equations in media are not fundamental anymore, and instead 
necessitate a complicated derivation from seemingly more fundamental {\it microscopic Maxwell equations}
by means of intricate averaging procedures. In particular, the field equations \eqref{eq_inhomo1}--\eqref{eq_inhomo2}
now have to be derived from the definitions of $\vec P$ and $\vec M$ as ``dipole densities''.
This in turn is very cumbersome and often involves assumptions about the behavior of the material (see e.g.~\cite[pp.~188ff.]{Wachter}). However, on the face of it, Eqs.~\eqref{eq_inhomo1}--\eqref{eq_inhomo2} for the fields $\vec P$ and $\vec M$ are simply the counterparts
of the respective equations~\eqref{eq_maxwell_1}--\eqref{eq_maxwell_2} for $\vec D$ and $\vec H$. Hence, it is  incomprehensible why the latter equations rank among the axioms, while the former have to be derived in the Standard Approach.

\end{enumerate}

\noindent
These considerations show that the Standard Approach is in need of 
field equations for the longitudinal part of the magnetization $\vec M$ and the transverse part of the polarization $\vec P$.
\medskip

\subsubsection{Ambiguous source splitting} \label{sec_amb}

Another problem of the Standard Approach lies in the splitting of the sources into ``free'' and ``bound'' parts, the nature of which is not clearly specified.
In fact, there is not even a consensus about this splitting anyway, and one can find nearly as many different interpretations as there 
are textbooks on classical electrodynamics. We will nonetheless try to group them into certain classes:

\renewcommand\labelitemi{\small\raisebox{1pt}{$\bullet$}\,}

\begin{itemize}[listparindent=\parindent,parsep=0pt,itemsep=1em]
\item Some textbooks split the sources {\it within the material} into ``free'' and ``bound'' parts. These are then associated with different charge carriers, which can either ``move almost freely within the crystal lattice'' or are ``bound within atoms and molecules'' \cite[p.~186]{Wachter} (see also
 \cite[p.~251]{Jackson}, \cite[p.~199]{Blundell} and \cite[p.~270]{Honer}).
In semiconductor physics, the free charges are sometimes associated with 
 the ``conduction electrons and valence band holes'', whereas ``[t]he bound charges are cemented into the lattice structure; the atomic charges comprising the lattice itself, and the inner electrons tightly localized at the atomic cores'' \cite[p.~9]{Perkowitz} 
 (see also \cite[p.~203]{Cardona} or \mbox{\cite[p.~19]{Roemer}).}

\item By contrast, the traditional textbook by D.\,J.~Griffiths identifies ``free'' contributions with ``those sources we control directly'', while bound are the ones
``over which you exert no direct control'' \cite[p.~328]{Griffiths}. For the current, an even more elaborate
splitting into ``free'', ``bound'' and ``polarization'' parts is used \cite[p.~329]{Griffiths}. On another occasion, we then learn that
``[t]he field due to magnetization of the medium is just the field produced by these bound currents'', while the free current corresponds to ``everything else''.
However, ``it is simply a {\itshape convenience} to separate the current into
these two parts because they {\itshape got} there by quite different means: the free current is there
because somebody hooked up a wire to a battery---it involves actual transport of charge; the
bound current is there because of magnetization---it results from the conspiracy of many
aligned atomic dipoles'' \cite[p.~269, emphasis in original]{Griffiths}. Inter alia, this leads to the replacement of Eq.~\eqref{eq_inhomo2} in this article by the more complicated Eq.~(7.50) in Ref.~\cite{Griffiths}:
\begin{equation}
 \nabla \times \vec M(\vec x, t) + \partial_t \vec P(\vec x, t) = \vec j_{\rm b}(\vec x, t) + \vec j_{\rm p}(\vec x, t) \,, \label{eq_Griffiths}
\end{equation}
which involves an extra current $\vec j_{\rm p}$ (``polarization current'').

 Similarly, Yu.\,A.~Il'inskii and L.\,V.~Keldysh (who are, in fact, rather critical of the Standard Approach)
confirm that the current is split into the ``current of free charges (conduction electrons, etc.)'', 
the ``polarization current [\ldots] due to the motion of bound charges'' and the ``magnetization vortex current'' \cite[p.~14]{IlinskiiKeldysh}.

\item Other textbooks replace Eq.~\eqref{eq_inhomo2} by a simplifed version without the displacement part (see e.g.~\cite[Eq.~(B.17)]{Blundell}), i.e.,
\begin{equation}
\nabla\times\vec M(\vec x,t)=\vec j_{\rm b}(\vec x,t)\,.
\end{equation}
Another version which instead neglects $\vec M$ in Eq.~\eqref{eq_inhomo2} in the sense of
\begin{equation}
\partial_t\vec P(\vec x,t) = \vec j_{\rm p}(\vec x,t)
\end{equation}
can also be found often in the literature (see e.g.~\cite[p.~7]{Keldysh}, \cite[p.~16]{IlinskiiKeldysh}), and together the above
equations would precisely reimply Eq.~\eqref{eq_Griffiths}.
Correspondingly, it is often said that the first equation applies to a ``magnetization current'' $\vec j_{\rm m}$, while the second one
describes the ``polarization current'' $\vec j_{\rm p}$ (see \cite[Eq.~(22.4)]{Martin}, \cite[p.~16]{IlinskiiKeldysh} or \cite[Eqs.~(6.109) and (6.116)]{FitzpatrickED}).

The textbook on quantum magnetism by W.~Nolting and A.~Ramakanth \cite[pp.~3ff.]{Nolting} states that the free current originates ``from the free charge carriers (electrons)'',
while the bound current originates ``from the localized ions'', and the bound current 
is then further split into a polarization and a magnetization current. This should be contrasted, however, with the statement of D.\,J.~Griffiths \cite[p.~329]{Griffiths} that
``[t]his polarization current [fulfilling $\vec j_{\rm p}=\partial_t\vec P$] has nothing whatever to do with the bound current $\vec j_{\rm b}$\hh.
The latter is associated with magnetization of the material and involves the spin and orbital motion of electrons; $\vec j_{\rm p}$\hh,
by contrast, is the result of the linear motion of charge when the electric polarization changes [notation adapted].''

\item The famous optics textbook by M.~Born and E.~Wolf even associates the electric and magnetic polarizations via  Eqs.~\eqref{eq_inhomo1}--\eqref{eq_inhomo2}
 with the {\it free} (and not with the {\it bound}) charges and currents \cite[Sec.~2.2.1, \linebreak Eqs.~(7)--(8)]{BornWolf}.

\item Some textbooks use a terminology whose consistency 
with the traditional splitting into ``free'' and ``bound'' sources is not always clear. For example, the textbook by F. Melia \cite[p.~14]{Melia} decomposes 
the sources ``in matter'' into ``free'' and ``bound'' contributions, but then goes on to say that the free sources ``give rise to $\vec E$ and $\vec B$ in vacuum'', 
while the bound sources represent ``the response of the medium to the presence of the fields''. G.~Brooker splits the source terms into ``accessible'' and ``polarization'' parts \cite[p.~2]{Brooker},
where the `accessible' charge density is ``the space average of those charges that we choose not to pair off into dipoles'',
and ``[t]he description `accessible' arises because these charges are not (necessarily) bound inside atoms or molecules;
a conduction current inside a metal is the commonest example of an `accessible' current.''
Correspondingly, M.~Dressel and G.~Gr\"{u}ner split the sources into ``bound'' and ``conductive'' contributions from the very outset \cite[Eq.~(2.2.3)]{Dressel}.
The textbook by A.~Kovetz uses a splitting into ``free'' and ``response'' parts,
but then associates the ``response'' with bound charges \mbox{and currents \cite[pp.~76ff.]{Kovetz2000}.}

The classic textbook by W.\,K.\,H.~Panofsky and M.~Phillips \cite[p.~129]{Panofsky} uses yet another subtle distinction:
``Currents may be classified in
two categories: true currents that may be identified with the motion of true
charges, and other currents which are associated with the medium itself.
This separation [\ldots] will lead us to consider two types of magnetic fields, one derived from true currents and the other derived from the
combined effects of all the currents whatever may be their origin. It is
this latter field, namely, the field of magnetic induction $\vec B$, that can be
considered to be the space-time average of the interatomic fields.''
Correspondingly, in addition to ``true'' currents, ``polarization'' currents and ``magnetization'' currents are introduced, where the latter are defined as 
``stationary currents that flow within regions that are inaccessible to observation but which might give rise to net
boundary or volume currents, due to imperfect orbit cancellation on an atomic scale''. Finally mentioned are also ``convective'' currents, whose expla\-{}nation is this: ``If a material medium in motion contains charges of various
types, additional currents will be obtained which arise from convective
effects. These convective currents will be derived from the motion of
both true and polarization charges contained in the medium.''

The modern textbook by R.~Fitzpatrick states \cite[p.~216]{FitzpatrickED}
 that $\rho_{\rm b}(\vec x)$ ``is attributable to {\itshape bound charges} (i.e., charges which
 arise from the polarization of neutral atoms), and is usually distinguished
 from the charge density $\rho_{\rm f}(\vec x)$ due to {\itshape free charges,} which typically represents a net surplus or deficit of electrons in the medium''.
 This textbook also introduces (p.~239) a ``true'' current density in addition to the magnetization and the polarization currents, which is explained as
``that part of the current density which is due to the movement
of free charges''. However, ``[i]t must be emphasized that all three terms
represent real physical currents, although only the first term [true current] is due to
the motion of real charges (over more than molecular dimensions).''

Equally interesting is this approach \cite[p.~305]{Kantorovich}: ``The average current can be represented as a sum of three terms. First of all, there
is a macroscopic current $\vec j$ due to external sources (charges). Secondly, there is a
current due to the fictitious charge (the so-called polarisation current) which is equal
to $\partial\vec P/\partial t$. Finally, there is an additional current related to the magnetic moment $\vec M$ per
unit volume. This term is especially important e.g.~in superconductors.''

However, the textbook \cite[p.~615]{Marder} confirms that the traditional splitting is actually this:
``Charge is divided into two groups, bound and free. The bound charges
produce dielectric behavior, while the free charges participate in conductivity. In
addition, materials have a magnetic permeability $\mu$ that relates the microscopic
field $\vec B$ to a macroscopic field $\vec H$ [notation adapted].'' Ultimately, however, the critical position
of \cite{IlinskiiKeldysh} is joined by stating that ``[t]he divisions between bound and free charge are not fundamental'' and that ``[t]here is nothing necessary about such a division [\ldots]'' \cite[p.~615]{Marder}.

\item Finally, we also mention that according to the so-called ``premetric approach'' to electrodynamics \cite{Truesdell, Hehl1, Hehl_depl}, the introduction of $\{\vec D,\vec H\}$ in addition to $\{\vec E,\vec B\}$ is not due to the presence of materials,
but instead these quantities refer to fundamentally different fields even on the microscopic level (namely to ``excitations'' and ``field strengths'', respectively).
Similarly, in string theoretical contexts one sometimes assumes that the macroscopic Maxwell equations 
``are as fundamental as the original [i.e.~microscopic] Maxwell equations, if not more so'' \mbox{\cite[p.~434]{Zwiebach}.}

\end{itemize}

\noindent
In addition, we note that with all these different interpretations of the source splitting, 
it becomes completely unclear which fields describe a response and which describe a cause. For example, the textbook \cite[Eq.~(1.0.1)]{Dressel} 
associates the displacement field $\vec D$ with the response of the solid. Similarly, in Ref.~\cite[p.~9]{Keldysh}
the displacement field $\vec D$ is regarded as ``the response to the electric field'' (see also \cite[Eq.~(7.2)]{Melrose}). 
The textbook \cite[p.~24]{Raab} even calls $\vec D$ and $\vec H$ the ``response fields'', and another textbook
\cite[Eq.~(20.34)]{Marder}  states that ``the electric displacement is produced in response to {\it external} fields'' [emphasis added].
By contrast, according to the textbook \cite[p.~306]{Kantorovich}
``the fields $\vec D$ and $\vec H$ are determined by the macroscopic charges and currents. These fields can be associated with {\it external}
(with respect to the material) {\it sources}'' [emphasis in original]. Similarly, \cite[p.~493]{Martin} states that ``$\vec D$ is the field due only to external sources'', in particular,
``the value of $\vec D$ at any point is independent of the material and is the same as if the material were absent''.
On the other hand, according to \cite[p.~8]{FoxQO} $\vec D$ and $\vec H$ ``include the effects \mbox{of the medium''.}

Correspondingly, there is also an enormous confusion concerning the dielectric function, namely the question: is the dielectric function itself a causal (retarded) response function
or rather its inverse? Traditional textbooks usually treat $\varepsilon$ as the response function and apply the Kramers-Kronig relations to it, see e.g.~\cite[Sec.~7.10]{Jackson}.
There, it is also stated explicitly that: ``at time~$t$ only values of the electric field {\itshape prior} to that time enter in determining the displacement,
in accord with our fundamental ideas of causality in physical phenomena'' \cite[p.~332]{Jackson}.
Similarly, in Ref.~\cite[Sec.~6.1.3]{Cardona} it is
affirmed that ``$\varepsilon(\omega)$ describe[s] the linear response of a medium to an external [sic!] field''.
Conversely, the important textbook \cite[p.~193]{Giuliani} in electronic structure theory states that ``$\varepsilon(\vec k,\omega)$
is not, in general, a causal response function'', and correspondingly, ``the Kramers-Kronig dispersion relations apply to $1/\varepsilon(\vec k,\omega)$
but not to $\varepsilon(\vec k,\omega)$ [notation adapted]''. This is also confirmed in Ref.~\cite{DolgovArticle}, 
and in the similar vein are the comments in Ref.~\cite[p.~285]{Kantorovich}.
In other words, some textbooks say that it is precisely the inverse of $\varepsilon$ 
which is the response function, and correspondingly the field $\vec D$ has to be considered the ``cause'' or ``perturbation'', but not the ``effect'' or ``response''. 

In summary, the problematic features of the Standard Approach concerning the splitting of the source terms are:
(i) competing---partly contradicting, partly synonymous, partly incomprehensible---designations, (ii) inter\-{}mingling of different contradistinctions
(external vs.~internal, microscopic vs. macroscopic, free vs.~bound, electronic vs.~ionic,
true vs.~fictitious, observable vs.~inobservable, conductive vs.~isolating, charged vs.~neutral, etc.).
These considerations show that the Standard Approach is in need of an unambiguous source-splitting prescription.
In particular, with regard to applications in ab initio materials physics, it should be clear what this source splitting means on the level of the
many-body Schr\"{o}dinger equation for electrons and nuclei (see in this context Ref.~\cite{EffWW}). To this problem we now turn.

\begin{landscape}

\begin{table}[t]
\begin{center}
\renewcommand{\arraystretch}{1.6}
\begin{tabular}{p{3.5cm}p{3.5cm}p{3.0cm}p{4.6cm}}
\toprule
$\nabla\cdot \vec E = \rho\tot/\varepsilon_0$ & $\nabla\times\vec E = -\partial_t\vec B$ & $\nabla\cdot\vec B = 0$ & $\nabla\times \vec B = \mu_0 \h \vec j\tot+\partial_t \vec E/c^2$ \\[3pt]
\midrule
$\nabla\cdot\vec D=\rho_{\rm f}$ & $\nabla\times\vec D= \,\vec ?$ & $\nabla\cdot\vec H=\,\vec ?$ &
$\nabla\times\vec H=\vec j_{\rm f}+\partial_t\vec D$ \\[3pt]
\midrule
$\nabla\cdot \vec P=-\rho_{\rm b}$ & $\nabla\times\vec P=\,\vec ?$ & $\nabla\cdot\vec M=\,\vec ?$ & 
$\nabla\times\vec M=\vec j_{\rm b} - \partial_t \vec P$ \\[3pt]
\bottomrule
\end{tabular}

\caption{Field equations of the Standard Approach. \label{tab_standard_1}}

\bigskip
\bigskip
\bigskip
\bigskip
\begin{tabular}{p{3.5cm}p{3.5cm}p{3.0cm}p{4.6cm}}
\toprule
$\vec E_{\rm L} = (\vec E\tot)_{\rm L}$ & $\vec E_{\rm T} = (\vec E\tot)_{\rm T}$ & $\vec B_{\rm L} = (\vec B\tot)_{\rm L}$ & $\vec B_{\rm T} = (\vec B\tot)_{\rm T}$ \\
\midrule
$\vec D_{\rm L} = \varepsilon_0 \h (\vec E_{\rm f})_{\rm L}$ & $\vec D_{\rm T} = \ \vec ?$ & $\vec H_{\rm L} = \ \vec ?$ & $\vec H_{\rm T} = (\vec B_{\rm f})_{\rm T} \h /\mu_0$\\
\midrule
$\vec P_{\rm L} = -\varepsilon_0 \h (\vec E_{\rm b})_{\rm L}$ & $\vec P_{\rm T} = \ \vec ?$ & $\vec M_{\rm L} = \ \vec ?$ & $\vec M_{\rm T} = (\vec B_{\rm b})_{\rm T} \h/\mu_0$\\
\bottomrule
\end{tabular}

\caption{Field identities of the Standard Approach.\label{tab_standard_2}}
\end{center}
\end{table}

\end{landscape}

\section{Functional Approach to electrodynamics of media} \label{sec_functional}

In contrast to the Standard Approach, the Functional Approach to be developed in this section is an inherently microscopic theory. In this context, we stress that the problem with 
electrodynamics of materials does not lie in the impossibility of solving the microscopic Maxwell equations (cf.~\cite[\mbox{pp.~1f.}]{Nolting}).
Quite to the contrary, the general analytical solution to the initial value problem
of Maxwell's equations {\it with given sources} is actually well known (see the short discussion in \cite[Sec.~3.2]{ED1}). 
Instead, the problem therefore lies in the calculation of the source terms in Maxwell's equations,
i.e., of the currents and charges induced under an applied external perturbation.
On the other hand, calculating these induced sources from first principles requires a microscopic theory.
Hence, we are in need of a microscopic theory of electrodynamics in materials, which reflects both the
experimental situation and the common practice in computational materials physics. This theory is provided by the Functional Approach to electrodynamics of media, 
which has been developed systematically by the authors of the present article in Ref.~\cite{ED1} and extended to the relativistic domain in Ref.~\cite{EDOhm}. 
In this section, we will motivate this Functional Approach from first principles and discuss its conceptual and practical implications.

\subsection{Derivation from first principles} \label{subsec_derivations}

In the following, we stick to the conventions of Ref.~\cite{ED1}. In particular, we use SI units \cite{SI}. Moreover, we choose the Minkowski metric as
 \begin{equation} \label{metric}
 \eta_{\mu\nu} = \eta^{\mu\nu} = \mathrm{diag}(-1,\,1,\,1,\,1) \,,
 \end{equation}
and we sum over all doubly appearing indices. We define response functions in such a way that they are compatible with the relativistic volume element,
\begin{equation}
 \int \! \de^4 x = \int \! \de^3 \vec x \int \! c \, \de t \,,
\end{equation}
and, correspondingly, we use the relativistic Dirac delta
\begin{equation}
 \delta^4(x - x') = \delta^3(\vec x -\vec x') \, \delta(c \h t - c \h t') \,,
\end{equation}
in the definition of Green functions and other integral kernels.

\subsubsection{Unique source splitting}

In Sec.~\ref{subsec_problems} we have shown that one of the foremost problems of the Standard Approach to electrodynamics in media lies
in the somewhat mysterious splitting of the source terms in Maxwell's equations into ``free'' and ``bound'' contributions. 
Therefore, we start the development of the Functional Approach with the motivation of a different splitting, which is suitable for a microscopic theory
but as such applies to the macroscopic domain as well: the splitting into {\it induced} and {\it external} fields.

For this purpose, we first note that in the absence of materials, the constitutive equations \eqref{eq_MatRel10}--\eqref{eq_MatRel20} 
with $\chi_{\rm e} = \chi_{\rm m} = 0$ and Eqs.~\eqref{eq_PED}--\eqref{eq_MBH} to-\linebreak gether imply the simple relations
\begin{align}
\vec D(\vec x,t)&=\varepsilon_0\vec E(\vec x,t)\,,\label{eq_MatRel1}\\[5pt]
\vec H(\vec x,t)&=\mu_0^{-1}\vec B(\vec x,t)\,.\label{eq_MatRel2}
\end{align}
Plugging these into the ``macroscopic'' Maxwell equations \eqref{eq_maxwell_3}--\eqref{eq_maxwell_2} leads to
\begin{align}
\nabla\cdot\vec E(\vec x,t)&=\rho(\vec x,t)/\varepsilon_0\,,\label{eq_Maxfund1}\\[3pt]
\nabla\times\vec E(\vec x,t)&=-\partial_t\vec B(\vec x,t)\,,\label{eq_Maxfund2}\\[3pt]
\nabla\cdot\vec B(\vec x,t)&=0\,,\label{eq_Maxfund3}\\[2pt]
\nabla\times\vec B(\vec x,t)&=\mu_0 \h \vec j(\vec x,t)+\varepsilon_0 \h \mu_0 \h \partial_t\vec E(\vec x,t)\,,\label{eq_Maxfund4}
\end{align}
which is formally identical to Eqs.~\eqref{eq_MaxTot1}--\eqref{eq_MaxTot4} for the total electric and magnetic fields in materials. 
Although Eqs.~\eqref{eq_Maxfund1}--\eqref{eq_Maxfund4} are sometimes referred to as ``Maxwell equations in vacuo'', this labeling is actually more than misleading.
In fact, there is now a general consensus that these equations
are {\it microscopically} always valid---i.e., even in the presence of materials---provided one includes in the source terms $\{\rho, \, \vec j\}$
all microscopic contributions to the charge and current densities (see in particular \cite[Sec.~2.3]{Zangwill}, as well as 
\cite{Martin, Fliessbach, SchafWegener, Nolting, Bertlmann, Sexl, Misner, Itzykson}). Put differently, electric and magnetic fields {\it always} 
obey Eqs.~\eqref{eq_Maxfund1}--\eqref{eq_Maxfund4}, and the converse also holds true: whenever two vector fields $\vec E$ and $\vec B$ 
obey these equations with some specific charge and current densities $\rho$ and $\vec j$, these vector fields are
none other than the electric and magnetic fields generated by these very sources.

Simple and intuitive as this may seem at first glance,
it has a profound interpretational impact on the very concept of {\it electrodynamics of materials}: From the macroscopic
point of view, the influence of the medium seems to consist in a {\it modification of the fundamental equations}, and correspondingly
one has to distinguish between electrodynamics {\it in} media and electrodynamics {\it in} vacuo. By contrast,
from the microscopic point of view, the fundamental equations are always the same, but electrodynamics {\it of} media
consists in the inclusion of internal charge and current densities $\{\rho_{\rm int}, \h \vec j_{\rm int}\}$ associated with the material.
In other words, on the microscopic level the source terms $\{\rho, \h \vec j\}$ split into external and internal contributions (see e.g.~\cite[p.~493]{Martin}):
\begin{align}
\rho(\vec x,t)&=\rho\ext(\vec x,t)+\rho_{\rm int}(\vec x,t)\,,\label{eq_micSource1}\\[5pt]
\vec j(\vec x,t)&=\vec j\ext(\vec x,t)+\vec j_{\rm int}(\vec x,t)\,.\label{eq_micSource2}
\end{align}
Here, the {\it internal sources} are those which are generated by the material itself, while
the {\itshape external sources} are those which do not belong to the material (see \cite{ED1} for a more detailed explanation). In particular, we remark that the external 
sources are not necessarily located only ``outside'' the material probe: for example, an impurity inside a 
crystal or a charged particle moving through the medium as in the case of Cherenkov radiation can also be regarded as external sources.

As a matter of principle, the internal sources are defined as quantum mechanical 
expectation values (see e.g.~\cite[Chap.~3]{Giuliani}, \cite[Sec.~6.2]{Bruus},  
\cite[Sec.~2.6.1]{SchafWegener} and \cite[Sec.~7.4]{Altland}),
\begin{align}
\rho_{\rm int}(\vec x,t)&=\langle\Psi(t) \h | \h \hat\rho(\vec x) \h |\h \Psi(t)\rangle\,,\\[5pt]
\vec j_{\rm int}(\vec x,t)&=\langle\Psi(t) \h | \,\h \hat{\!\vec j}(\vec x) \h |\h \Psi(t)\rangle\,,
\end{align}
where $\hat \rho(\vec x)$ and $\hat{\!\vec j}(\vec x)$ are suitably defined charge and current operators (see App.~\ref{app_elmKubo})
evaluated in a many-body quantum state $|\Psi(t)\rangle$ which describes the material on the microscopic level. 
For example, $|\Psi(t)\rangle$ might be the solution of the many-body Schr\"{o}dinger equation 
\begin{equation}
{\rm i}\hbar \, \frac{\partial}{\partial t} \h |\Psi(t)\rangle=\hat H |\Psi(t)\rangle\,,
\end{equation}
where $\hat H$ denotes the fundamental Hamiltonian of a system of electrons and nuclei forming a crystalline solid (see \cite[Eq.~(2.1)]{EffWW}). In principle, though, mixed states like a
thermodynamical ensemble are of course also possible.

In a typical experimental setup, however, one is hardly able to detect these internal sources directly. Instead,
one acts on the material by an {\it external perturbation} and then measures the electromagnetic {\it response}.
On the theoretical side, this means that the internal sources further split into
\begin{align}
\rho_{\rm int}(\vec x,t)&=\rho_0(\vec x)+\rho_{\rm ind}(\vec x,t)\,,\label{eq_intSplit1}\\[5pt]
\vec j_{\rm int}(\vec x,t)&=\vec j_0(\vec x)+\vec j_{\rm ind}(\vec x,t)\,,\label{eq_intSplit2}
\end{align}
where the first terms on the right hand side describe the (typically, but not necessarily, static) microscopic charge and current densities
of the material {\it in the absence of an external perturbation}. Concretely, one may think of 
$\rho_0$ and $\vec j_0$ as ground-state (or thermal) expectation values, i.e.,
\begin{align}
\rho_0(\vec x)&=\langle\Psi_0 \h |\h \hat\rho(\vec x)\h |\h \Psi_0\rangle\,,\\[3pt]
\vec j_0(\vec x)&=\langle\Psi_0 \h |\h \,\hat{\!\vec j}(\vec x)\h |\h \Psi_0\rangle\,,
\end{align}
where $|\Psi_0\rangle$ denotes the ground state of an unperturbed many-body Hamiltonian $\hat H_0$\hh. 
By contrast, the respective second terms in Eqs.~\eqref{eq_intSplit1}--\eqref{eq_intSplit2}
are {\it induced} by the perturbation and hence describe the reaction or response of the 
system to that very perturbation. For its part, the latter can be identified with the external electromagnetic fields produced in 
an experiment to act on the material and, at least partially, controlled by the experimenter.

Microscopic electrodynamics of media is therefore centered around the interplay of {\it external} and {\it induced fields}:
In experiments, one considers material probes under the influence of external perturbations, and
in ab initio calculations, one considers microscopic many-body systems in external electromagnetic fields.
Correspondingly, with the external and induced fields we re-introduce the {\itshape total sources} as
\begin{align}
\rho\tot(\vec x,t)&=\rho\ext(\vec x,t)+\rho_{\rm ind}(\vec x,t)\,,\label{eq_totSource1}\\[5pt]
\vec j\tot(\vec x,t)&=\vec j\ext(\vec x,t)+\vec j_{\rm ind}(\vec x,t)\,,\label{eq_totSource2}
\end{align}
which correspond to the sum of the external and the induced (not internal) sources. 
We stress that this microscopic splitting of the source terms, i.e.~Eqs.~\eqref{eq_totSource1}--\eqref{eq_totSource2},
reflects both the experimental and the theoretical situation. Although rather uncommon in traditional textbooks on electromagnetism, 
it is  well known in electronic structure physics (see e.g.~\cite[p.~191]{Giuliani} \linebreak and \cite[p.~493]{Martin}), 
solid state physics (\cite[Chap.~10]{Kittel}, \cite[Eq.~(17.29)]{Ashcroft}, \cite[p.~285]{Kantorovich}), condensed matter 
quantum field theory (\cite[Sec.~6.4]{Bruus}, \cite[Eq. (4.77)]{MartinRothen}), plasma physics \cite[Sec.~1.5.1]{Melrose1Book}, physics of radiation processes \cite[p.~11]{Melrose},
and semiconductor theory 
(see Refs.~\cite[Sec.~2.6]{SchafWegener}, or \cite{Adler,Hanke,Strinati}).
Similarly, in Ref.~\cite[p.~305]{Kantorovich} we find the statement that the bound charges actually coincide with the internal charges.
Finally, we particularly recognize the German textbook by T.~Flie\ss bach \cite{Fliessbach}, 
whose approach (see in particular Sec.~27 and Eqs.~(27.3)) in this respect 
coincides with ours, and who also clearly states (pp.~261f.) that the traditional splitting into ``bound'' and ``free'' 
contributions does neither reflect the experimental situation nor the common practice in modern ab initio materials physics.

\subsubsection{Complete field equations}

The splitting of the source terms, Eqs.~\eqref{eq_totSource1}--\eqref{eq_totSource2}, 
translates directly into an analogous splitting of the electromagnetic fields themselves:
\begin{align}
\vec E\tot(\vec x, t)&=\vec E\ext(\vec x, t)+\vec E\ind(\vec x, t)\,,\label{eq_totEfield}\\[5pt]
\vec B\tot(\vec x, t)&=\vec B\ext(\vec x, t)+\vec B\ind(\vec x, t)\,.\label{eq_totBfield}
\end{align}
On the other hand, if we replace the obsolete splitting of the source terms \eqref{eq_totalold1}--\eqref{eq_totalold2} into ``free'' and ``bound'' parts by the more fundamental microscopic splitting \eqref{eq_totSource1}--\eqref{eq_totSource2} into external and induced parts, then the above equations \eqref{eq_totEfield}--\eqref{eq_totBfield} are analogous to the well-known equations \eqref{eq_PED}--\eqref{eq_MBH}, provided one sets
\begin{align}
 \vec P(\vec x,t) & = -\varepsilon_0\vec E\ind(\vec x, t) \,, \label{eq_PE} \\[2pt]
 \vec D(\vec x, t) & = \varepsilon_0\vec E\ext(\vec x, t) \,, \label{eq_DE} \\[2pt]
 \vec E(\vec x, t) & = \vec E\tot(\vec x, t) \,, \label{eq_EE}
\end{align}
and
\begin{align}
 \vec M(\vec x, t) & = \vec B\ind(\vec x, t)/\mu_0 \,, \label{eq_MB} \\[2pt]
 \vec H(\vec x, t) & = \vec B\ext(\vec x, t)/\mu_0 \,, \label{eq_HB} \\[2pt]
 \vec B(\vec x, t) & = \vec B\tot(\vec x, t) \,. \label{eq_BB}
\end{align}
These {\it Fundamental Field Identifications} are the starting point of the {\it Functional Approach} to electrodynamics of media.
In fact, $\{\vec D,\vec H\}$ now simply coincide with the external electromagnetic fields (i.e., those fields which are not generated by the medium). 
Similarly, $\{\vec P,\vec M\}$ coincide
with the electromagnetic fields generated by the induced sources (i.e., the degrees of freedom which constitute the medium) under the action of the external fields. Note, in particular, that the fundamental constants $\varepsilon_0$ and $\mu_0$ are not material properties of the vacuum but conversion factors.
All fields are now uniquely defined by their respective Maxwell equations, which are given explicitly by Eqs.~\eqref{eq_MaxTot1}--\eqref{eq_MaxTot4} 
for the total fields and by
\begin{align}
\nabla\cdot\vec D(\vec x, t)&=\rho\ext(\vec x, t)\,,\label{eq_MaxDH1}\\[3pt]
\nabla\times\vec D(\vec x, t)&=-\partial_t\vec H(\vec x, t)/c^2\,,\label{eq_MaxDH2}\\[3pt]
\nabla\cdot\vec H(\vec x, t)&=0\,,\label{eq_MaxDH3}\\[3pt]
\nabla\times\vec H(\vec x, t)&=\vec j\ext(\vec x, t)+\partial_t\vec D(\vec x, t)\label{eq_MaxDH4}\,,
\end{align}
as well as by
\begin{align}
\nabla\cdot\vec P(\vec x, t)&=-\rho\ind(\vec x, t)\,,\label{eq_MaxPM1}\\[3pt]
\nabla\times\vec P(\vec x, t)&=\partial_t\vec M(\vec x, t)/c^2\,,\label{eq_MaxPM2}\\[3pt]
\nabla\cdot\vec M(\vec x, t)&=0\,,\label{eq_MaxPM3}\\[3pt]
\nabla\times\vec M(\vec x, t)&=\vec j\ind(\vec x, t)-\partial_t\vec P(\vec x, t)\,.\label{eq_MaxPM4}
\end{align}
for the external and the induced fields, respectively. We particularly emphasize that our equation $\nabla \cdot \vec M = 0$ is usually not accepted in the Standard
Approach because it had been assumed to be in conflict with the case of a homogeneously magnetized body 
or with a constant spin density. For a detailed discussion of these problems, we refer the interested reader to App.~\ref{app_const} and Sec.~\ref{subsubsec_spinmagn}, respectively. Meanwhile, we shortly summarize what we have achieved thus far.

\subsubsection{Summary and supporting arguments}

The basic tenets of the microscopic approach to electrodynamics of materials can be summarized by the following fundamental principles:
\begin{enumerate}[listparindent=\parindent,parsep=0pt,itemsep=1em]
 \item[(i)] {\itshape Unique splitting of the sources.} The Functional Approach replaces the traditional splitting into ``free'' and ``bound'' contributions by the splitting
 into external and induced fields. This splitting is common practice in modern {\itshape ab initio} approaches to materials physics, 
 as it is independent of {\itshape a priori} assumptions about the material. Furthermore,
 this splitting also directly matches the experimental situation, where one measures the response of a 
 material probe under externally applied perturbations \cite{Fliessbach}. By contrast, the traditional splitting into 
 ``free'' and ``bound'' sources cannot be upheld microscopically. This is indeed well known in the condensed matter physics community.
 For example, L.\,V.~Keldysh states clearly \cite[p.~8]{Keldysh}: ``there is no qualitative criterion
 that would enable us to distinguish them [sc., the bound current from the magnetization current and from the current of free charges] 
 from one another locally at each point''. Moreover, ``in processes similar to the photoeffect, the electron passes from the bound to the free state,
 so that the contribution of these processes cannot be included either in $\vec j_{\rm f}$ or $\vec j_{\rm b}$\hh, \mbox{strictly speaking.''}
 
 \item[(ii)] {\itshape Complete field equations.}
 Consequently, all fields are now determined uniquely by their respective Maxwell equations (with external, induced
 and total sources, respectively). In particular, we are given explicit equations for the longitudinal parts of $\vec H$ and $\vec M$ (Eqs.~\eqref{eq_MaxDH3} and \eqref{eq_MaxPM3}) as well as for the transverse parts of $\vec D$ and $\vec P$ (Eqs.~\eqref{eq_MaxDH2} and \eqref{eq_MaxPM2}). By contrast, an alleged ``gauge freedom'' of the {\it observable} fields $\{\vec D, \vec H, \vec P, \vec M\}$ is impossible.

 On the field theoretical side, apart from the new splitting of the sources,
 the only new  equations of the Functional Approach are in fact \eqref{eq_MaxDH2}--\eqref{eq_MaxDH3} and \eqref{eq_MaxPM2}--\eqref{eq_MaxPM3}.
 We stress that these equations do not replace older equations of the Standard Approach, but  they remove an indeterminacy (see Tables \ref{tab_standard_1} and \ref{tab_standard_2}). In particular, with the re-identifications \eqref{eq_EE} and \eqref{eq_BB}, the equations~\eqref{eq_MaxTot1}--\eqref{eq_MaxTot4}
 of the Standard Approach remain valid. The field identifications of the Functional Approach 
 and the corresponding field equations are summarized in Tables \ref{tab_functional_1} and \ref{tab_functional_2}. 
 These may be compared to the respective Tables \ref{tab_standard_1} and \ref{tab_standard_2} for the Standard Approach.

 \item[(iii)] {\itshape Microscopic validity.} On the interpretational level, the Functional Approach is completely different from the Standard Approach:
 As all fields are now electromagnetic in nature, with their respective field equations valid on the microscopic level as well,
 there is no need anymore to derive electrodynamics in media by macroscopic averaging procedures.
\end{enumerate}

\smallskip \noindent
Ultimately, the fundamental identifications \eqref{eq_PE}--\eqref{eq_BB} are the axioms of the Functional Approach. Nonetheless, we will show in Sec.~\ref{sec_thermodynamics} that they can be derived straightforwardly
from the definitions of polarization and magnetization used in classical thermodynamics. Independently of this aspect, we now give some additional
arguments in favor of Eqs.~\eqref{eq_PE}--\eqref{eq_BB}.

First, we stress that the field identifications are in accord with what many treatises affirm anyway. For example, in Sec.~\ref{subsec_problems} we have shown
that even some traditional textbooks (such as \cite{Griffiths,Melia}) divide the electromagnetic fields not so much into those
corresponding to ``free'' and ``bound'' charges and currents within the material, but rather into those which are ``under control'' of the experimenter and those which are not.
This distinction is apparently so fundamental that the famous textbook by L.\,D.~Landau and E.\,M.~Lifshitz \cite{Landau} introduces it even {\it in addition}
to the ordinary distinction of \{$\vec D,\vec P,\vec E,\vec H, \vec M,\vec B$\} (see their fields $\vec{\mathfrak E}$ and $\vec{\mathfrak H}$ introduced in 
\S\,11 \linebreak and \S\,32, respectively). 

Then, there are authors \cite{Dolgov} who explicitly
state that due to the gauge freedom in $\vec D$ and $\vec H$ ``the definitions of dielectric function and magnetic permeability [\ldots] prove ambiguous'' (p.~228)
and therefore go on to ``identify the induction $\vec D$ and the magnetic field $\vec H$ 
with the external fields [\ldots] defined by Maxwell's equations in empty space'' (p.~230, notation adapted).
Correspondingly, Gauss' law for $\vec D$ and $\vec P$ in terms of $\rho\ext$ and $\rho\ind$ can be found in Ref.~\cite[Eqs.~(5.187) and (5.188)]{Kantorovich}, while
Amp\`{e}re's law for $\vec H$ and $\vec D$ in terms of $\vec j\ext$ is written down in Ref.~\cite[Eq.~(24.5)]{Marder}.

Furthermore, the Fundamental Field Identifications also seem obvious from the study of superconductivity, where usually $\vec H$ denotes the external or 
``applied'' magnetic field, while $\vec B$ describes the magnetic field ``inside'' the superconductor (see e.g.~\cite[p.~468]{MahanNutshell}). 
When the latter then turns out to be zero by the 
Mei\ss ner--Ochsenfeld effect, this obviously implies that the ``magnetization'' $\vec M$ also 
describes a magnetic field which just cancels the externally applied field (cf.~\cite[p.~192 and p.~295]{Ibach}). 
Needless to say that~this interpretation neatly squares with our interpretation of $\vec M,\vec H$ and $\vec B$ as
induced, external, and total magnetic fields, respectively. Correspondingly, according to Ref.~\cite[p.~270]{Schwabl} the field $\vec H$ is the magnetic field ``produced by some external sources''
(see also \cite[p.~644]{GrossoParravicini}). 

We also note that without the division into external and induced field quantities, it would be completely unclear why the linear coefficients in the corresponding constitutive relations (e.g.~in Eqs.~\eqref{eq_MatRel10} or \eqref{eq_MatRel20}) should be given by retarded response functions (cf.~\cite{Giuliani, Kantorovich, DolgovArticle} and Sec.~\ref{sec_response}).

Furthermore, it would be unclear why in special relativity the ``dipole densities'' $\vec P$ and $\vec M$ behave precisely as electric and magnetic fields under Lorentz transformations, as it is assumed in the derivation of the Minkowski transformation
formulae for the constitutive laws in moving media (see the original work \cite{Minkowski} or the textbooks 
\cite[Eqs.~(5.127)--(5.128)]{Toptygin}, \cite[Sec.~22.8]{Zangwill},  \cite[Eqs.~(2.132) and (2.138)]{Roemer}, \cite[p.~234]{Bredov}, as well as the discussion in Ref.~\cite{EDLor}).

In actual fact, it has been noted already by several textbooks---mainly in condensed matter physics---that our fundamental 
field identifications are somehow evident (see e.g.~\cite[footnote 24 on p.~99]{Bruus}, \cite[footnote 21 on p.~338]{Ashcroft}, \cite[p.~230]{Dolgov} or \cite[Sec.~4.3.2]{MartinRothen}). 
In particular, the textbook \cite[footnote 14 on p.~33]{SchafWegener} directly identifies $\vec D=\varepsilon_0 \h \vec E\ext$\hh. Similarly, Ref.~\cite[p.~493]{Martin} (see above p.~18) hits the nail on the head. 

Finally, we remark there is even a textbook on electronic structure theory \cite[App.~A.2]{Kaxiras}
which rests its case entirely on the Fundamental Field Identifications, although without stating that this is something of a novelty.

\subsection{General Conclusions}\label{subsec:generalConclusions}

Astonishingly, from the sheer re-interpretation of electrodynamics of media in terms of induced and external fields,
a number of paradigmatic conclusions can be drawn. These are well known in the ab initio community, but not in the general
textbook literature on electrodynamics. In the remainder of this section, we will explicitly spell out these paradigmatic conclusions.

\subsubsection{Electromagnetic response functions} \label{sec_response}

First, we stress that {\it independently} of possible macroscopic averaging procedures, electrodynamics of media as a response theory
is based on the interpretation of {\it induced quantities as functionals of external perturbations} (see e.g.~\cite[Sec.~1.4]{IlinskiiKeldysh}).
This postulate is the mathematical reflection of the intuitive idea that the induced fields 
are {\it caused by}---and hence their concrete behavior {\itshape depends on}---the external fields.

Concretely, as the induced electromagnetic fields are necessarily generated by the induced charge and current sources, the functional dependence
\begin{equation}
 j^\mu\ind = j^\mu\ind[A^\nu\ext]
\end{equation}
of the induced four-current
$j^\mu\ind = (c\rho\ind \h , \,\vec j\ind)^{\rm T}$ on the external four-potential
$A^\nu\ext = (\varphi\ext/c \h , \, \vec A\ext)^{\rm T}$
completely determines the electromagnetic response of any material and hence all electromagnetic material properties.  The reason for this is that the external four-potential contains the whole information about the applied perturbation, and the induced four-current contains the whole information about the induced electromagnetic fields.
In particular, {\it linear} response theory corresponds to the first order expansion of this functional, \smallskip
\begin{equation}
j\ind^\mu(x)=\int\!\de^4 x'\,\chi\indices{^\mu_\nu}(x,x') \h A^\nu\ext(x')\,, \smallskip \label{eq_fundRespRel} \smallskip
\end{equation}
and the formulae of linear response theory are identities between first order functional derivatives \cite{ED1}. 
For the integral kernel $\chi\indices{^\mu_\nu}(x, x')$ we propose the name {\itshape fundamental response tensor} (see Sec.~\ref{subsec_urr}).

The principle of causality implies that the response tensor fulfills the constraint \smallskip
\begin{equation}
\chi\indices{^\mu_\nu}(\vec x,t; \h \vec x'\mh ,t')=0 \qquad \textnormal{if } \, t < t' \,. \smallskip \vspace{2pt}
\end{equation}
This retardation condition means that the induced field at a certain space-time point $(\vec x,t)$ 
depends on its causing field at another point $(\vec x'\mh, t')$ only if the latter lies in the past of the former point.
Within the context of special relativity, this dependence is even restriced to the backward light cone (``causal past''), i.e., to those
space-time points $(\vec x'\mh, t')$ for which $t > t'$ {\itshape and} $|\vec x-\vec x'|^2\leq c^2(t-t')^2$. 
Consequently, as a matter of principle, retarded response functions relate fields at different space-time points. 
Hence, they have to be considered as {\it nonlocal integral kernels} depending on {\it two} space-time points.

In other words, in the Maxwell equations {\it every ``material constant'' is actually a non-local response function}:
Maxwell's equations in media involving these ``material constants'' are actually {\it integro-differential equations}.
This latter conclusion changes the whole way in which these Maxwell equations in media are to be manipulated mathematically and has particularly important consequences for the derivation of the refractive index \cite{Refr}.

\subsubsection{Electromagnetic material constants} \label{sec_mc}

While electromagnetic {\it response functions} necessarily constitute non-local integral kernels, 
the electromagnetic {\it material constants} (such as the dielectric constant $\varepsilon_{\rm r} \approx 81$ for water \cite[Table 3.2]{Hecht})
are their corresponding zero-\linebreak wavevector and zero-frequency limits (see \cite{Adler,Hanke,Wiser} 
or \cite[Sec.~28]{Fliessbach}, \cite[pp. 430f.]{Bechstedt}, \cite[Sec.~3.1]{Dolgov}).
Therefore, material constants only relate {\itshape constant} fields, i.e., fields which do not vary in space and time and whose Fourier 
transforms have only zero-wavevector and zero-frequency components (as in many models dealing with the idealized parallel-plate capacitor).
By contrast, {\it dynamical}, i.e.~position- and time-dependent fields, are never re\-{}lated by constants.

For concreteness, consider the density response function $\upchi \equiv \chi\indices{^0_0}/c^2$ which mediates between the induced charge density $\rho \equiv \rho_{\rm ind}$ and the external scalar potential $\varphi \equiv \varphi_{\rm ext}$\h, i.e.,
\begin{equation} \label{density_response}
 \rho(\vec x, t) = \int \! \de^3 \vec x' \int \! c\, \de t' \,\h \upchi(\vec x, \vec x'; \h t - t') \, \varphi(\vec x'\mh, t') \,.
\end{equation}
As always, we have assumed {\itshape temporal homogeneity,} which implies that the response function depends only on the difference of the two time arguments, or in Fourier space on only one frequency. By Fourier transformation, Eq.~\eqref{density_response} is then equivalent to
\begin{equation}  \label{here_1}
 \rho(\vec k, \omega) = \int \! \de^3 \vec k' \,\h\upchi(\vec k, \vec k' ; \h \omega) \h\hh \rho(\vec k', \omega) \,,
\end{equation}
where the response function transforms as (cf.~\cite[Sec.~2.1]{ED1})
\begin{equation}
 \upchi(\vec k, \vec k' \h ; \h \omega) = \frac{1}{(2\pi)^3} \int \! \de^3 \vec x \int \! \de^3 \vec x' \mh \int \! c \, \de \tau \,\h \e^{\i\omega \tau} \h \e^{-\i\vec k \cdot \vec x} \, \upchi(\vec x, \vec x'; \h \tau) \, \e^{\i \vec k' \cdot \vec x'} \,.
\end{equation}
For {\itshape homogeneous systems}, the response function depends essentially only on the difference of its spatial arguments, or in Fourier space on only one spatial wavevector, i.e., \smallskip
\begin{equation}
 \upchi(\vec k, \vec k' \h ; \h \omega) = \upchi(\vec k, \omega) \h \delta^3(\vec k - \vec k') \,. \smallskip
\end{equation}
The corresponding response law then simplifies to
\begin{equation} \label{hom_law}
 \rho(\vec k, \omega) = \upchi(\vec k, \omega) \h \varphi(\vec k, \omega) \,.
\end{equation}
As a matter of principle, an appropriate limiting value of $\upchi(\vec k, \omega)$ for $\vec k \to 0$ and $\omega \to 0$ would then represent a {\itshape material constant}. 
It relates the spatially homogeneous and static component of the induced field to the respective component of the external perturbation.
 
We note, however, that in general the zero-wavevector and zero-frequency limit does not necessarily have to exist or to yield a non-vanishing result,
and if it exists for one response function, it does not have to exist for another. Finally, the way in which the origin in Fourier space is approached may also be relevant.
For example, a conductor does not have a finite dielectric constant (cf.~\cite[Sec.~3.4.4]{Giuliani}), but it does of course have a dielectric response tensor.
On the other hand, for a dielectric material (i.e., a material with a finite dielectric constant $\varepsilon_{\rm r}(\vec 0,0)$), 
the (proper) conductivity has to vanish for $\omega \to 0$ (see \cite[Exercise 3.11]{Giuliani}), and, moreover, 
the density response function has to vanish for $|\vec k| \to 0$ \h as \h$\mathcal O(|\vec k|^2)$\hh. The last fact can be seen directly from the standard relation for the (longitudinal) dielectric function \cite[Eq.~(7.41)]{ED1}
\begin{equation}
\varepsilon^{-1}_{\rm r}(\vec k,\omega)=1+v(\vec k) \h\hh \upchi(\vec k,\omega) = 1 + \frac{\upchi(\vec k, \omega)}{\varepsilon_0 \hh |\vec k|^2} \,,
\end{equation}
where $v(\vec k)$ is the Coulomb interaction kernel in Fourier space. Furthermore, taking into account that the integrated induced charge density vanishes,
\begin{equation}
\int\!\de^3\vec x\,\rho\ind(\vec x,t)=0\,, 
\end{equation}
on account of charge conservation, i.e.,
\begin{equation}
\int\!\de^3\vec x\,\rho_{\rm int}(\vec x,t)=\int\!\de^3\vec x\,\rho_{\rm int, 0}(\vec x)\,,
\end{equation}
we obtain the general condition on the density response function,
\begin{equation}
\upchi(\vec k=0, \h \omega)=0\,,
\end{equation}
Hence, in this case the corresponding ``material constant'' is always (i.e.~for all materials) zero.

In crystalline systems, one usually does not assume the full spatial homogeneity, but only the invariance of the response functions under lattice 
translations. This means (suppressing the time dependencies)
\begin{equation}
 \upchi(\vec x, \vec x') = \upchi(\vec x + \vec a, \h \vec x' + \vec a)
\end{equation}
for any vector $\vec a$ in the direct crystal lattice (see e.g.~\cite{Kittel,Ashcroft}). In momentum space, one shows easily that this implies
\begin{equation}
\chi(\vec k,\vec k')=\e^{{\rm i}(\vec k-\vec k')\hh\cdot\hh \vec a} \, \chi(\vec k,\vec k')\,,
\end{equation}
and hence $\chi(\vec k,\vec k')$ can be non-zero only if for every direct lattice vec-\linebreak tor~$\vec a$ the wavevectors fulfill
\begin{equation}
(\vec k-\vec k')\cdot\vec a=2\pi n\,,
\end{equation}
where $n$ is an integer. This means that the wavevectors $\vec k$ and $\vec k'$ differ by a reciprocal lattice vector (see 
e.g.~\cite[p.~287]{Kantorovich} or \cite[App.~A.1]{Schober}). Thus, they can be decomposed as
\begin{align}
\vec k  &=\vec k_0 +\vec G\,,\\[5pt]
\vec k' &=\vec k_0 +\vec G'\,,
\end{align}
where $\vec k_0$ lies in the first Brillouin zone, while $\vec G,\vec G'$ are reciprocal lattice vectors. For the integral kernel $\upchi$, this implies the general form
\begin{equation}
\upchi(\vec k_0 + \vec G, \h \vec k_0' + \vec G')=\delta^3(\vec k_0-\vec k_0')\,\hh\upchi^{\vec G \vec G'}\mh(\vec k_0)\,.
\end{equation}
The response law \eqref{here_1} now translates into
\begin{equation}
\rho(\vec k_0+\vec G)=\sum_{\vec G'} \h \upchi^{\vec G\vec G'}\mh(\vec k_0) \, \varphi(\vec k_0+\vec G')\,.
\end{equation}
If both $\rho(\vec k)$ and $\varphi(\vec k)$ are supported in the first Brillouin zone, then we recover again Eq.~\eqref{hom_law}, provided that we identify
\begin{equation} \label{macr_resp}
 \upchi(\vec k_0) \equiv \upchi^{\vec 0 \vec 0}(\vec k_0) \,.
\end{equation}
In other words, for wavevectors much smaller than the smallest reciprocal lattice vector, the crystal appears to be homogeneous.
Since the smallest reciprocal lattice vector is typically of the order $\pi/a$, where $a$ is the lattice constant, we recover
the intuitive fact that the crystal appears to be homogeneous if probed at wavelengths much larger than the lattice constant.

Correspondingly, the {\it macroscopically averaged fields} are defined for a crystalline system as those fields ``which only contain the $\vec G=\vec G'= \vec 0$ components''
\cite[p.~430]{Bechstedt} (see also \cite[p.~287]{Kantorovich}). Clearly, such fields are related by a ``macroscopic response function'' as given e.g.~in Eq.~\eqref{macr_resp}. 
Taking as another example the inverse (longitudinal) dielectric function, we obtain
\begin{equation}
(\varepsilon^{-1}_{\rm r})(\vec k, \omega)=\left. (\varepsilon_{\rm r}^{-1})^{\vec G\vec G'}\mh(\vec k,\omega) \h \right|_{\vec G\hh=\hh\vec G'=\hh\vec 0}\,, \vspace{2pt}
\end{equation}
where it is understood that $\vec k$ is restricted to the first Brillouin zone. Furthermore, from this function one takes the zero-wavevector 
limit (to obtain, for example, a purely frequency-dependent refractive index, cf.~\cite{Refr}):
\begin{equation}
\varepsilon^{-1}_{\rm r}(\omega)=\lim_{|\vec k|\rightarrow 0} \h \varepsilon^{-1}_{\rm r}(\vec k;\omega)\,.
\end{equation}
Finally, in order to retrieve from this so-called ``macroscopic dielectric function'' a material {\it constant}, one evaluates it at zero frequency,
\begin{equation}
 \varepsilon^{-1}_{\rm r} = \lim_{\omega \to 0} \h \varepsilon^{-1}_{\rm r}(\omega) \,.
\end{equation}
This procedure is nowadays routinely followed in ab initio calculations of the dielectric constant 
(see e.g.~\cite[Eq.~(3.1)]{Hanke} and \cite[p.~3]{KresseShishkin}). Similar equations for the magnetic material constants
are also standard in ab initio materials physics (see e.g.~\cite[Eq.~(3.183)]{Giuliani}).
We note, however, that a certain complication
occurs from the fact that, in general, setting $\vec G = \vec G' = \vec 0$ in the response function $\varepsilon_{\rm r}^{-1}$ does not com\-{}mute
with the transition to the dielectric function $\varepsilon_{\rm r}$\h.
In other words, it makes a difference whether one first inverts the full, microscopic dielectric function (as an integral kernel)
and then performs the limit $\vec G, \vec  G' \to \vec 0$, or whether one first performs this limit and then inverts the resulting dielectric response 
(as a multiplicative function, i.e., by $\varepsilon_{\rm r}^{-1}(\vec k,\omega)=1/\varepsilon_{\rm r}(\vec k,\omega)$).
This problem goes under the name of ``local field corrections'' and has been treated in the classical work
\cite{Adler} (see also \cite[Sec.~18.3.1, in particular Eq.~(18.21)]{Bechstedt}).

We conclude that, as matter of principle, ``macroscopic electrodynamics'' is not different from the more fundamental
microscopic electrodynamics. Instead, the macroscopic fields just correspond to the small-wavevector (or long-wavelength) components of the microscopic
fields, and consequently, their response relations are given by suitable small-wavevector limits of the  microscopic response 
functions. In particular, electromagnetic material constants are obtained by taking all wavevectors and frequencies to zero.
By contrast, for the conceptual set-up of electrodynamics in media
and for the concrete calculation of the response functions from first principles (by means of the Kubo formalism, see Sec.~\ref{subsubsec_Kubo}), these macroscopic fields {\itshape do not play any r\^{o}le.}
In this context, we recognize once more the profound work of L.\,V.\,Keldysh, who states clearly \cite[pp.~5f.]{Keldysh}:
``Thus, for modern condensed matter physics the dielectric constant is a function of the frequency and wavevector,
which describes the response of the medium to any field, both macroscopic and microscopic. The Faraday--Maxwell dielectric constant
is the limiting value of this function for fields slowly varying in space and time.''
To this we only add that, strictly speaking, it is the limiting value for {\itshape constant} fields, i.e.~fields not varying at all in space and time.

\subsubsection{Universal response relations} \label{subsec_urr}

Our next conclusion concerns the very nature of electromagnetic response properties.
Within the linear r\'{e}gime, the response of any material to external electromagnetic perturbations is governed by its 
{\it fundamental response tensor}, which is defined as the functional derivative (cf.~\cite{Bechstedt,ED1,Altland,Melrose1Book,Strinati})
\begin{equation} \label{fund_resp_tensor}
\chi\indices{^\mu_\nu}(x,x') = \frac{\delta j^\mu_{\rm ind}(x)}{\delta A^\nu_{\rm ext}(x')} \,.
\end{equation}
In fact, the corresponding expansion \eqref{eq_fundRespRel} is the most general constitutive relation which only assumes the linearity of the material, but otherwise
{\it includes all effects of inhomogeneity, anisotropy, relativistic retardation and magneto-electric cross coupling}.
In particular, this implies that linear response theory is Lorentz covariant, whereby it is understood that response functions 
themselves obey transformation laws (see \cite{EDOhm,Melrose73} for applications).

Now, since the induced four-current has to fulfill the continuity equation,
\begin{equation}
\partial_\mu \h\hh j^\mu_{\rm ind}(x) = 0 \,,
\end{equation}
and has to be invariant under gauge transformations of the external four-potential, 
\begin{equation}
A^\mu_{\rm ext}(x)\mapsto A^\mu_{\rm ext}(x)+\partial^\mu \mh f(x) \,, \smallskip
\end{equation}
it follows that the fundamental response tensor of any physical system satisfies the constraint equations (cf.~\cite{ED1, Altland, Melrose1Book, Fradkin}):
\begin{align}
\partial_\mu \h \chi\indices{^\mu_\nu}(x,x')&=0\,, \label{eq_constraints_1} \\[5pt]
\partial'^{\nu}\chi\indices{^\mu_\nu}(x,x')&=0\,.\medskip \label{eq_constraints_2}
\end{align}
These constraints can be used to deduce the general form of the Lorentz-covariant response tensor (see \cite{Melrose1Book} or \cite[Eq.~(5.12)]{ED1}), which is given by
\begin{equation}\label{generalform1}
\chi^\mu_{~\nu}(\vec k,\vec k';\omega)=
\left( \!
\begin{array}{rr} -\lar{\frac{c^2}{\omega^2}} \, \vec k^{\rm T} \, \tsr{\chi}\,\h \vec k' & \lar{\frac{c}{\omega}} \, \vec k^{\rm T} \, \tsr{\chi}\, \\
[10pt] -\lar{{\frac{c}{\omega}}} \,\h \tsr{\chi}\,\h \vec k' & \, \tsr{\chi}\, 
\end{array} \right).
\end{equation}
Therefore, {\it there are at most 9 independent linear electromagnetic response functions for any material}. In particular, 
the cartesian $(3\times 3)$ {\itshape current response tensor,}
\begin{equation}
\tsr\chi(\vec x,t;\vec x',t')=\frac{\delta\vec j_{\rm ind}(\vec x,t)}{\delta\vec A_{\rm ext}(\vec x,t)}\,, \smallskip
\end{equation}
already completely describes the linear response of any material. In other words, 
 there is only a single microscopic $(3 \times 3)$ response tensor which encapsulates all information about the electromagnetic 
behavior of the material on the linear level. 
Furthermore, as the spatial current response is proportional to the conductivity tensor by the standard relation (cf.~\cite{Giuliani,Dolgov,ED1,EDOhm,Nam})
\begin{equation}
\tsr{\chi}(\vec x,\vec x';\omega)={\rm i}\omega \h \tsr{\sigma}(\vec x,\vec x';\omega)\,,\label{eq_standardRel}
\end{equation}
the conductivity tensor also contains the whole information about all linear response properties \cite{ED1, Melrose, Smith}.

For the sake of completeness, we now provide a fully relativistic and {\itshape gauge-independent} derivation of the Universal Response Relation \eqref{eq_standardRel} (thereby generalizing the derivation in Ref.~\cite{ED1} which assumes the temporal gauge). For this purpose, we start from the definition of the conductivity as the {\itshape total functional derivative} (see \cite[Secs.~4.2 and 6.1]{ED1}) of the induced current $\vec j \equiv \vec j_{\rm ind}$ with respect to the external electric field $\vec E \equiv \vec E_{\rm ext}$\h,
\begin{equation}
 \sigma_{k\ell}(x, x') \equiv \frac{\de j_k(x)}{\de E_\ell(x')} \,,
\end{equation}
and we employ the functional chain rule (see \cite{ED1}) in the form
\begin{equation} \label{zwischen_1}
\sigma_{k\ell}(x,x') = \int\!\de^4 y \, \h \frac{\delta j_k(x)}{\delta A^0(y)} \, \frac{1}{c} \frac{\de \varphi(y)}{\de E_\ell(x')} + \int\!\de^4 y \, \h \frac{\delta j_k(x)}{\delta A_j(y)} \,\frac{\de A_j(y)}{\de E_\ell(x')} \,,
\end{equation}
where $A^\mu = (\varphi/c, \vec A) \equiv A^\mu_{\rm ext}$ is the external four-potential.
Next, we perform a Fourier transformation with respect to the time variables (using homogeneity in time) and express $\chi_{k0}$ in 
the first term through Eq.~\eqref{generalform1} as
\begin{equation}
 \chi_{k0}(\vec x, \vec y; \omega) = \frac{c}{\i\omega} \h \frac{\partial}{\partial y_j} \, \chi_{kj}(\vec x, \vec y; \omega) \,.
\end{equation}
Then, we obtain
\begin{align}
& \sigma_{k\ell}(\vec x, \vec x'; \omega) = \int \! \de^3 \vec y  \, \left\{ \left( \frac{1}{\mathrm i \omega} \h \frac{\partial}{\partial y_j} \, \chi_{kj}(\vec x, \vec y; \omega) \right) \frac{\de\varphi(\vec y, \omega)}{\de E_\ell(\vec x', \omega)} \right. \\[5pt] \nonumber
 & \hspace{4.3cm} + \left. \chi_{kj}(\vec x, \vec y; \omega) \, \frac{\de A_j(\vec y, \omega)}{\de E_\ell(\vec x', \omega)} \right\} \\[8pt]
&=\frac{1}{\mathrm i \omega} \h \int \! \de^3 \vec y \,\h \chi_{kj}(\vec x, \vec y; \omega) \,
\frac{\de}{\de E_\ell(\vec x', \omega)} \left\{-\frac{\partial}{\partial y_j} \h \varphi(\vec y, \omega) + \mathrm i \omega A_j(\vec y, \omega)\right\} ,
\end{align}
where in the last step we have used a partial integration with respect to the $y_j$ variable.
The term in curly brackets equals $E_j(\vec y, \omega)$, hence we obtain
\begin{align}
 \sigma_{k\ell}(\vec x, \vec x'; \omega) & = \frac{1}{\mathrm i \omega} \int \! \de^3 \vec y \,\h \chi_{kj}(\vec x, \vec y; \omega) \, \delta_{j\ell} \, \delta^3(\vec y - \vec x') \\[5pt]
 & = \frac{1}{\mathrm i \omega} \, \chi_{k\ell}(\vec x, \vec x'; \omega) \,,
\end{align}
which is the desired relation.

While the fact that the electromagnetic response of any material is contained in a single response tensor (such as the conductivity) goes against the intuition from the na\"{i}ve picture presented in the traditional textbooks on macroscopic electrodynamics, 
it is actually well known in condensed matter physics. For example, O.\,V.~Dolgov
and E.\,G.~Maksimov \cite[p.~224]{Dolgov} make it clear that ``the properties of a system in an external electromagnetic field can be uniquely
described with the aid of the relation between the induced current [\ldots] and the electric field'', i.e., by the conductivity. Put differently, all linear response functions (including magneto-electric coupling coefficients) 
can be expressed analytically in terms of the conductivity tensor
by means of {\it universal} (i.e.~material-independent) {\it response relations}. 
These Universal Response Relations have been derived in Ref.~\cite[Sec.~6]{ED1}. In the Fourier domain, they read explicitly (cf.~\cite[Eqs.~(6.37)--(6.40)]{ED1})
\begin{align}
& (\chi_{EE})_{ij}(\vec k, \vec k'; \omega) \label{eq_URR1} \\[1pt] \nonumber
& \qquad = \mathbbmsl D_0(\vec k, \omega) \left( \delta_{im}\! -\! \frac{c^2 k_i k_m}{\omega^2} \right) \mathrm i \omega \h \sigma_{mj}(\vec k, \vec k'; \omega)\,, \\[6pt]
& (\chi_{EB})_{ij}(\vec k, \vec k'; \omega) \label{eq_URR2} \\[1pt] \nonumber
& \qquad =  \mathbbmsl D_0(\vec k, \omega) \left( \delta_{im}\! -\! \frac{c^2 k_i k_m}{\omega^2} \right)\mathrm i \omega \h \sigma_{mn}(\vec k, \vec k'; \omega)\left( -\epsilon_{n\ell j} \h \frac{\omega k'_\ell}{c \hh |\vec k'|^2} \right) , \\[6pt]
& (\chi_{BE})_{ij}(\vec k, \vec k'; \omega) \label{eq_URR3} \\[1pt] \nonumber
& \qquad = \mathbbmsl D_0(\vec k, \omega) \left( \epsilon_{ikm} \h \frac{c k_k}{\omega} \right) \mathrm i \omega \h \sigma_{mj}(\vec k, \vec k'; \omega)\,, \\[6pt]
& (\chi_{BB})_{ij}(\vec k, \vec k'; \omega) \label{eq_URR4} \\[1pt] \nonumber
& \qquad = \mathbbmsl D_0(\vec k, \omega) \left( \epsilon_{ikm} \h \frac{c k_k}{\omega} \right)  \mathrm i \omega \h \sigma_{mn}(\vec k, \vec k'; \omega)\left( -\epsilon_{n\ell j} \h \frac{\omega k'_\ell}{c \hh |\vec k'|^2} \right).
\end{align}
Here, $\mathbbmsl D_0$ denotes the retarded Green function of the d'Alembert operator, which is given in Fourier space by
\begin{equation}
\mathbbmsl D_0(\vec k, \omega) = \frac{c^2\mu_0}{-(\omega + \i \eta)^2+c^2|\vec k|^2} \,,
\end{equation}
where the infinitesimal $\eta$ in the frequency domain ensures the retardation in the time domain \cite[Sec.~3.1]{ED1}. Importantly, in the above formulae,
\begin{align}
 \tsr \chi_{EE}(\vec k, \vec k'; \omega) & \equiv \frac{\de \vec E\ind(\vec k, \omega)}{\de \vec E\ext(\vec k', \omega)} \\[5pt]
 & = \frac{\delta \vec E\ind(\vec k, \omega)}{\delta \vec E\ext(\vec k', \omega)} + \frac{\delta \vec E\ind(\vec k, \omega)}{\delta \vec B\ext(\vec k', \omega)} \, \frac{\delta \vec B\ext(\vec k', \omega)}{\delta \vec E\ext(\vec k', \omega)}
\end{align}
denotes the {\itshape total} functional derivative of the induced electric field with re-\linebreak spect 
to the external electric field (see \cite[Sec.~4.2]{ED1}). Similarly,
\begin{align}
 \tsr \chi_{EB}(\vec k, \vec k'; \omega) & \equiv \frac 1 c \h \frac{\de \vec E\ind(\vec k, \omega)}{\de \vec B\ext(\vec k', \omega)} \\[5pt]
 & = \frac 1 c\h \frac{\delta \vec E\ind(\vec k, \omega)}{\delta \vec B\ext(\vec k', \omega)} + \frac 1 c\h \frac{\delta \vec E\ind(\vec k, \omega)}{\delta \vec E\ext(\vec k', \omega)} \, \frac{\delta \vec E\ext(\vec k', \omega)}{\delta \vec B\ext(\vec k', \omega)}
\end{align}
denotes the {\itshape total} functional derivative with respect to the external magnetic field, etc. These total functional derivatives directly correspond to the physically observable response functions (see the discussion in \cite[Sec.~6.1]{ED1}). For example, the dielectric tensor and the (relative) magnetic permeability can be expressed in terms of these as
\begin{align}
(\tsr\varepsilon_{\rm r})^{-1}&=\tsr 1+\tsr\chi\EE\,, \label{rel_1} \\[2pt]
\tsr\mu_{\rm r}&=\tsr 1+\tsr\chi\BB\,. \label{rel_2}
\end{align}
We stress again that the above relations \eqref{eq_URR1}--\eqref{eq_URR4} and \eqref{rel_1}--\eqref{rel_2} between linear electromagnetic response functions are valid for any material and in particular include all possible effects of inhomogeneity, anisotropy, relativistic retardation and magnetoelectric cross-coupling. On the other hand, 
all standard relations between linear electromagnetic response functions can be rederived as special cases of the Universal Response Relations (see \mbox{\cite[Sec.~7]{ED1}).}

We close this subsection with the following remark: As the physical response functions necessarily
correspond to total functional derivatives, part of the {\itshape magnetic} reaction is already contained in the {\itshape electric} response function, and vice versa.
A na\"{i}ve expression of the induced electric and magnetic fields in terms of the external fields, as often used in the context of bianisotropic media, therefore leads to an overcounting.
This raises the delicate question of how the induced electric and magnetic fields can actually be expanded in terms of the physical response functions.
For the answer to this question, the interested reader is referred to \cite[Sec.~6.6]{ED1}, where it is shown that there exist
three different but equivalent field expansions on the fundamental level.

\subsubsection{Kubo formalism}\label{subsubsec_Kubo}

Up to now, our considerations have been completely classical: Although they have shown that by the Universal Response Relations 
all linear electromagnetic response properties are implicitly determined by the current response tensor,
they have left open the question of how this quantity can actually be calculated.
Here, ab initio materials physics is based on the {\it Kubo formalism} (see e.g.~\cite{Giuliani, Bruus, Altland,  Strinati, KeldyshKirzhnitz}, and
in particular \cite[Chap.~7, App.~B]{GrossoParravicini} for the connection to Fermi's Golden Rule)
developed by the Japanese physicist Ry\={o}go Kubo (1920-1995) \cite{Kubo}. In this subsection, we shortly
summarize its main formulae as applied to the fundamental response tensor, 
and we highlight its conceptual and practical importance for electrodynamics of media.

The fundamental response tensor can be calculated from the electromagnetic {\itshape Kubo formula} 
(cf.~\cite[Sec.~3.4]{Giuliani}, \cite[Chap.~6]{Bruus}, \cite[Sec.~7.4]{Altland}; see also App.~\ref{app_elmKubo}) as follows:
\begin{equation} \label{kubo_formula}
\begin{aligned}
 \chi\indices{^\mu_\nu}(x,x') & = \frac{\i}{c \h \hbar} \, \varTheta(t-t') \, \big\langle \big[\,\h\hat{\! j}^\mu(x),\,\h\hat{\! j}_\nu(x')\big]\big\rangle \\[3pt]
 & \quad \, +\frac{e}{m} \h \sum_{i=1}^3 \delta\indices{^\mu_i} \, \delta_{i\,\nu} \, \delta^4(x-x') \, \big \langle \hh \hat \rho (x) \big\rangle \,,
\end{aligned}
\end{equation}
or more explicitly,
\begin{align}
\chi\indices{^0_0}(\vec x,t;\vec x',t')&=-\frac{\i c}{\hbar} \, \varTheta(t-t') \, \big\langle \big[\h\hat\rho(\vec x,t),\h\hat\rho(\vec x',t')\big]\big\rangle\,, \label{kubo_for_1} \\[8pt]
\chi\indices{^k_0}(\vec x,t;\vec x',t')&=-\frac{\i}{\hbar} \, \varTheta(t-t') \, \big\langle \big[\,\h\hat{\! j}_k(\vec x,t),\h\hat \rho(\vec x',t')\big]\big\rangle\,, \label{kubo_for_2} \\[8pt]
\chi\indices{^0_\ell}(\vec x,t;\vec x',t')&=\frac{\i}{\hbar}\, \varTheta(t-t') \, \big\langle \big[\h\hat \rho(\vec x,t),\,\h\hat{\! j}_\ell(\vec x',t')\big]\big\rangle\,, \label{kubo_for_3} \\[8pt]
\chi\indices{^k_\ell}(\vec x,t;\vec x',t')&=\frac{\i}{c \h \hbar} \, \varTheta(t-t') \, \big\langle \big[\,\h\hat{\! j}_k(\vec x,t),\,\h\hat{\! j}_\ell(\vec x',t')\big]\big\rangle \nonumber \\[5pt]
 & \quad \, + \frac{e}{mc} \, \delta_{k\ell} \, \delta^3(\vec x - \vec x') \h \delta(t - t') \, \big \langle \hh \hat \rho (\vec x, t) \big\rangle\,. \label{kubo_for_curr}
\end{align}
Here, for the sake of concreteness, we have assumed that we deal with an electronic many-particle system such that $e$ is the elementary charge, $(-e)$ the charge of an electron
and $m$ the electron mass. Correspondingly, all expectation values refer to the unperturbed {\itshape reference state}, 
which in ab initio electronic structure physics is mostly identified with the ground state of $N$ electrons.
In the last equation, the term proportional to the density 
stems from the fact that the current itself also depends explicitly on the vector potential. At the same time, 
this term guarantees that the constraint equations \eqref{eq_constraints_1}--\eqref{eq_constraints_2} are fulfilled (see \cite{Fradkin}). 
In the case of the Schr\"{o}dinger field (see App.~\eqref{app_pauli} for the generalization to the Pauli field), 
the charge and current density operators are given by (see e.g.~\cite[Sec.~1.4.3]{Bruus})
\begin{align}
\hat\rho(\vec x,t)&=(-e) \, \sum_s \hat\psi_s^\dagger(\vec x,t) \h \hat\psi_s(\vec x,t)\,,\label{eq_gaugeinvCh} \\[3pt]
\hat{\!\vec j}(\vec x,t)&=\frac{(-e)}{2m} \h \sum_s\bigg( \hat\psi_s^\dagger(\vec x,t) \h \bigg(\frac{\hbar}{\i} \h \nabla\hat\psi_s(\vec x,t)\bigg) \nonumber \\[1pt]
& \hspace{2.6cm} -
\bigg(\frac{\hbar}{\i} \h \nabla\hat\psi_s^\dagger(\vec x,t)\bigg)\hh \hat\psi_s(\vec x,t)\mh\bigg)\,, \label{eq_gaugeinvCurr}
\end{align}
where $\hat\psi_s(\vec x,t)$ denotes the Schr\"{o}dinger field operator \cite{Giuliani,Bruus,MartinRothen,Hedin69,Schweber} in the interaction picture
(i.e., the time dependence is induced by the unperturbed Hamiltonian). For later purposes, 
we have also included the spin index $s \in \{\uparrow, \downarrow\}$. 

We remark that in most cases the Kubo formula is not evaluated directly. Instead, as the four-current is bilinear in the field operators,
its products can be expressed in terms of the four-point Green function (see \cite{Strinati,Fetter} or \cite[Sec.~7.4.1]{Altland}). In this context, 
we further note that all standard approximations in many-body physics, such as the Hartree, the Hartree-Fock and the GW approximation, 
can be interpreted as a hierarchy of approximations for that very four-point Green function (see~\cite{Starke}).

We conclude this subsection by stressing that the Kubo formalism is {\itshape inseparable} from electrodynamics of media
in that it constitutes both its practical and its conceptual basis:
Practically, the Kubo formalism provides for the link between electrodynamics and quantum field theory, and it gives the concrete 
formulae for the actual calculation of the response functions.
In particular, this shows that response functions cannot be interpreted as free parameters 
being only accessible from experiment. Conceptually, the Kubo formalism implies that linear electromagnetic material properties are to be defined as
response functions mediating between induced and external fields. Furthermore, 
the Kubo formalism shows that response functions are (i) microscopic by their very nature,
(ii) non-local integral kernels, and (iii) free from any ``gauge freedom'' in their definition.

\subsubsection{Spin magnetization} \label{subsubsec_spinmagn}

As a matter of principle, {\itshape the magnetization is always generated by a microscopic current.}
In the case of the so-called orbital magnetization, the corresponding current operator is given
by the quantized Schr\"{o}dinger current defined in the last subsection. It is, however, well known that apart from the ``orbital motion'' of the charge carriers,
there is yet a second source of magnetism stemming from the spin of the particles (the electrons in most cases).
Na\"{i}vely, one might be tempted to think that this magnetization is of a completely different nature and is hence outside the scope of the Kubo formula.
Furthermore, one might object to the Functional Approach that a hypothetical constant spin density (which should at least be allowed as a reasonable ``macroscopic'' approximation for a finite body) would be at variance with the fundamental equation $\nabla\cdot\vec M=0$ (see App.~\ref{app_const}).
The problem of incorporating \linebreak the magnetization generated by the spin degrees of freedom into the Functional Approach
therefore deserves a careful discussion.

In fact, the spin magnetization is actually not outside the scope of the Kubo formalism.
Instead, it can be included by a {\itshape spin contribution to the electric current} 
(which is distinct from the {\itshape current of spin} considered 
in \linebreak recent applications \cite{Rashba03, Murakami, Shitade09, Komnik2010, Ezawa2010, Sugimoto, Enss15,Vignale2015, Yin,Hamada,Milletari2016}). 
This spin contribution to the electric current \linebreak is of the following form 
(see e.g.~\cite[Eq.~(2.4)]{Kirzhnitz}, \cite[Sec.~20.8.6]{Zangwill}, \cite[Eq.~(6.1.3)]{Schwabl}, \cite[Chap.~XX, \S\,29]{Messiah}, \cite{Nowakowski}, 
or Eq.~\eqref{eq_spincurr} in App.~\ref{app_elmKubo}):
\begin{equation}
\vec j_{\rm s}(\vec x,t)=(\nabla\times\vec S)(\vec x,t)\,, \label{eq_spincurrent}
\end{equation}
where the {\itshape electromagnetic spin density} is defined as
\begin{equation}
\vec S(\vec x,t)=\frac{(-e)}{m}  \, \Psi^\dagger(\vec x,t) \h  \bigg(\h \frac{\hbar}{2} \h \vec\sigma \bigg)  \Psi(\vec x,t)\,,\label{eq_spindensity}
\end{equation}
for electrons with charge $(-e)$ and mass $m$. Here, $\boldsymbol\sigma=(\sigma_1,\sigma_2,\sigma_3)$ denotes the vector of Pauli matrices, and the quantities
\begin{equation}
 \Psi(\vec x,t) = \left( \! \begin{array}{l} \psi_{\uparrow}(\vec x, t) \\[5pt] \psi_{\downarrow}(\vec x, t) \end{array} \! \right)
\end{equation}
denote Pauli spinors. As always, the corresponding current operators follow from these classical field expressions
by replacing the classical fields with the respective field operators (see e.g.~\cite[Chap.~1]{Bruus}).
The resulting expression is then to be added to the usual Schr\"{o}dinger current defined in Eq.~\eqref{eq_gaugeinvCurr}. Hence, 
the electric current is composed of two contributions: the ordinary ``orbital'' current 
already defined in the last subsection, Eq.~\eqref{eq_gaugeinvCurr}, and the spinorial current introduced 
in this subsection, Eqs.~\eqref{eq_spincurrent}--\eqref{eq_spindensity} 
(see App.~\ref{app_pauli}, Eq.~\eqref{eq_gaugeinvCurr1}--\eqref{eq_spincurr} for the complete expression).

The magnetization generated by the spin degrees of freedom now simply corresponds to the magnetic field
of the respective spin contribution to the microscopic current. For the sake of simplicity, let us consider the case of a static spin density $\vec S(\vec x)$.
Its current then generates the vector potential
\begin{equation}
\vec A_{\rm ind}(\vec x)=\frac{\mu_0}{4\pi}\int\!\de^3\vec x'\,\frac{\vec j_{\rm s}(\vec x')}{|\vec x-\vec x'|}\,,
\end{equation}
and by $\vec B=\nabla\times\vec A$, the resulting magnetic field reads
\begin{equation}
\vec B_{\rm ind}(\vec x)=\mu_0\hh\nabla\times\left(\frac{1}{4\pi}\int\!\de^3\vec x'\,\frac{\nabla\times\vec S(\vec x')}{|\vec x-\vec x'|}\right)\,.
\end{equation}

\pagebreak\noindent
Comparison with the Helmholtz vector theorem (see Eqs.~\eqref{eq_HVT_1}--\eqref{eq_HVT}) and the 
Fundamental Field Identification $\vec B_{\rm ind}=\mu_0\vec M$ shows that this is equivalent to the simple equation
\begin{equation}
\vec M(\vec x)=\vec S_{\rm T}(\vec x)\,, \smallskip
\end{equation}
i.e.,~the magnetization generated by the spin corresponds to the transverse part of the spin density. Consequently, even for a hypothetical constant spin density
the equation \mbox{$\nabla\cdot\vec M=0$} is fulfilled. This shows that the incorpora-\linebreak tion of spin into the Functional Approach does not pose any problems. 

Finally, we remark that with the inclusion of a spinorial contribution in the electromagnetic current, it is possible
to recover both the Landau orbital diamagnetism {\it and} the Pauli spin paramagnetism from the current response function of the free electron gas (see \cite[Eqs.~(4.29) and (4.50)]{Giuliani}). In addition, the splitting of the current operator in the Kubo formula---which is bilinear in the current---into an orbital and a spin contribution  implies that there are generally also spin-orbital cross contributions to the electromagnetic response (see e.g.~\cite{Vorontsov,Kulakowski,KoshinoCross,NomuraCross} in the case of the magnetic susceptibility).

\begin{landscape}

\begin{table}[t]
\begin{center}
\renewcommand{\arraystretch}{1.6}
\begin{tabular}{p{3.5cm}p{4.3cm}p{3.0cm}p{4.6cm}}
\toprule
$\nabla\cdot \vec E = \rho\tot/\varepsilon_0$ & $\nabla\times\vec E = -\partial_t\vec B$ & $\nabla\cdot\vec B = 0$ & $\nabla\times \vec B = \mu_0 \h \vec j\tot+\partial_t \vec E/c^2$ \\[3pt]
\midrule
$\nabla\cdot\vec D=\rho_{\rm ext}$ & $\nabla\times\vec D= -\partial_t \vec H / c^2$ & $\nabla\cdot\vec H=0$ &
$\nabla\times\vec H=\vec j_{\rm ext}+\partial_t\vec D$ \\[3pt]
\midrule
$\nabla\cdot \vec P=-\rho_{\rm ind}$ & $\nabla\times\vec P=\partial_t \vec M / c^2$ & $\nabla\cdot\vec M=0$ & 
$\nabla\times\vec M=\vec j_{\rm ind} - \partial_t \vec P$ \\[3pt]
\bottomrule
\end{tabular}

\caption{Field equations of the Functional Approach. \label{tab_functional_1}}

\bigskip
\bigskip
\bigskip
\bigskip
\begin{tabular}{p{4cm}p{3cm}}
\toprule
$\vec E = \vec E\tot$ & $\vec B = \vec B\tot$ \\
\midrule
$\vec D = \varepsilon_0 \h \vec E_{\rm ext}$ & $\vec H = \vec B_{\rm ext}/\mu_0$ \\
\midrule
$\vec P = -\varepsilon_0 \h \vec E_{\rm ind}$ & $\vec M = \vec B_{\rm ind}/\mu_0$ \\
\bottomrule
\end{tabular}

\caption{Field identitifications of the Functional Approach.\label{tab_functional_2}}
\end{center}
\end{table}

\end{landscape}

\section{Field identifications from thermodynamics} \label{sec_thermodynamics}

At the heart of the Functional Approach to electrodynamics of media lie the Fundamental Field Identifications, Eqs.~\eqref{eq_PE}--\eqref{eq_BB}.
As explained already, these equalities form the axioms of the Functional Approach, on which the further development of the theory crucially hinges.
Once they are accepted, all concrete formulae follow by more or less trivial manipulations, which are, in any case, of a purely mathematical nature
and hence indubitable. Consequently, any possible dissension could only concern these Fundamental Field Identifications. Unfortuntately, in their quality as axioms
they cannot be proven, at least not within their own framework. They can, however, be made plausible by the conclusions one draws from them. In the preceding section,
the central argument in this respect has been the agreement with the common practice in ab initio materials physics. 
Furthermore, we have shown in Ref.~\cite[Sec.~7]{ED1} that all well-known relations between linear electromagnetic response functions can 
be rederived as limiting cases of the Universal Response Relations (see Sec.~\ref{subsec_urr}), which in turn follow directly from the Fundamental Field Identifications.
By contrast, in this section we will take recourse to a completely different field of reseach, namely to classical thermo\-{}dynamics. 
Concretely, we will show that the Fundamental Field Identifications
follow directly from the definitions of polarization and magnetization as used in this branch of physics.

\subsection{Short review of thermodynamics}

For the convenience of the reader, we first provide a brief summary of the main formulae of classical thermodynamics as needed in this section. 
For details, we refer to Refs.~\cite{Blundell,Schwabl,HuangStat,Chodhur,Honerkamp}.

Thermodynamical systems can be characterized by their {\itshape internal energy} $E$, their {\itshape temperature} $T$ and a set of {\itshape external parameters} $(X_1,\ldots, X_n)$.
In equilibrio, the possible combinations of $(E,T,X_1, \ldots, X_n)$ are restricted to fulfill an {\itshape equation of state}, which is of the form
\begin{equation}
f(E,T,X_1, \ldots, X_n)=0\,,
\end{equation}
with a real function $f$. This equation defines (locally) the manifold $\mathcal M$ of equilibrium states, which is an $(n+1)$-dimensional hypersurface in $\mathbbm R^{n+2}$. Any function
\begin{equation}
 g:\mathcal{M}\rightarrow\mathbbm{R} \,,
\end{equation}
is called a {\it state variable}. In particular, the energy itself can be regarded as a state variable in the sense of a function
mapping each point $(E, T, X_1, \ldots, \linebreak X_n) \in \mathcal M$ to its first argument. 
Parametrizing $\mathcal M$ locally by the variables $(T, X_1, \ldots, X_n)$, this state variable can be expressed as
\begin{equation}
 E =E(T,X_1, \ldots, X_n) \,.
\end{equation}
This equation is called the {\it caloric equation of state.} Similarly, the temperature $T$ and the external parameters $(X_1, \ldots, X_n)$ can also be regarded as state variables.

It is a fundamental tenet of classical thermodynamics that there exist further state variables 
$(S, B_1, \ldots, B_n)$, in terms of which the differential of the energy can be written as
\begin{equation} \label{Gibbsian}
\de E=T\de S - \sum_{i=1}^n B_i \h\hh \de X_i\,.
\end{equation}
These $B_i$ are called {\itshape generalized forces}, while $S$ is called the {\it entropy}, and the above equation \eqref{Gibbsian} is known as the {\it Gibbsian fundamental form.}
The two contributions in Eq.~\eqref{Gibbsian} now correspond to the exchange of {\itshape heat} and {\it work}, which would not be the case for the first and second term
in the analogous expansion 
\begin{equation}
\de E=\frac{\partial E}{\partial T} \, \de T +  \sum_{i=1}^n \frac{\partial E}{\partial X_i} \, \de X_i \smallskip
\end{equation}
of the caloric equation of state.

It is noteworthy that the Gibbsian fundamental form does not depend on the local coordinates on the manifold. 
Instead, it relates in a coordinate-independent way the differential of the energy to the differentials of 
the state variables $(S,X_1,\ldots, X_n)$. Generally, if we have a set of functions $f,x_1,\ldots,x_k,y_1,\ldots, y_k$ on a manifold $\mathcal{M}$ 
with $\mbox{dim}~\mathcal{M}=k$, such that the differential of $f$ is related to the differentials of $y_i$ by
\begin{equation}
\de f=\sum_{i=1}^k x_i \h \de y_i\,,
\end{equation}
then the $y_i$ are called {\it natural coordinates} for $f$. Thereby, we assume that the differentials $\de y_i$ are linearly independent.
In a natural coordinate system given by the niveau hypersurfaces of the $y_i$\hh, the function
$f$ is parametrized as $f=f(y_1,\ldots y_k)$, and hence we have $x_i=\partial f/\partial y_i$\hh.  
Consequently, the entropy together with the external parameters constitute natural coordinates for the energy, \smallskip
\begin{equation}
E=E(S,X_1, \ldots, X_n) \,, \smallskip
\end{equation}
and for this parametrization we get
\begin{align}
\frac{\partial E}{\partial S} & = T\,,\\[5pt]
\frac{\partial E}{\partial X_i} & = -B_i\,.
\end{align}
The corresponding derivatives of the generalized forces,
\begin{align}
\chi_{ij}(S, X_1, \ldots, X_n) & := \frac{\partial B_i(S, X_1, \ldots, X_n)}{\partial X_j} \\[5pt]
& \phantom{:} = -\frac{\partial^2 E(S, X_1, \ldots, X_n)}{\partial X_i \h \partial X_j} \,, 
\end{align}
are the {\it adiabatic susceptibilities}, which naturally depend on the entropy and the external parameters.
By contrast, if we use $(T, X_1, \ldots, X_n)$ as local coor\-{}dinates (as in the caloric equation of state) 
and also express the entropy in terms of these coordinates, we obtain
\begin{equation}
 E(T, X_1, \ldots, X_n) = E(S(T, X_1, \ldots, X_n), X_1, \ldots, X_n),
\end{equation}
and consequently,
\begin{align}
\frac{\partial E}{\partial T}&=T\,\frac{\partial S}{\partial T}\,,\\[5pt]
\frac{\partial E}{\partial X_i}&=T\h\frac{\partial S}{\partial X_i}-B_i\,.
\end{align}
Similarly, from the Gibbsian fundamental form \eqref{Gibbsian} we read off the differential of the entropy  as
\begin{equation}
\de S=\frac{1}{T} \, \de E + \sum_{i=1}^n \frac{B_i}{T} \, \de X_i\,. \smallskip
\end{equation}
This means, the entropy is given in its natural coordinates by
\begin{equation} 
S=S(E, X_1, \ldots, X_n) \,,
\end{equation}
and hence,
\begin{align}
\frac{\partial S}{\partial E}&=\frac{1}{T}\,,\\[5pt]
\frac{\partial S}{\partial X_i}&=\frac{B_i}{T}\,.
\end{align}
Next, we introduce the {\it Helmholtz free energy} as the state variable
\begin{equation}
F=E-TS\,.\label{eq_DefF}
\end{equation}
Note that at temperature zero, which is the relevant case for electronic structure theory, this coincides with the energy, i.e.,
\begin{equation} \label{eq_zero}
 E=F \quad \textnormal{for } \, T = 0 \,.
\end{equation}
For the differential of the Helmholtz free energy, we find 
\begin{align}
\de F & = \de E - S \h \de T - T \h \de S \\[5pt]
& = -S \h \de T-\sum_{i=1}^n B_i \h \de X_i\,. \label{dHelm}
\end{align}
Therefore, in the natural coordinates,
\begin{equation}
 F = F(T, X_1, \ldots, X_n) \,,\label{eq_natCoordF}
\end{equation}
and for this parametrization the partial derivatives yield
\begin{align}
 \frac{\partial F}{\partial T} & = -S \,,\label{eq_PartialFTS} \\[5pt]
 \frac{\partial F}{\partial X_i} & = -B_i \,.
\end{align}

\pagebreak \noindent
Finally, the (negative) second derivatives of the Helmholtz free energy,
\begin{align}
\chi_{ij}(T,X_1, \ldots, X_n) & :=\frac{\partial B_i(T, X_1, \ldots, X_n)}{\partial X_j} \\[5pt]
 & \phantom{:} = -\frac{\partial^2 F(T,X_1, \ldots, X_n)}{\partial X_i \h \partial X_j} \,,
\end{align}
give the {\it isothermal susceptibilities}, which naturally depend on the temperature and the external parameters.

\subsection{Connection to quantum mechanics}

We relate thermodynamics to quantum mechanics by identifying the internal energy $E$ at the temperature $T$ and the
external parameters $(X_1, \ldots, X_n)$ with the expectation value
\begin{equation} \label{qm}
E=\langle\hat H\rangle \equiv \Tr \h (\hat \rho \h \hat H )
\end{equation}
in the {\itshape canonical ensemble} described by the density matrix
\begin{equation}
\hat\rho=
\frac{\e^{-\hat H/k_{\rm B} \hspace{-0.5pt} T}}{Z} \,, \smallskip
\end{equation}
where the Hamiltonian depends on the external parameters through
\begin{equation} \label{eq_genH}
\hat H(X_1,\ldots,X_n)=\hat H_0 - \sum_{i=1}^n X_i \h \hat B_i\,.
\end{equation}
In these equations, $H_0$ denotes a reference Hamiltonian describing 
the isolated system. The $B_i$ are hermitean operators representing the generalized forces, which are now interpreted as the 
external perturbations of the system. Furthermore, we have introduced the {\itshape partition function}
\begin{equation} \label{def_part}
Z(T, X_1, \ldots, X_n)=\Tr \mh \big(\e^{-\hat H(X_1,\ldots,X_n)/ k_{\rm B} \hspace{-0.5pt} T} \hh\big) \,,
\end{equation}
which formally depends on the natural variables of the Helmholtz free energy (see Eq.~\eqref{eq_natCoordF}). Now, Eq.~\eqref{qm} yields the energy as
\begin{equation} \label{comp_1}
 E = \frac{\Tr \mh \big(\hat H \h \e^{-\beta \hat H}\hh\big) }{\Tr \mh \big(\e^{-\beta \hat H}\hh\big)} = -\frac{\partial}{\partial \beta} \h \ln \h \Tr \mh \big(\e^{-\beta \hat H} \hh\big) = -\frac{\partial \ln Z}{\partial \beta} \,,
\end{equation}
where we have introduced the inverse temperature
\begin{equation}
 \beta = \frac{1}{k_{\rm B} T} \,.
\end{equation}
On the other hand, on account of Eqs.~\eqref{eq_DefF} and \eqref{eq_PartialFTS}, the internal energy is related to the free energy by
\begin{equation} \label{comp_2}
 E = F - T \h \frac{\partial F}{\partial T} = F + \beta \h \frac{\partial F}{\partial \beta} = -\frac{\partial}{\partial \beta} \, (-\beta F)\,.
\end{equation}
The comparison of Eqs.~\eqref{comp_1} and \eqref{comp_2} yields
\begin{equation}
F(T, X_1, \ldots, X_n)=-k_{\rm B} T\ln Z(T, X_1, \ldots, X_n) + C(X_1, \ldots, X_n) \,,\label{eq_FEPF_prel}
\end{equation}
where $C$ is yet an arbitrary function of the external parameters. By considering the zero-temperature limit $T \to 0$, we now show that this function has to be zero.
In fact, using a complete set of orthonormal eigenvectors $|\Psi_s\rangle$ of $\hat H$ with corresponding eigenvalues $E_s$\hh,
\begin{equation}
 \hat H  \h \ket{\Psi_s} = E_s \h \ket{\Psi_s} \,,
\end{equation}
we can write the partition function as
\begin{equation}
 Z = \Tr \mh \big(\e^{-\beta \hat H} \hh\big)=\sum_{s} \langle\Psi_s \mid \e^{-\beta \hat H} \mid\Psi_s\rangle=
 \sum_{s} \e^{-\beta E_s}  \,, \label{eq_RepZ}
\end{equation}
where in fact $E_s=E_s(X_1,\ldots,X_n)$ 
by the general form \eqref{eq_genH} of the Hamil\-{}tonian.
From this, by factoring out the contribution of the ground-state energy, we find 
\begin{equation}
Z=\e^{-\beta E_0}\left(1+\e^{-\beta(E_1-E_0)}+\e^{-\beta(E_2-E_0)}+\ldots\,\right),
\end{equation}
and hence
\begin{equation}
-k_{\rm B}T\ln Z = E_0 -k_{\rm B}T \ln\mh\left(1+\e^{-\beta(E_1-E_0)}+\e^{-\beta(E_2-E_0)}+\ldots\,\right).
\end{equation}
Assuming for simplicity that the ground state is non-degenerate, such that $E_s>E_0$ for $s \not = 0$, the exponentials all go to zero in the limit $\beta\to\infty$, implying that
\begin{align}
-\lim_{T\to 0}k_{\rm B}T\ln Z(T, X_1, \ldots, X_n) & = E_0(X_1,\ldots,X_n) \\[2pt]
 & \equiv E(T=0,X_1,\ldots,X_n)\,.
\end{align}
On the other hand, by the definition \eqref{eq_DefF}, the free energy $F$ also approaches $E$ in the zero-temperature limit.
Hence, taking $T \to 0$ in Eq.~\eqref{eq_FEPF_prel} implies that $C(X_1, \ldots, X_n) \equiv 0$. Thus, we obtain the important result
\begin{equation}
F(T, X_1, \ldots, X_n)=-k_{\rm B} T\ln Z(T, X_1, \ldots, X_n) \,.\label{eq_FEPF}
\end{equation}
From this, we now further show that the partial derivatives $\partial F/\partial X_i$ yield the thermal expectation
values of the generalized forces. In fact, from the representation \eqref{eq_RepZ} of
the partition function, we obtain via Eq.~\eqref{eq_FEPF} the relation (with $\partial_i \equiv \partial/\partial X_i$)
\begin{equation}
\partial_i F= -\frac 1 \beta \h \frac 1 Z \, \partial_i Z = \frac{1}{Z}\sum_s \h (\partial_i E_s) \, {\rm e}^{-\beta E_s}\,.
\end{equation}
With the Hellmann--Feynman theorem (see e.g.~\cite[pp.~56--59]{Martin} or \cite[Sec.~5.1]{EffWW}), we further obtain
\begin{equation}
\partial_i E_s = \partial_i \h\hh \langle \Psi_s \mid \hat H \mid \Psi_s \rangle = \langle \Psi_s \mid \partial_i \hat H \mid \Psi_s \rangle = -\langle \Psi_s \mid \hat B_i \mid \Psi_s \rangle  \,,
\end{equation}
and consequently,
\begin{align}
\partial_i F &= -\frac{1}{Z} \sum_s \langle \Psi_s \mid \hat B_i \mid \Psi_s \rangle \, \e^{-\beta E_s} \\[3pt]
& = -\frac{1}{Z} \sum_s \langle \Psi_s \mid \hat B_i\, \e^{-\beta\hat H} \mid \Psi_s \rangle \\[5pt]
& = -\Tr \h ( \hat \rho \h \hat B_i) \,.
\end{align}
Thus, we have shown the general relation
\begin{equation}
 \frac{\partial F}{\partial X_i} = -\langle \hat B_i \rangle \,,  \label{use_this}
\end{equation}
which will prove useful in the following. Finally, we consider the entropy: From the relation
\begin{equation}
S=\frac{1}{T} \h (E-F)
\end{equation}
and \smallskip
\begin{equation}
\ln \h \hat\rho=-\beta\hat H-(\ln Z) \h \hat{\mathrm I}\,, \smallskip
\end{equation}
where $\hat{\mathrm I}$ denotes the identity operator, we obtain
\begin{align}
\langle \h \ln \hat\rho \h \rangle&=-\beta\h  \langle\hat H\rangle-\ln Z =-\beta \h (E - F)\,,
\end{align}
which implies that
\begin{equation}
S(T, X_1, \ldots, X_n)=-k_{\rm B}\h \langle\h \ln\hat\rho\h \rangle=-k_{\rm B}\Tr\h(\hat\rho\ln\hat\rho) \,.
\end{equation}
This is the famous {\it von-Neumann entropy}. 

\subsection{Proof of Fundamental Field Identifications} \label{sec_thpol}

In the thermodynamic context, polarization and magnetization are usually 
defined as derivatives of the Helmholtz free energy with respect to {\it external} \cite[p.~6]{Chodhur}
electric and magnetic fields (see e.g.~\cite[Eqs.~(11.4), (32.4)]{Landau}, \cite[Eq. (6.1.14b)]{Schwabl}, or \cite[Eq.~(15.3a)]{GrossoParravicini}). Hence,
\begin{align}
 \vec P & = -\frac{1}{V} \h \frac{\partial F(T,\vec E\ext,\vec B\ext)}{\partial\vec E_{\rm ext}}\,, \label{thdef_1} \\[5pt]
 \vec M & = -\frac{1}{V} \h \frac{\partial F(T,\vec E\ext,\vec B\ext)}{\partial \vec B_{\rm ext}}\,, \label{thdef_2}
\end{align}
where $V$ denotes the volume of the system. The differential of the free energy then reads (cf.~Eq.~\eqref{dHelm})
\begin{equation}
\de F = -S \h \de T - V\vec P\cdot\de\vec E_{\rm ext} - V\vec M\cdot\de\vec B_{\rm ext} \,.
\end{equation}
In this treatment of the electric and magnetic fields $\vec E\ext,\vec B\ext$
as external parameters, it is inherent that they are mere constants and hence do not constitute fields.
Fortunately, however, the generalization of Eqs.~\eqref{thdef_1}--\eqref{thdef_2} to spatially varying fields is straightforward. 
In terms of functional derivatives, it reads (cf.~\cite[Eqs.~(11.3) and (32.3)]{Landau})
\begin{align}
 \vec P(\vec x) & = -\frac{\delta F[\h T,\vec E\ext,\vec B\ext]}{\delta \vec E_{\rm ext}(\vec x)}\,, \label{thdef_1gen} \\[5pt]
 \vec M(\vec x) & = -\frac{\delta F[\h T,\vec E\ext,\vec B\ext]}{\delta \vec B_{\rm ext}(\vec x)}\,, \label{thdef_2gen}
\end{align}
such that the differential of the free energy is now given by
\begin{equation} \label{eq_ExpFreeEn_local}
\delta F = -S \h \de T - \int\!\de^3\vec x\,\h \vec P(\vec x)\cdot\delta\vec E_{\rm ext}(\vec x) - \int \! \de^3\vec x\,\h\vec M(\vec x)\cdot\delta\vec B_{\rm ext}(\vec x) \,.
\end{equation}
In this subsection, we will show that these thermodynamic definitions of the electric polarization and magnetization are consistent with the Fundamental Field Identifications \eqref{eq_PE}--\eqref{eq_BB}.

For this purpose, we consider a system described by a reference Hamiltonian $\hat H_0$ and couple this to an external, 
static electromagnetic field represented by the four-potential $A^\mu=(\varphi\ext/c,\h\vec A\ext)$. The total Hamiltonian is given by
$\hat H = \hat H_0 + \hat H_{\rm int}$\,,
where we assume an interaction Hamiltonian
\begin{equation}
 \hat H_{\rm int} = \hat H_{\rm int}[ \h \varphi_{\rm ext}, \vec A_{\rm ext} \hh] \,,
\end{equation}
which satisfies the standard relation (see e.g.~\cite[Eq.~(1.96)]{Bruus} or App.~\ref{app_pauli}),
\begin{equation} \label{intHam}
\delta \hat H_{\rm int} = \int\! \de^3\vec x \, \big( \h \hat \rho_{\rm ind}(\vec x)\h \delta\varphi_{\rm ext}(\vec x)-\,\hat{\!\vec j}_{\rm ind}(\vec x)\cdot\delta\vec A\ext(\vec x)\big) \,.
\end{equation}
Here, $j^\mu = (c \rho, \h \vec j)$ denotes the four-current density of the system. Note~that we have replaced the {\itshape internal} sources
by their {\itshape induced} counterparts by subtracting the internal sources in the absence of the external perturbation (see \cite[Sec.~5.1]{ED1}). 
This replacement only amounts to a constant shift of the interaction Hamiltonian without any bearing on the dynamics of the system. 
In a relativistic notation, we can write Eq.~\eqref{intHam} equivalently as (recall our metric convention \eqref{metric})
\begin{equation} \label{HjA}
 \delta\hat H_{\rm int} = -\mh\int\! \de^3 \vec x\ \hat{\! j}_\mu(\vec x) \h \delta A^{\mu}(\vec x) \,.
\end{equation}
Comparing this expression with Eq.~\eqref{eq_genH} shows that the four-potential corresponds to the external parameters in the general formalism, 
while the four-current corresponds to the generalized forces. In particular, the general relation \eqref{use_this} now translates into
\begin{equation} 
 \frac{\delta F}{\delta A^\mu(\vec x)}= -\langle \,\h \hat{\! j}_\mu(\vec x) \rangle \equiv -j_\mu(\vec x) \,,
\end{equation}
which can be spelled out in components as
\begin{align}
 \frac{\delta F}{\delta \varphi_{\rm ext}(\vec x)} & = \rho_{\rm ind}(\vec x) \,, \label{toplug_1}\\[5pt]
 \frac{\delta F}{\delta A_{\rm ext}^k(\vec x)} & = -j^k_{\rm ind}(\vec x) \,. \label{toplug_2}
\end{align}
In order to find the corresponding derivatives with respect to the external
electric and magnetic fields, we invoke the functional chain rule:
\begin{align}
P_i(\vec x) & = -\mh \frac{\delta F}{\delta E^i_{\rm ext}(\vec x)}=
-\int\!\de^3\vec x'\,\h\frac{\delta F}{\delta\varphi_{\rm ext}(\vec x')} \,
\frac{\delta\varphi_{\rm ext}(\vec x')}{\delta E^i_{\rm ext}(\vec x)}\,,\label{eq_Pfunc} \\[5pt]
M_i(\vec x) & = -\mh \frac{\delta F}{\delta B^i_{\rm ext}(\vec x)}=
-\int\! \de^3\vec x'\,\h\frac{\delta F}{\delta A^k_{\rm ext}(\vec x')} \,
\frac{\delta A^k_{\rm ext}(\vec x')}{\delta B^i_{\rm ext}(\vec x)}\,.\label{eq_Mfunc}
\end{align}
Here, we have used that in a static situation we have the simple relations
\begin{align}
\vec E & \equiv\vec E\,[\varphi]=-\nabla\varphi \,, \label{m1} \\[5pt]
\vec B & \equiv\vec B\,[\vec A]=\nabla\times\vec A\,, \label{m2}
\end{align}
i.e., the {\itshape static} electric field depends only on the scalar potential (while the magnetic field {\itshape always} depends exclusively on the vector potential).
On the other hand, static fields $\varphi(\vec x')$ and $\vec A(\vec x')
$ can be represented by their sources through the following equations (see e.g.~\cite[p.~163, Theorem 2.15]{Wachter}):
\begin{align}
\varphi(\vec x')&=\frac{1}{4\pi\varepsilon_0}\int\!\de^3\vec x\,\frac{\rho(\vec x)}{|\vec x-\vec x'|}\,, \label{gl1} \\[5pt]
\vec A(\vec x')&=\frac{\mu_0}{4\pi}\int\!\de^3\vec x\,\frac{\vec j(\vec x)}{|\vec x-\vec x'|}\,, \label{gl2} 
\end{align}
where the vector potential fulfills the Coulomb gauge condition, $\nabla\cdot\vec A=0$, on account of the continuity equation (which reduces to $\nabla\cdot\vec j=0$ in the static case).
Hence, with the static, inhomogeneous Maxwell equations 
\begin{align}
\rho(\vec x)&=\varepsilon_0 \h \nabla\cdot\vec E(\vec x)\,, \label{sta_1} \\[5pt]
\vec j(\vec x)&=\nabla\times\vec B(\vec x)/\mu_0\,, \label{sta_2}
\end{align}
we obtain
\begin{align}
\varphi(\vec x')&=\frac{1}{4\pi}\int\!\de^3\vec x\,\frac{\nabla\cdot\vec E(\vec x)}{|\vec x-\vec x'|} \,,\\[5pt]
\vec A(\vec x')&=\frac{1}{4\pi}\int\!\de^3\vec x\,\frac{\nabla\times\vec B(\vec x)}{|\vec x-\vec x'|} \,.
\end{align}
By partial integrations, this is equivalent to
\begin{align}
\varphi(\vec x')&=\frac{1}{4\pi}\int\!\de^3\vec x\,\frac{\vec x-\vec x'}{|\vec x-\vec x'|^3} \cdot \vec E(\vec x) \,, \\[5pt]
\vec A(\vec x')&=\frac{1}{4\pi}\int\!\de^3\vec x\,\frac{\vec x-\vec x'}{|\vec x-\vec x'|^3} \times \vec B(\vec x)\,.
\end{align}
These are the inverse relations of Eqs.~\eqref{m1}--\eqref{m2}. For the functional derivatives of the potentials with respect to the electric 
and magnetic fields, we therefore find
\begin{align} 
\frac{\delta \varphi(\vec x')}{\delta E_i(\vec x)}&=\frac{1}{4\pi} \, \frac{x_i - x'_i}{|\vec x - \vec x'|^3}\,, \label{eq_third_prev} \\[5pt]
\frac{\delta A_{k}(\vec x')}{\delta B_i(\vec x)}&= \frac{1}{4\pi} \, \epsilon_{k\ell i} \, 
\frac{x_\ell - x_\ell'}{|\vec x - \vec x'|^3}\,.\label{eq_third}
\end{align}
By putting these results and Eqs.~\eqref{toplug_1}--\eqref{toplug_2} into Eqs.~\eqref{eq_Pfunc}--\eqref{eq_Mfunc}, we obtain
\begin{align}
\vec P(\vec x) & =
-\frac{1}{4\pi} \int\!\de^3\vec x'\,\rho\ind(\vec x')\, \frac{\vec x-\vec x'}{|\vec x-\vec x'|^3}\,, \label{c_1} \\[5pt]
\vec M(\vec x) & =
\frac{1}{4\pi}\int\!\de^3\vec x'\,\vec j\ind(\vec x')\times\frac{\vec x-\vec x'}{|\vec x-\vec x'|^3}\,. \label{c_2}
\end{align}
On the other hand, Eqs.~\eqref{gl1}--\eqref{gl2} combined with Eqs.~\eqref{m1}--\eqref{m2} imply that the induced fields are given in terms of the induced sources by
\begin{align}
\vec E_{\rm ind}(\vec x)&=\frac{1}{4\pi\varepsilon_0}\int\!\de^3\vec x'\,\rho_{\rm ind}(\vec x')\, \frac{\vec x-\vec x'}{|\vec x-\vec x'|^3}\,, \label{func_der_1} \\[5pt]
\vec B_{\rm ind}(\vec x)&=\frac{\mu_0}{4\pi}\int\!\de^3\vec x'\,\vec j_{\rm ind}(\vec x')\times\frac{\vec x-\vec x'}{|\vec x-\vec x'|^3}\,. \label{func_der_2}
\end{align}
The comparison of these equations therefore shows that
\begin{align}
\vec P(\vec x)& =-\varepsilon_0 \h \vec E_{\rm ind}(\vec x)\,,\\[5pt]
\vec M(\vec x)& =\vec B_{\rm ind}(\vec x)/\mu_0\,.
\end{align}
We conclude that the thermodynamic definitions of polarization and magnetization reproduce their fundamental
identifications with induced electric and magnetic fields.

\subsection{Susceptibilities and response functions}\label{app_thermo4}

In the last subsection, we have shown that the thermodynamic definitions of polarization and magnetization are in accord with our
general identifications \eqref{eq_PE}--\eqref{eq_BB}. For most practical calculations, however, these definitions themselves are not as important as the ensuing determination of the respective susceptibilities,\footnote{Note that Eq.~\eqref{defchie} differs from the standard definition of the electric susceptibility $\chi_{\rm e}$ given by Eq.~\eqref{eq_MatRel10} or equivalently,
\begin{equation*}
 \tsr{\chi}_{\rm e} = \frac{1}{\varepsilon_0} \h \frac{\partial \vec P}{\partial \vec E_{\rm tot}} \,.
\end{equation*}
The latter quantity actually corresponds to the {\itshape proper} electric susceptibility, 
which is related to the (direct) electric susceptibility $\chi_{\rm p}$ through
\begin{equation*}
 \tsr\chi{}_{\rm p}^{-1} - \tsr{\chi}{}_{\rm e}^{-1} = \tsr 1 \,,
\end{equation*}
For details on proper response functions, see \cite{Giuliani} as well as \cite[Sec.~5.2]{ED1} and \cite[Sec.~2.3]{Refr}.}
\begin{align}
\tsr\chi_{\rm p}&:=\frac{1}{\varepsilon_0} \h \frac{\partial\vec P}{\partial\vec E\ext}=-\frac{1}{\varepsilon_0 V} \h \frac{\partial^2F}{\partial\vec E\ext^2}\,, \label{defchie} \\[5pt]
\tsr\chi_{\rm m}&:=\mu_0 \h \frac{\partial\vec M}{\partial\vec B\ext}=-\frac{\mu_0}{V} \h \frac{\partial^2F}{\partial\vec B\ext^2}\,. \label{defchim}
\end{align}
Here, we have used a compact vector notation, which is equivalent to the component form
\begin{equation}
(\chi_{\rm p})_{ij} = \frac 1{\varepsilon_0} \h \frac{\partial P_i}{\partial E_{\rm ext}^j} = -\frac{1}{\varepsilon_0 V} \h \frac{\partial^2 F}{\partial E_{\rm ext}^i \h \partial E_{\rm ext}^j} \,, \smallskip \vspace{2pt}
\end{equation}
and similarly for the magnetic susceptibility.
In fact, such quantities as the Larmor diamagnetism, the Landau diamagnetism or the Pauli paramagnetism are obtained in the standard calculations as second derivatives of a thermodynamic potential, and hence they correspond to generalized susceptibilities 
(see e.g.~\cite[Secs.~3.2 and 3.4]{Nolting}).
Now, the local generalizations of the above definitions \eqref{defchie}--\eqref{defchim} are again straightforward,
\begin{align}
\tsr\chi_{\rm p}(\vec x,\vec x')&=\frac{1}{\varepsilon_0} \h \frac{\delta\vec P(\vec x)}{\delta\vec E\ext(\vec x')} = -\frac{1}{\varepsilon_0} \h \frac{\delta^2 F}{\delta \vec E_{\rm ext}(\vec x) \, \delta \vec E_{\rm ext}(\vec x')} \,, \label{eq_def_chie} \\[5pt]
\tsr\chi_{\rm m}(\vec x,\vec x')&=\mu_0 \h \frac{\delta\vec M(\vec x)}{\delta\vec B\ext(\vec x')} = -\mu_0 \h \frac{\delta^2 F}{\delta \vec B_{\rm ext}(\vec x) \, \delta \vec B_{\rm ext}(\vec x')} \,. \label{eq_def_chim}
\end{align}
However, it is a priori not clear how these time-independent {\itshape susceptibilities} \linebreak are related to dynamical, i.e.~time-dependent {\itshape response functions.} In particular, the question arises of how the formulae \eqref{eq_def_chie}--\eqref{eq_def_chim} from thermodynamics square with the Kubo formalism, which in ab initio materials physics forms the basis for microscopic calculations of linear electromagnetic response properties.

In this final subsection, we show the consistency of these two approaches. Concretely, 
we will show that the generalized susceptibilities \eqref{eq_def_chie}--\eqref{eq_def_chim} evaluated at 
temperature zero precisely agree with the {\itshape instantaneous} (i.e. zero-frequency) limits of the corresponding response functions, hence
\begin{align}
\tsr\chi_{\rm p}(\vec x,\vec x')&=-\tsr{\chi}\EE(\vec x,\vec x';\omega=0)\,, \label{toprove_1} \\[5pt]
\tsr\chi_{\rm m}(\vec x,\vec x')&=\tsr\chi\BB(\vec x,\vec x';\omega=0)\,. \label{toprove_2}
\end{align}
The quantities on the right hand side are defined as the Fourier transforms of the corresponding response functions in the time domain, i.e.,
\begin{equation}
\tsr\chi(\vec x,\vec x';\omega)=\int\! c \, \de \tau \,\e^{{\rm i}\omega\tau}\,\tsr\chi(\vec x,\vec x';\tau)\,,
\end{equation}
where $\tau = t - t'$ denotes the difference of the two time arguments. Note that for instantaneous response functions, there is no difference between the total and the partial functional derivatives, because the electric and magnetic fields decouple. (This also follows directly by comparing Eqs.~(5.50)--(5.53) in Ref.~\cite{ED1} with the corresponding Eqs.~(6.37)--(6.40) there.) 

In the following, we will state our assertion in a more general framework, which is described in detail in App.~\ref{subsec_genKubo}. In brief, we consider the time-dependent Hamiltonian \smallskip
\begin{equation}
\hat H(t)=\hat H(X(t)) \equiv \hat H(X_1(t), \ldots, X_n(t)) \,, \smallskip
\end{equation}
which is assumed to coincide for $t < t_0$ with the unperturbed Hamiltonian $\hat H(X \equiv 0) = \hat H_0$\hh, and, furthermore, we consider the special observables
\begin{equation}
\hat B_i(X)=-\frac{\partial\hat H(X)}{\partial X_i}\,.\label{eq_defObs}
\end{equation}
We are interested in the variation of their expectation value,
\begin{equation}
 B_i(t) = \langle \Psi(t) \mid \hat B_i(X(t)) \mid \Psi(t) \rangle \,,
\end{equation}
with respect to the reference state $|\Psi(t = t_0)\rangle = |\Psi_0\rangle$, which we define as the ground state of $\hat H_0$\hh.
The response function
\begin{equation}
 \chi_{ij}(t - t') := \frac{\delta B_i(t)}{\delta X_j(t')} \h \bigg|_{X \h \equiv \h 0} \label{defchiij}
\end{equation}
is then given by the generalized Kubo formula (see Eq.~\eqref{eq_GenKuboForm}),
\begin{equation}
\begin{aligned}
 \chi_{ij}(t - t') & = \delta(c\h t - c\h t') \, \big\langle \Psi_0 \hh \big| \h \partial_j \hat B_i \h \big| \Psi_0 \big\rangle \\[5pt]
  & \quad \, + \frac{\i}{c\h \hbar} \, \varTheta(t - t') \, \big\langle \Psi_0 \hh  \big| \big[ (\hat B_i)_{\mathrm I}(t), (\hat B_j)_{\mathrm I}(t') \big] \big| \Psi_0 \big\rangle\,, \label{eq_GenKuboForm1}
\end{aligned}
\end{equation}
where it is understood that the right hand side has to be evaluated at vanishing parameters, $X \equiv 0$. Note that this response function indeed depends only on the time difference $t - t'$, which can be shown easily using the time evolution in the interaction picture of the operators $\hat B_i$ and $\hat B_j$ inside the commuator. We will now first show the following general identity, and then apply it to the electric and magnetic susceptibilities.

\bigskip
\noindent
{\bfseries Theorem.} The instantaneous limit of the response function $\chi_{ij}$ equals the negative second 
derivative of the energy with respect to the corresponding external parameters, i.e.,
\begin{equation}
\chi_{ij}(\omega=0) \h = \h -\frac{\partial^2 E(X)}{\partial X_i \h \partial X_j} \, \bigg|_{X \h \equiv \h 0} \,,
\end{equation}
where $E(X)$ is defined for each $X$ as the energy in the ground state $\ket{\Psi_0(X)}$ of the Hamiltonian $\hat H(X)$, i.e.,
\begin{equation}
E(X) = \langle\Psi_0(X) \mid \hat H(X) \mid \Psi_0(X)\rangle \,,
\end{equation}
and where the frequency-dependent response function (with $\tau = t - t'$)
\begin{equation}
\chi_{ij}(\omega)=\int \! c\,\de \tau\,\e^{{\rm i}\omega\tau} \h \chi_{ij}(\tau)
\end{equation}
is the Fourier transform of Eq.~\eqref{eq_GenKuboForm1} with respect to the time variables.

\bigskip
\noindent
{\bfseries Proof.} We use the spectral representation of the second term in the Kubo formula \eqref{eq_GenKuboForm1}. This is obtained by introducing a complete set of orthonormal eigenvectors of the unperturbed Hamiltonian,
\begin{equation}
 \hat H(X \equiv 0) \h\h \ket{\Psi_s} = E_s \h \ket{\Psi_s} \,,
\end{equation}
and inserting the identity operator,
\begin{equation}
 \hat{\mathrm I} = \sum_s \ket{\Psi_s} \bra{\Psi_s} \,, \vspace{-2pt}
\end{equation}
between the two operators $\hat B_i$ and $\hat B_j$ inside the commutator. This yields the expression in the time domain,
\begin{align}
 \chi_{ij}(\tau) & = \delta(c \h \tau) \, \langle \Psi_0 \hh \mid  \h \partial_j \hh \hat B_i \h \mid  \Psi_0 \rangle 
\\[5pt] \nonumber 
  & \quad \, + \frac{\i}{c \h \hbar} \, \varTheta(\tau) \, \sum_{s \not = 0} \Big( \e^{-\i \tau(E_s - E_0)/\hbar} \, \langle \Psi_0 \mid \hat B_i \mid \Psi_s \rangle \langle \Psi_s \mid \hat B_j \mid \Psi_0 \rangle \\[-5pt] \nonumber & \hspace{3.5cm} - \e^{\i \tau (E_s - E_0)/\hbar} \, \langle \Psi_0 \mid \hat B_j \mid \Psi_s \rangle \langle \Psi_s \mid \hat B_i \mid \Psi_0 \rangle \Big) \,,
\end{align}
where the contribution from $s = 0$ vanishes. By Fourier transformation, this is equivalent to
\begin{align}
 \chi_{ij}(\omega) & = \langle \Psi_0 \hh \mid   \h \partial_j \hh \hat B_i \h \mid   \Psi_0 \rangle \\[5pt] \nonumber
 & \quad \, - \sum_{s \not = 0} \bigg( \frac{\langle\Psi_0\mid \hat B_i\mid \Psi_s\rangle\langle\Psi_s\mid \hat B_j\mid \Psi_0\rangle}{\hbar\omega-(E_s-E_0)+\i\eta}
 -\frac{\langle\Psi_0\mid \hat B_j\mid \Psi_s\rangle\langle\Psi_s\mid \hat B_i\mid \Psi_0\rangle}{\hbar\omega+(E_s-E_0)+\i\eta} \bigg) \,,
\end{align}
where the infinitesimal $\eta > 0$ corresponds  to the retardation in the time domain. Evaluating this expression at $\omega = 0$ yields the instantaneous response function,
\begin{align}
 & \chi_{ij}(\omega = 0) = \\ \nonumber
 & \langle \Psi_0 \hh \mid  \h \partial_j \hh \hat B_i \h \mid  \Psi_0 \rangle + 
 2 \, \mathfrak{Re} \h \bigg( \sum_{s \not = 0} \frac{\langle\Psi_0\mid \hat B_i\mid \Psi_s\rangle\langle\Psi_s\mid \hat B_j\mid \Psi_0\rangle}{E_s-E_0} \bigg) \,.
\end{align}
Inserting the definition \eqref{eq_defObs} of the observables further yields
\begin{align}
 & \chi_{ij}(\omega = 0) = \label{eq_3} \\ \nonumber
 & -\langle \Psi_0 \hh \mid  \h \partial_i \h \partial_j \hat H \h \mid  \Psi_0 \rangle - 2 \, \mathfrak{Re} \h \bigg( \sum_{s \not = 0} \frac{\langle\Psi_0\mid \partial_i \hat H\mid \Psi_s\rangle\langle\Psi_s\mid \partial_j \hat H\mid \Psi_0\rangle}{E_0-E_s} \bigg) \,,
\end{align}
where the the right-hand side of the equation has to be evaluated at $X \equiv 0$ (which is implicit in our notation). 
By the second-order Hellmann--Feynman theorem, which we have stated in Ref.~\cite[Sec.~5.1]{EffWW}, Eq.~\eqref{eq_3} is precisely equivalent to \smallskip
\begin{equation}
 \chi_{ij}(\omega = 0) \h = \h -\partial_i \h \partial_j \h \langle \Psi_0 \mid \hat H \mid \Psi_0 \rangle \hh \big|_{X \h \equiv \h 0} \,, \smallskip \vspace{2pt}
\end{equation}
which was the assertion. \qed

\bigskip
It remains to deduce the formulae \eqref{toprove_1}--\eqref{toprove_2} from the above theorem. For this purpose, 
we first apply the theorem to the fundamental response tensor \eqref{fund_resp_tensor}. Thus, we consider a general Hamiltonian
\begin{equation}
 \hat H[A] = \hat H_0 + \hat H_{\rm int}[A]
\end{equation}
depending on an external four-potential $A \equiv A_{\rm ext}$\hh, while
the observable is the induced four-current $j \equiv j_{\rm ind}$\hh. The latter is the negative derivative 
of the Hamiltonian with respect to the external four-potential (cf.~Eq.~\eqref{HjA}),
\begin{equation}
 \,\hat{\!j}^\mu(\vec x) = -\frac{\delta \hat H_{\rm int}[A]}{\delta A_\mu(\vec x)} \,.
\end{equation}
For concreteness, we may think of $\hat H[A]$ as the Pauli Hamiltonian given by Eqs.~\eqref{pauli_1}--\eqref{pauli_2}.
Our theorem now shows that the fundamental response tensor evaluated at zero frequency equals the negative second derivative of the energy with respect to the four-potential, i.e.,
\begin{equation} \label{zw_1}
 \chi\indices{^\mu_\nu}(\vec x, \vec x'; \omega = 0) \h = \h -\frac{\delta^2 E[A]}{\delta A_\mu(\vec x) \h \delta A^\nu(\vec x')} \h \bigg|_{A \h \equiv \h 0} \,.
\end{equation}
Here, $E[A]$ is the ground-state energy given by
\begin{equation}
 E[A] = \big\langle \Psi_0[A] \h\hh \big| \hh \hat H_0 + \hat H_{\rm int}[A] \, \big| \Psi_0[A] \h \big\rangle \,, \smallskip
\end{equation}
which at zero temperature coincides with the free energy,
\begin{equation}
 E[A] = F[A] \qquad (T = 0) \,.
\end{equation}
On the other hand, the static electric and magnetic response functions are given by functional chain rules as
\begin{align}
 & \big(\chi\EE\big)_{ij}(\vec x,\vec x';\omega = 0) \label{zw_2} \\[3pt] \nonumber
 & \qquad = \int \! \de^3 \vec y \int \! \de^3 \vec y' \, \frac{\delta E_i(\vec x)}{\delta \rho(\vec y)} \, \frac{1}{c^2} \, \chi\indices{^0_0}(\vec y, \vec y'; \omega = 0) \, \frac{\delta \varphi(\vec y')}{\delta E_j(\vec x')} \,, \\[8pt] 
 & \big(\chi\BB\big)_{ij}(\vec x,\vec x';\omega = 0) \label{zw_3} \\[3pt] \nonumber
 & \qquad = \int \! \de^3 \vec y \int \! \de^3 \vec y' \, \frac{\delta B_i(\vec x)}{\delta j_k(\vec y)} \, \chi_{k\ell}(\vec y, \vec y'; \omega = 0) \, \frac{\delta A_\ell(\vec y')}{\delta B_j(\vec x')} \,.
\end{align}
Here, we have used that a static electric field $\vec E$ depends only on the charge density $\rho$, while a static magnetic field $\vec B$ depends 
only on the current density~$\vec j$. Similarly, in the static case, the scalar potential $\varphi$ depends only on $\vec E$, while the vector potential $\vec A$  depends only on $\vec B$. The explicit expressions of the functional derivatives have been obtained already in the previous subsection (see Eqs.~\eqref{eq_third_prev}--\eqref{eq_third} and \eqref{func_der_1}--\eqref{func_der_2}):
\begin{align}
 \frac{\delta E_i(\vec x)}{\delta \rho(\vec x')} & = \frac{1}{4\pi\varepsilon_0} \h \frac{x_i - x_i'}{|\vec x - \vec x'|^3} = \frac{1}{\varepsilon_0} \h \frac{\delta \varphi(\vec x')}{\delta E_i(\vec x)} \label{id_1} \,, \\[5pt]
 \frac{\delta B_i(\vec x)}{\delta j_k(\vec x')} & = \frac{\mu_0}{4\pi} \, \epsilon_{i k \ell} \, \frac{x_\ell - x_\ell'}{|\vec x - \vec x'|^3} = \mu_0 \h \frac{\delta A_k(\vec x')}{\delta B_i(\vec x)} \,. \label{id_2}
\end{align}
By substituting these identities in Eqs.~\eqref{zw_2}--\eqref{zw_3} and using the result \eqref{zw_1}, we obtain
\begin{align}
 & \big(\chi\EE\big)_{ij}(\vec x,\vec x';\omega = 0) = \\[3pt] \nonumber
 & \qquad = \frac{1}{\varepsilon_0} \int \! \de^3 \vec y \int \! \de^3 \vec y' \, \frac{\delta \varphi(\vec y)}{\delta E_i(\vec x)} \, \frac{\delta^2 F}{\delta \varphi(\vec y) \h \delta \varphi(\vec y')} \, \frac{\delta \varphi(\vec y')}{\delta E_j(\vec x')} \\[8pt]
 & \big(\chi\BB\big)_{ij}(\vec x,\vec x';\omega = 0) \,, \\[3pt] \nonumber
 & \qquad = -\mu_0 \int \! \de^3 \vec y \int \! \de^3 \vec y' \, \frac{\delta A_k(\vec y)}{\delta B_i(\vec x)} \, \frac{\delta^2 F}{\delta A_k(\vec y) \h \delta A_\ell(\vec y')} \, \frac{\delta A_\ell(\vec y')}{\delta B_j(\vec x')} \,.
\end{align}
Now, employing again the functional chain rule and comparing the resulting equations with the definitions \eqref{toprove_1}--\eqref{toprove_2} of the electric and magnetic susceptibilities, we finally obtain
\begin{align}
 (\chi\EE)_{ij}(\vec x, \vec x'; \omega = 0) & \h = \h \frac{1}{\varepsilon_0} \, \frac{\delta^2 F}{\delta E_i(\vec x) \h \delta E_j(\vec x')} \h = \h -(\chi_{\rm p})_{ij}(\vec x, \vec x') \,, \\[5pt]
 (\chi\BB)_{ij}(\vec x, \vec x'; \omega = 0) & \h = \h -\mu_0 \, \frac{\delta^2 F}{\delta B_i(\vec x) \h \delta B_j(\vec x')} \h = \h (\chi_{\rm m})_{ij}(\vec x, \vec x')\,,
\end{align}
which are the desired relations.

\bigskip
We close this section with a general remark about susceptibilities and response functions:
Our considerations have shown the general identity of thermodynamic susceptibilities at temperature $T = 0$ and 
instantaneous response functions (i.e.~$\omega=0$) with respect to the ground state of the unperturbed system.
In fact, this suffices for the reproduction of well-known standard results like the 
Landau diamagnetism or the Pauli paramagnetism within the Kubo formalism (see the calculation in Ref.~\cite[Secs.~4.4--4.5]{Giuliani}).
Unfortunately though, this result does not generalize directly to the case $T>0$, where the system is described by a thermal mixture of energy states. 
In this case, a perturbation does not only act on the time evolution of the state of the system but also on the thermal weighting factors. One then has to distinguish
strictly between isothermal and adiabatic susceptibilities, a distinction which as of yet has no counterpart in the Kubo formalism.
We will treat this issue in detail in a later publication. Meanwhile we recommend the illuminating discussion in Ref.~\cite[Sec.~3.2.11]{Giuliani}.

\section{Conclusion}

We have given a systematic critique of the Standard Approach to electrodynamics in media.
This critique has been based on the following central arguments: The Standard Approach 
(i) suffers from the incompleteness of its field equations,
(ii) uses an undefined source splitting,
(iii) ignores the measurable induced electromagnetic fields,
(iv) is unsuitable for, and in fact even inconsistent with the common practice in ab initio materials physics.
To overcome these problems, we have axiomatized this common practice into the Functional Approach,
which we propose as the new, inherently microscopic theory of electrodynamics in media.
We have then elaborated upon the most important conceptual conclusions to be drawn from the Functional Approach and, in particular, 
highlighted the Kubo formalism as its necessary complement for calculating microscopic response functions and macroscopic material constants from first principles.
Finally, we have shown the consistency of the Functional Approach with the definitions of polarization and magnetization as used in thermodynamics.

\section*{Acknowledgements}

This research was supported by the Austrian Science Fund (FWF) within the SFB ViCoM, Grant No. F41, 
and by the DFG Reseach Unit FOR 723. R.\,S.~thanks the Institute for Theoretical Physics at the TU Bergakademie Freiberg for its hospitality. We further thank Manfred Salmhofer and members of the Institute for Theoretical Physics at Heidelberg University for discussions.

\begin{appendices}
\section{Misleading designations in the literature} \label{app_const}

In the traditional textbook literature, the case of a constantly (or ``homogeneously'') polarized or magnetized body is often considered as a counter-example
to the interpretation of $\vec P$ and $\vec M$ as induced electric and magnetic fields (see e.g.~\cite[Secs.~4.3.2 and 6.3.2]{Griffiths}). Here we will show that this reasoning is based on a misleading notation, or more precisely, on the mistaken identity of two actually completely different quantities.

 Consider a body for which the polarization $\vec P$ or magnetization $\vec M$
is constantly given by $\vec P_0$ or $\vec M_0$ within a finite volume $V$ (corresponding to where the body is), and zero outside. Formally, this can be expressed as
\begin{align}
 \vec P(\vec x) & = \vec P_0 \, \chi_{V}(\vec x) \,, \\[3pt]
 \vec M(\vec x) & = \vec M_0 \, \chi_{V}(\vec x) \,,
\end{align}
where $\chi_{V}(\vec x)$ is the characteristic function supported on $V$,
\begin{equation}
 \chi_V(\vec x) = \left\{ \begin{array}{ll} 1 \,, & \textnormal{if } \vec x \in V , \\[5pt]
 0 \,, & \textnormal{otherwise.} \end{array} \right.
\end{equation}
With the vector identities
\begin{align}
\nabla\cdot(f\vec A)&=(\nabla f) \cdot \vec A + f\, \nabla\cdot\vec A\,,\\[3pt]
\nabla\times(f\vec A)&=(\nabla f)\times\vec A+f\, \nabla\times\vec A\,,
\end{align}
this leads to the rotations and divergences
\begin{align}
\nabla\cdot\vec P(\vec x)&=-\vec P_0\cdot\vec n(\vec x)\,, \label{divP} \\[3pt] 
\nabla\times\vec P(\vec x)&=\vec P_0\times\vec n(\vec x)\,, \label{rotP} \\[3pt]
\nabla\cdot\vec M(\vec x)&=-\vec M_0\cdot\vec n(\vec x)\,, \label{divM} \\[3pt]
\nabla\times\vec M(\vec x)&=\vec M_0\times\vec n(\vec x)\,, \label{rotM}
\end{align}
where we have introduced the \mbox{{\itshape surface normal}} $\vec n(\vec x)$ (cf.~Ref.~\cite[Sct.~4.2]{EDFresnel}) as a vector-valued distribution defined by \smallskip
\begin{equation}
 \vec n(\vec x) = -\nabla \chi_{V}(\vec x) \,. \smallskip
\end{equation}
By using Gauss' law, one can show that for any scalar function $f$,
\begin{align}
 \int_{\mathbbm R^3} \! \de^3 \vec x \, f(\vec x) \, \vec n(\vec x)  &= \int_V \de^3 \vec x \,\h \nabla f(\vec x) \\[5pt]
  & = \int_{\partial V} \de \vec A(\vec x) \, f(\vec x) \\[5pt]
  & = \int_{{\mathbbm R}^3} \de^3 \vec x' \h f(\vec x')  \int_{\partial V} \de \vec A(\vec x) \, \delta^3(\vec x-\vec x') \,,
\end{align}
and hence we can interpret
\begin{equation} \label{surf_Dirac}
 \vec n(\vec x') = \int_{\partial V} \de \vec A(\vec x) \, \delta^3(\vec x - \vec x')
\end{equation}
as the vectorial Dirac delta distribution of the surface $\partial V$.\footnote{In fact, in adapted coordinates one shows easily that
\begin{equation}
 \de\vec A_{\partial V}(\vec x) = - \nabla\chi_V(\vec x) \h \de^3\vec x \,,
\end{equation}
such that Gauss' law in the form
\begin{equation}
 \int_V\de^3\vec x \,\nabla\cdot\vec E(\vec x) = \int_{\partial V} \! \de\vec A(\vec x)\cdot\vec E(\vec x) 
\end{equation}
follows directly from the trivial identity
\begin{equation}
 \int_V\de^3\vec x \,\nabla\cdot\vec E(\vec x) = \int_{{\mathbbm R}^3} \! \de^3\vec x \,\chi_{V}(\vec x)\h \nabla\cdot\vec E(\vec x) 
\end{equation}
by means of a partial integration.}

The equations \eqref{rotP}--\eqref{divM} show that the thus-defined static field $\vec P(\vec x)$ is not curl free, while the field $\vec M(\vec x)$ is not divergence free. 
The standard text\-{}book problem now consists in the determination of the electric field $\vec E(\vec x)$ and the magnetic field $\vec B(\vec x)$ generated by 
the constant polarization and magnetization, respectively (see e.g.~\cite[Example 4.2 and Example 6.1]{Griffiths}).

In actual fact, the electric field of a constantly polarized body simply corresponds to the static
electric field generated by the {\itshape surface charge density}
\begin{equation}
\rho_{\partial V}(\vec x):=-\nabla\cdot\vec P(\vec x)\,.
\end{equation}
As opposed to the polarization, the desired electric field 
therefore obeys the 
usual equations for a static electric field, i.e.,
\begin{align}
\nabla\cdot\vec E(\vec x)&=\rho_{\partial V}(\vec x)/\varepsilon_0\,, \label{inhom_max_e} \\[5pt]
\nabla\times\vec E(\vec x)&=0\,. \label{hom_max_e}
\end{align}
Similarly, the magnetic field generated by the constantly magnetized body corresponds to the static magnetic field generated by the {\itshape surface current density} \smallskip
\begin{equation}
\vec j_{\partial V}(\vec x) := \nabla\times\vec M(\vec x)\,, \smallskip
\end{equation}
and therefore obeys the equations
\begin{align}
\nabla\cdot\vec B(\vec x)&=0\,, \label{hom_max_b} \\[5pt]
\nabla\times\vec B(\vec x)&=\mu_0 \h \vec j_{\partial V}(\vec x)\,. \label{inhom_max_b}
\end{align}
This means, the searched-for electric field $\vec E(\vec x)$ differs from the polarization $\vec P(\vec x)$ 
only in that its rotation vanishes identically, while the divergences of the two fields coincide (up to the conversion factor).
Similarly, the searched-for magnetic field $\vec B(\vec x)$ differs from the magnetization $\vec M(\vec x)$ only in that its divergence vanishes identically,
while the rotations coincide. In other words,
the electric field of a constantly polarized body simply corresponds to the {\itshape longitudinal part} of the polarization, while the magnetic field of a constantly magnetized body corresponds to the {\itshape transverse part} of the magnetization:
\begin{align}
\vec E(\vec x)&=-\vec P\L(\vec x) / \varepsilon_0 \,,\\[5pt]
\vec B(\vec x)&=\mu_0 \h \vec M\T(\vec x)\,.
\end{align}
From these considerations, we conclude that the relations between the quantities $\vec P$ and $\vec M$ on the one hand, and $\vec E$ and $\vec B$ on the other hand, have nothing to do with response relations. In particular, {\itshape there is no proportionality} between these fields in the sense of
\begin{align}
 \vec P & \stackrel{?}{=} \varepsilon_0 \h\hh \chi_{\rm e} \h \vec E \,, \\[3pt]
 \vec M & \stackrel{?}{=} \widetilde \chi_{\rm m} \, \vec B / \mu_0 \,,
\end{align}
with some material-specific constants $\chi_{\rm e}$ and $\widetilde \chi_{\rm m}$\hh. This becomes immediately clear from the fact that $\vec E$ and $\vec B$---in contrast to $\vec P$ and $\vec M$---are neither constant outside nor inside the finite body. Consequently, the whole problem has actually nothing to do with electrodynamics of materials.
Instead, the constant polarization and magnetization are obviously {\it fictitious fields}, which 
only serve the purpose of determining the surface charges and currents, which in turn 
determine the electric and magnetic fields by the usual formulae known from electrostatics and magnetostatics.

Unfortunately, in this already confusing situation, it is also common practice to define so-called {\it auxiliary} (i.e.~fictitious) fields, 
unfortunately {\it denoted} by $\vec D$ and $\vec H$, such that
\begin{align}
\vec D&=\varepsilon_0\h \vec E+\vec P\,,\\[5pt]
\vec B&=\mu_0\h (\vec H+\vec M)\,,
\end{align}
which is {\itshape formally analogous} to Eqs.~\eqref{eq_PED}--\eqref{eq_MBH}.
These auxiliary fields are used for the concrete calculation of $\vec E$ and $\vec B$. Of course, 
they just correspond to the {\itshape transverse} and {\itshape longitudinal} parts of the polarization and magnetization, respectively, i.e.,
\begin{align}
\vec D(\vec x)&=\vec P\T(\vec x)\,,\\[5pt]
\vec H(\vec x)&=-\vec M\L(\vec x)\,.
\end{align}
However, these auxiliary fields have again nothing to do with their counterparts $\vec D$ and $\vec H$ used in electrodynamics of materials, which would be related to $\vec E$ and $\vec B$ by material-dependent response relations.

We conclude that these misleading designations have caused a confusion in the traditional textbook literature regarding the nature of the electric and magnetic polarizations, and in particular hindered their natural identification with induced electric and magnetic fields.

\section{Dipole densities as fields} \label{app_dip}
\subsection{Definition of point dipoles}

As a matter of principle, dipole moments are not objects in space, hence it is impossible to associate to a dipole moment ``its'' position.
Instead, given localized (i.e.~sufficiently fast decaying) charge and current densities $\rho(\vec x,t)$ and $\vec j(\vec x,t)$,
we can associate with them {\itshape their} electric and magnetic dipole moments $\vec p(t)$ and $\vec m(t)$ by means of
\begin{align}
\vec{p}(t)&=\int\!\de^3\vec{x}\,\h (\vec{x} - \vec x_0) \,\rho(\vec{x},t)\,, \label{eq_pointdipole1}\\[5pt]
\vec{m}(t)&=\frac{1}{2}\int\! \de^3\vec{x}\,\h (\vec{x} - \vec x_0) \times\vec{j}(\vec{x},t)\,, \label{eq_pointdipole2}
\end{align}
where $\vec x_0$ denotes a {\it reference position}. The electric and magnetic dipole moments naturally occur in the expansion of the static potentials
\begin{align}
\varphi(\vec{x})&=\frac{1}{4\pi\varepsilon_0}\int\!\de^3\vec{x}'\,\frac{\rho(\vec{x}')}{|\vec{x}-\vec{x}'|}\,, \label{via_1} \\[5pt]
\vec{A}(\vec{x})&=\frac{\mu_0}{4\pi}\int\!\de^3\vec{x}'\,\frac{\vec{j}(\vec{x}')}{|\vec{x}-\vec{x}'|}\,, \label{via_2}
\end{align}
around the reference vector $\vec x_0$ (see e.g.~\cite[Sec.~2.5.2]{Wachter}):
\begin{align}
\varphi(\vec{x})&=\frac{1}{4\pi\varepsilon_0} \h \frac{q}{|\vec{x} - \vec x_0|}+\frac{1}{4\pi\varepsilon_0}\frac{\vec{p}\cdot(\vec{x} - \vec x_0)}{|\vec{x} - \vec x_0|^3}+\ldots\,, \label{multipole_1} \\[5pt]
\vec{A}(\vec{x})&=\frac{\mu_0}{4\pi} \h \frac{\vec{m}\times(\vec{x} - \vec x_0)}{|\vec{x} - \vec x_0|^3}+\ldots\,. \label{multipole_2}
\end{align}
In these ``multipole expansions'', $q$ denotes the total charge,
\begin{equation}
 q = \int \! \de^3 \vec x \, \rho(\vec x) \,.
\end{equation}
Correspondingly, the second term in Eq.~\eqref{multipole_1} and the first term in Eq.~\eqref{multipole_2} are referred to as {\itshape dipolar contributions.}

Charge and current densities whose potentials are exclusively given by such dipolar contributions are called {\itshape electric} or {\itshape magnetic point dipoles.} 
These densities are singular ({\itshape distributions} in a mathematical sense) and read
\begin{align}
\rho_{\vec x_0}(\vec x) &= -\vec p_0 \cdot \nabla_{\mh\vec x} \h\hh \delta^3(\vec x - \vec x_0) \,, \label{el_dip} \\[6pt]
\vec{j}_{\vec x_0}(\vec{x}) &= -\vec m_0 \times \nabla_{\mh\vec x} \h\hh \delta^3(\vec x - \vec x_0) \,, \label{ma_dip}
\end{align}
for {\itshape point dipoles} $\vec p_0$ or $\vec m_0$ {\itshape located at $\vec x_0$\hh.} 
In other words, these charge and current densities lead via Eqs.~\eqref{via_1}--\eqref{via_2} exactly to the purely dipolar potentials
\begin{align}
\varphi(\vec{x})&=\frac{1}{4\pi\varepsilon_0} \, \frac{\vec{p}_0\cdot(\vec{x}-\vec x_0)}{|\vec{x}-\vec x_0|^3}\,,\label{eq_potelPD}\\[5pt]
\vec{A}(\vec{x})&=\frac{\mu_0}{4\pi} \, \frac{\vec{m}_0\times(\vec{x}-\vec x_0)}{|\vec{x}-\vec x_0|^3}\,.\label{eq_potmagPD}
\end{align}
Conversely, the definitions of the dipole moments for given charge and current densities, Eqs.~\eqref{eq_pointdipole1}--\eqref{eq_pointdipole2},
precisely yield the vector-valued parameters $\vec p_0$ and $\vec m_0$
in the case of the above singular densities corresponding to point dipoles, Eqs.~\eqref{el_dip}--\eqref{ma_dip}.

Intuitively, we can think of the electric point dipole as being produced by the singular charge density of two equal but opposite charges $q$ and $-q$ 
with distance vector $\vec \ell$, in the limit $|\vec \ell| \to 0$ where the product of charge and distance stays constant, i.e.~$\vec p_0=q \h \vec \ell$\h. Explicitly, this means
\begin{equation}
 \rho_{\vec x_0}(\vec x) =\lim_{|\vec \ell|\rightarrow 0} \h q(|\vec \ell|) \h \big(\delta^3(\vec x-(\vec x_0+\vec \ell))-\delta^3(\vec x-\vec x_0)\big)\,,
\end{equation}
where $\vec \ell$ is directed parallel to $\vec p_0$\hh, and
\begin{equation}
 q(|\vec \ell|) = \frac{|\vec p_0|}{|\vec \ell|} \,.
\end{equation}
Similarly, the magnetic
point dipole can be produced by an electric current~$I$ circulating in a loop with radius $R$ around $\vec x_0$ in the plane perpendicular to $\vec m_0$\hh, 
in the limit $R \to 0$ where now the product of current and area stays constant in the sense of $|\vec m_0| = \pi R^2 I$.
Explicitly, this means
\begin{equation}
\vec{j}_{\vec x_0}(\vec{x}) =\lim_{R\rightarrow 0} \h I(R) \oint_{\mathcal C(R)} \mh \de\vec x'\,\delta^3(\vec x-\vec x')\,,
\end{equation}

\pagebreak \noindent
where \smallskip
\begin{equation}
 I(R) = \frac{|\vec m_0|}{\pi R^2} \,, \medskip
\end{equation}
and where $\mathcal C(R)$ is the positively oriented boundary of a disk with radius $R$, center $\vec x_0$ and surface normal parallel to $\vec m_0$\hh. Concretely, 
if we choose the coordinate system such that $\vec x_0 = 0$ coincides with the origin  and $\vec m_0 = |\vec m_0|\h\hh (0, 0, 1)^{\rm T}$ is parallel to the $z$-axis, then $\mathcal C(R)$ can be parametrized as
\begin{equation}
 \vec x' = (R \cos \varphi, \h R \sin \varphi, \h 0)^{\rm T} \,, \qquad 0 \leq \varphi < 2\pi \,.
\end{equation}
The charge and current densities of electric and magnetic point dipoles have the following interesting features: The total force
\begin{equation} \label{eq_1}
\vec F(\vec x_0) =\int \! \de^3\vec x\,\big( \hh\rho_{\vec x_0}(\vec x) \h\vec E(\vec x)+\vec j_{\vec x_0}(\vec x)\times\vec B(\vec x)\big)\,,
\end{equation}
exerted on them by static external fields $\vec E(\vec x)$ and $\vec B(\vec x)$ turns out to be 
\begin{equation} \label{eq_2}
\vec F(\vec x_0) = \nabla_{\mh\vec x_0} \big( \h \vec p_0\cdot\vec E(\vec x_0)+\vec m_0\cdot\vec B(\vec x_0)\big)\,,
\end{equation}
and can thus be written as minus the gradient of a ``potential energy''
\begin{equation} \label{Uxo}
U(\vec x_0) =-\vec p_0\cdot\vec E(\vec x_0)-\vec m_0\cdot\vec B(\vec x_0)\,.
\end{equation}
Note that for this formula to be true, it is essential that the rotation of the electric field vanishes, $\nabla \times \vec E = 0$. 
The equality between \eqref{eq_1} and \eqref{eq_2} can then be shown using the general vector identity \cite{Jackson}
\begin{equation}
 \vec Y \times (\nabla \times \vec Z) + \vec Z \times (\nabla \times \vec Y) + (\vec Y \cdot \nabla) \vec Z + (\vec Z \cdot \nabla) \vec Y = \nabla (\vec Y \cdot \vec Z) \,,
\end{equation}
which for $\vec Y(\vec x_0) \equiv \vec E(\vec x_0)$ and $\vec Z(\vec x_0) \equiv \vec p_0$ implies
\begin{equation}
 (\hh\vec p_0 \cdot \nabla_{\mh\vec x_0}) \h \vec E(\vec x_0) = \nabla_{\mh\vec x_0} ( \vec p_0 \cdot \vec E(\vec x_0)) \,.
\end{equation}
We remark that the result \eqref{Uxo} for the potential energy can also be deduced directly from the standard interaction Hamiltonian
\begin{equation} \label{hint}
H_{\rm int}=\int\!\de^3\vec x\,\big( \hh \rho_{\vec x_0}(\vec x) \h \varphi(\vec x)-\vec j_{\vec x_0}(\vec x)\cdot\vec A(\vec x) \big),
\end{equation}
where the static potentials are defined by
\begin{align}
 \vec E(\vec x) & = -\nabla \varphi(\vec x) \,, \label{stat_pot_1} \\[5pt]
 \vec B(\vec x) & = \nabla \times \vec A(\vec x) \,. \label{stat_pot_2}
\end{align}
Indeed, by inserting the definitions \eqref{el_dip}--\eqref{ma_dip} into Eq.~\eqref{hint} and performing partial integrations, we obtain
\begin{equation}
 H_{\rm int} = -\vec p_0\cdot\vec E(\vec x_0)-\vec m_0\cdot\vec B(\vec x_0) \,,
\end{equation}
which coincides with Eq.~\eqref{Uxo}.

Furthermore, we remark that Eq.~\eqref{hint} is analogous to the static interaction Hamiltonian for induced fields given by
\begin{equation}
H_{\rm int}=\int\!\de^3\vec x\,\big(\hh\rho_{\rm ind}(\vec x)\h \varphi\ext(\vec x)-\vec j_{\rm ind}(\vec x)\cdot\vec A\ext(\vec x)\big)\,.\label{eq_fundIntHam}
\end{equation}
In fact, after insertion of the static Maxwell equations,
\begin{align}
\rho_{\rm ind}(\vec x)&=-\nabla\cdot\vec P(\vec x)\,,\\[5pt]
\vec j\ind(\vec x)&=\nabla\times\vec M(\vec x)\,,
\end{align}
and subsequent partial integrations, Eq.~\eqref{eq_fundIntHam} is equivalent to
\begin{equation}
H_{\rm int}=-\int\!\de^3\vec x\,\big(\vec P(\vec x)\cdot\vec E\ext(\vec x)+\vec M(\vec x)\cdot\vec B\ext(\vec x)\big)\,.\label{eq_apprIntHam}
\end{equation}
This last expression is analogous to the ``potential energy'' of the point dipoles, Eq.~\eqref{Uxo}. (It is also analogous to the expansion of the free energy, 
Eq.~\eqref{eq_ExpFreeEn_local}.) At the same time, the above form \eqref{eq_apprIntHam} of the more fundamental interaction Hamiltonian \eqref{eq_fundIntHam} explains why in many approximate calculations it suffices to consider such a ``dipolar coupling'' instead of the more fundamental coupling of the four-current to the external four-potential.

Finally, we note another analogy: The static point-dipole formulae \eqref{el_dip}--\eqref{ma_dip} generalize to time-dependent point dipoles (located at $\vec x_0$) as
\begin{align}
\rho(\vec x,t)&=-\vec{p}_0(t)\cdot\nabla_{\mh\vec x}\h\hh \delta^3(\vec{x}-\vec x_0)\,,\label{eq_pointdipoledens1}\\[5pt]
\vec{j}(\vec{x},t)&=-\vec{m}_0(t)\times\nabla_{\mh\vec x} \h\hh \delta^3(\vec{x}-\vec x_0)+ \partial_t \h \vec p_0(t) \, \delta^3(\vec x-\vec x_0)\,,\label{eq_pointdipoledens2}
\end{align}
where the last contribution is necessary to ensure the continuity equation,
\begin{equation}
\partial_t \hh \rho(\vec x,t)+\nabla\cdot\vec j(\vec x,t)=0\,.
\end{equation}
With these singular charge and current distributions, we further introduce the {\itshape point-dipole densities} as
\begin{align}
\vec p(\vec x,t)&=\vec p_0(t) \, \delta^3(\vec x-\vec x_0)\,, \label{sing_p} \\[5pt]
\vec m(\vec x,t)&=\vec m_0(t) \, \delta^3(\vec x-\vec x_0)\,. \label{sing_m}
\end{align}
These singular densities fulfill the equations
\begin{align}
\nabla\cdot\vec p(\vec x,t)&=-\rho(\vec x,t)\,,\\[5pt]
\nabla\times\vec m(\vec x,t)&=\vec j(\vec x,t)-\partial_t \h \vec p(\vec x,t)\,,
\end{align}
which are analogous to the inhomogeneous Maxwell equations for the induced fields, Eqs.~\eqref{eq_MaxPM1} and \eqref{eq_MaxPM4}. 
For these singular point-dipole densities, the transverse part of $\vec p(\vec x,t)$ and the longitudinal part of $\vec m(\vec x, t)$ 
are determined by the explicit equations \eqref{sing_p}--\eqref{sing_m}.

\subsection{Transition to continuous distributions}

The above considerations have dealt with hypothetical point dipoles only. In the Standard Approach to electrodynamics in media, however,
the polarization and magnetization are supposed to describe (macroscopically) continuous dipole densities, which at the same time are considered
as bona fide fields depending on space-time points $(\vec x,t)$.
The only straightforward way to derive these from the above formulary seems to be the replacement of the
singular point-dipole densities \eqref{sing_p}--\eqref{sing_m} with continuous distributions $\vec P(\vec x, t)$ and $\vec M(\vec x, t)$.
For the sake of clarity, we restrict ourselves again to the static case, where these hypothetical, continuous point-dipole densities 
would correspond to the ``polarization charge density'' and the ``magnetization current density'' given respectively by \vspace{-3pt}
\begin{align}
\rho_{\rm p}(\vec x) & =-\int \! \de^3\vec x'\,\vec P(\vec x')\cdot \nabla_{\mh\vec x} \h\hh \delta^3(\vec x-\vec x')\,, \label{el_dip_cont} \\[3pt]
\vec j_{\rm m}(\vec x) & =-\int \! \de^3\vec x'\,\vec M(\vec x')\times \nabla_{\mh\vec x} \h\hh \delta^3(\vec x-\vec x')\,. \label{ma_dip_cont}
\end{align}
These formulae obviously generalize the corresponding relations \eqref{el_dip}--\eqref{ma_dip} for singular point dipoles. However, by performing partial integrations, one sees that the above equations \eqref{el_dip_cont}--\eqref{ma_dip_cont} are equivalent to
\begin{align}
 \rho_{\rm p}(\vec x) & = -\nabla \cdot \vec P(\vec x) \,, \label{rhoP} \\[5pt]
 \vec j_{\rm m}(\vec x) & = \nabla  \times \vec M(\vec x) \,, \label{jM}
\end{align}
and hence they cannot be used to {\itshape define} $\vec P(\vec x)$ and $\vec M(\vec x)$ uniquely from any given distribution of charges and currents. Instead, these equations define only the longitudinal part of $\vec P(\vec x)$ and the transverse part of $\vec M(\vec x)$. 
This is in contrast to the case of point dipoles, where by Eqs.~\eqref{el_dip}--\eqref{ma_dip}, $\vec p_0$ \linebreak and $\vec m_0$ are uniquely determined once the singular densities $\rho_{\vec x_0}(\vec x)$ and $\vec j_{\vec x_0}(\vec x)$ are given.

Now, consider the potentials that would be created by the hypothetical, continuous point-dipole densities,
\begin{align}
\varphi(\vec{x})&=\frac{1}{4\pi\varepsilon_0}\int\!\de^3\vec x'\,\h\frac{\vec{P}(\vec x')\cdot(\vec{x}-\vec x')}{|\vec{x}-\vec x'|^3}\,,\\[5pt]
\vec{A}(\vec{x})&=\frac{\mu_0}{4\pi}\int\!\de^3\vec x'\,\h\frac{\vec{M}(\vec x')\times(\vec{x}-\vec x')}{|\vec{x}-\vec x'|^3}\,.
\end{align}
These equations are the continuous analoga of Eqs.~\eqref{eq_potelPD}--\eqref{eq_potmagPD}. However, by rewriting them as
\begin{align}
\varphi(\vec{x})&=-\frac{1}{4\pi\varepsilon_0}\int\!\de^3\vec x'\,\h\frac{(\nabla\cdot\vec{P})(\vec x')}{|\vec{x}-\vec x'|}\,,\\[5pt]
\vec{A}(\vec{x})&=\frac{\mu_0}{4\pi}\int\!\de^3\vec x'\,\h\frac{(\nabla\times\vec{M})(\vec x')}{|\vec{x}-\vec x'|}\,,
\end{align}
we see again that the electric dipole density $\vec P$ is only defined up to a transverse vector field, while the magnetic
dipole density $\vec M$ is only defined up to a longitudinal vector field.
Furthermore, from the static relations \eqref{stat_pot_1}--\eqref{stat_pot_2}
and the Helmholtz vector theorem,
\begin{align}
 & \vec F(\vec x) \equiv \vec F_{\mathrm{L}}(\vec x) + \vec F_{\mathrm{T}}(\vec x) \label{eq_HVT_1} \\[5pt]
 & = -\nabla\left(\frac{1}{4\pi}\int\!\de^3\vec x'\,\frac{\nabla\cdot\vec F(\vec x')}{|\vec x-\vec x'|}\right) \h +\h 
\nabla\times\left(\frac{1}{4\pi}\int\!\de^3\vec x'\,\frac{\nabla\times\vec F(\vec x')}{|\vec x-\vec x'|}\right) \,,\label{eq_HVT}
\end{align}
we obtain immediately the identities
\begin{align}
\vec P\L(\vec x)&=-\varepsilon_0 \h \vec E(\vec x)\,, \label{id_nochmal_1} \\[5pt]
\vec M\T(\vec x)&=\vec B(\vec x) / \mu_0 \,. \label{id_nochmal_2}
\end{align}
This means, the longitudinal part of the electric dipole density $\vec P_{\rm L}$ just 
coincides with the electric field $\vec E$ generated by the corresponding charge density \eqref{rhoP},
and similarly, the transverse part of the magnetic dipole density $\vec M_{\rm T}$ coincides with the magnetic field $\vec B$ generated by the current density \eqref{jM}. The identities \eqref{id_nochmal_1}--\eqref{id_nochmal_2} can of course also be deduced directly from Eqs. \eqref{rhoP}--\eqref{jM}, since the latter coincide with the static Maxwell equations for the electric and magnetic fields $\vec E$ and $\vec B$ (up to the conversion factors).

In summary, we precisely recover again the state of affairs which we had already encountered in the Standard Approach with its incomplete field equations. 
In particular, we have shown that the attempt to give a precise meaning to the electric and magnetic polarizations as  
``continuous dipole densities'' via the respective formulae for point dipoles 
simply leads back to the problem of insufficiently defined fields $\vec P(\vec x)$ and $\vec M(\vec x)$, 
whose longitudinal and transverse parts respectively coincide with electric and magnetic fields of suitably defined charge and current densities.

\section{Electromagnetic linear response}\label{app_elmKubo}

\subsection{Generalized Kubo formula} \label{subsec_genKubo}

Consider a Hamiltonian of the general form
\begin{equation} \label{gen_Ham}
\hat H(X_1,\ldots,X_n) = \hat H_0 + \hat H_{\rm int}(X_1,\ldots,X_n)\,,
\end{equation}
where $H_0$ represents the isolated system, while $H_{\rm int}$ is an external perturbation which is
supposed to depend on the parameters $X_j$ in such a way that $H_{\rm int}=0$ if all $X_j=0$. Defining the multi-index
$
 X = (X_1, \ldots, X_n),
$
this condition can be expressed as
\begin{equation}
 \hat H_{\rm int}(X \equiv 0) \h = \h 0 \,.
\end{equation}
Furthermore, let the parameters depend on time, $X_j=X_j(t)$,
such that
\begin{equation}
 X_j(t)=0 \ \, \forall j\,, \quad \textnormal{if} \ \, t<t_0 \,.
\end{equation}
Consequently, the resulting time-dependent Hamiltonian
\begin{equation}
\hat H(t) \equiv \hat H(X(t))
\end{equation}
coincides with the unperturbed Hamiltonian $\hat H_0$ for $t<t_0$\h, and $t_0$ can be regarded as the time where the external perturbation is switched on. We further define a time-dependent state $|\Psi(t)\rangle$
by the following initial value problem: For $t=t_0$ the state is supposed to coincide with an eigenstate of the unperturbed
Hamiltonian, for concreteness, say, the ground state $|\Psi_0\rangle$ of $\hat H_0$\hh. Hence, \smallskip
\begin{equation}
 |\Psi(t = t_0)\rangle \h = |\Psi_0\rangle \,. \smallskip \vspace{2pt}
\end{equation}
For $t > t_0$\h, the time evolution of $\ket{\Psi(t)}$ is defined by the full Schr\"{o}dinger equation, \smallskip
\begin{equation}
\i\hbar \, \partial_t|\Psi(t)\rangle=\hat H(X(t)) \h |\Psi(t)\rangle\,. \smallskip \vspace{2pt}
\end{equation}
Now, assume that we are given an observable $\hat O$ which may also depend on the same parameters as the Hamiltonian:
\begin{equation}
 \hat O = \hat O(X) \,.
\end{equation}
Its time-dependent expectation value,
\begin{equation}
O(t)=\langle\Psi(t)|\h\hat O(X(t))\h|\Psi(t)\rangle\,,
\end{equation}
is then a functional of the scalar functions $X_j(t')$: firstly, on account of the explicit parameter dependence of the observable, and secondly, 
through the implicit parameter dependence of the time evolution of $|\Psi(t)\rangle$.

The generalized Kubo formula is a statement about
the functional derivative of this expectation value $O(t)$ with respect to the parameter functions $X_j(t')$. Concretely, it says that this functional derivative
evaluated at vanishing parameter functions is given by
\begin{align} \label{eq_GenKuboForm}
 \frac{\delta O(t)}{\delta X_j(t')} \h \bigg|_{X \h \equiv \h 0} & =  \delta(c\h t - c\h t') \, \bigg\langle\! \Psi_0 \h \bigg|\, \frac{\partial\hat O}{\partial X_j}(X \equiv 0) \, \bigg| \h \Psi_0 \! \bigg\rangle \\[5pt] \nonumber
 & \quad \, -\frac{\rm i}{c \h \hbar}\h \varTheta(t-t') \, \bigg\langle\!\Psi_0\h\bigg|\, \bigg[ \, \e^{\i t \hat H_0/\hbar} \, \hat O(X \equiv 0) \, \e^{-\i t \hat H_0 / \hbar} \h, \\ \nonumber
 & \hspace{4.47cm} \e^{\i t' \mh \hat H_0 / \hbar} \, \frac{\partial \hat H_{\rm int}}{\partial X_j} (X \equiv 0) \, \e^{-\i t' \mh \hat H_0 / \hbar} \, \bigg] \, \bigg|\h\Psi_0\! \bigg\rangle\,.
\end{align}
Note the factor $1/c$, which appears in this formula because the functional derivative is defined through the expansion
\begin{equation}
 O(t) = \int \! c \, \de t' \, \bigg( \sum_j \h \frac{\delta O(t)}{\delta X_j(t')} \h \bigg|_{X \h \equiv \h 0} \h \bigg) \, X_j(t') + \mathcal O(X^2) \,.
\end{equation}
Equation \eqref{eq_GenKuboForm} can be written in a more familiar form by introducing the time evolution in the interaction picture,
\begin{equation}
\hat A_{\rm I}(t)=\e^{\i t \hat H_0 / \hbar} \h \hat A(t) \, \e^{-\i t \hat H_0 /\hbar}\,,
\end{equation}
by abbreviating
$\partial_j \hh \hat O \equiv \partial \hat O / \partial X_j$\h,
and by stipulating that all operators on the right hand side of the equation should be evaluated at $X \equiv 0$. Then the generalized Kubo formula can be written compactly as
\begin{equation} \label{eq_GenKuboForm_simpl}
\begin{aligned}
 \frac{\delta O(t)}{\delta X_j(t')} \h \bigg|_{X \h \equiv \h 0} & = \h \delta(c \h t - c \h t') \, \big\langle \Psi_0 \hh \big| \h \partial_j \hh \hat O \h \big| \Psi_0 \big\rangle \\
 & \quad \, {- \frac{\i}{c \h \hbar}} \h \varTheta(t - t') \, \big\langle \Psi_0 \hh  \big| \big[ \hat O_{\mathrm I}(t), (\partial_j \hat H_{\mathrm{int}})_{\mathrm I}(t') \big] \hh \big| \Psi_0 \big\rangle\,.
\end{aligned}
\end{equation}
The proof of this formula is a straightforward, Leibniz-rule based generalization
of the original proof for the ordinary Kubo formula (see e.g.~\cite[Chap.~6]{Bruus}) and will therefore be omitted.
Here, we only note that the generalized Kubo formula reduces to the ordinary Kubo formula under two simplifying assumptions: (i) the observable $\hat O$ does not depend on the perturbative parameters $X_j$\hh, 
and (ii) the interaction Hamiltonian depends linearly on them, such that
\begin{equation}
 \hat H_{\rm int}(X(t)) = -\sum_{j = 1}^k X_j(t) \hh \hat B_j \,,
\end{equation}
with some hermitean operators $\hat B_j$. In this case, Eq.~\eqref{eq_GenKuboForm_simpl} reduces to
\begin{equation}
 \frac{\delta O(t)}{\delta X_j(t')} \h \bigg|_{X \h \equiv \h 0} = \h \frac{\i}{c\h\hbar} \h \varTheta(t - t') \, \big\langle \Psi_0 \hh \big| \big[  \hat O_{\mathrm I}(t), (\hat B_j)_{\rm I}(t') \h \big] \big| \Psi_0 \big\rangle \,,
\end{equation}
which is the well-known Kubo formula. Although the generalized Kubo formula is a fairly obvious generalization of the standard Kubo formula, we felt the necessity of stating Eq.~\eqref{eq_GenKuboForm} explicitly
because it allows for a direct application to the electromagnetic current of the Pauli equation, as will be shown in the next subsection. Furthermore, our formula \eqref{eq_GenKuboForm} has been used in Sec.~\ref{app_thermo4} for the comparison of
response functions and susceptibilities.

\subsection{Electromagnetic current of Pauli spinors} \label{app_pauli}

This subsection is dedicated to the somewhat intricate problem of finding the correct expression 
for the electromagnetic current corresponding to the Pauli equation (see e.g.~\cite[Chap.~XX, \S\,29]{Messiah} or \cite[Sec.~1.4]{Bjorken}).
In fact, in the presence of an external four-potential,
\begin{equation}
 A^\mu\ext(\vec x, t) \equiv A^\mu(\vec x, t) = ( \varphi(\vec x, t)/ c, \h \vec A(\vec x, t))^{\rm T} \,,
\end{equation}
the charge and current densities of the Pauli spinor field,
\begin{equation}
 \Psi(x) = \left( \! \begin{array}{l} \psi_\uparrow(x) \\[5pt] \psi_\downarrow(x) \end{array} \! \right),
\end{equation}
are respectively given by
\begin{align}
\rho(\vec x,t)=&-e \, \sum_s \psi_s^*(\vec x,t) \h \psi_s(\vec x,t)\,, \label{eq_charge} \\[3pt]
{\!\vec j}(\vec x,t)=&-\frac{e}{2m} \h \sum_s \psi_s^*(\vec x,t) \h \bigg(\h\frac{\hbar}{\i} \h \nabla + e \h \vec A(\vec x,t)\mh\bigg) \hh \psi_s(\vec x,t)\label{eq_gaugeinvCurr1} \\
&-\frac{e}{2m} \h \sum_s \bigg(\bigg({-\frac{\hbar}{\i} \h \nabla} + e \h \vec A(\vec x,t)\mh\bigg) \hh \psi_s^*(\vec x,t)\bigg) \hh \psi_s(\vec x,t)\label{eq_gaugeinvCurr2}\\
&-\frac{e \hbar}{2m} \h\hh \nabla\times \bigg( \sum_{s, \h s'} \psi^*_s(\vec x,t) \h\hh \boldsymbol\sigma_{ss'} \h\hh \psi_{s'}(\vec x,t)\bigg) \,,\label{eq_spincurr}
\end{align}
where $\boldsymbol\sigma = (\sigma_1,\sigma_2,\sigma_3)^{\rm T}$ denotes the vector of Pauli matrices.
As always, the corresponding charge and current operators are obtained by replacing the classical spinor fields with their operator counterparts,
\begin{equation}
 \psi(\vec x, t) \mapsto \hat\psi(\vec x, t) \,.
\end{equation}
The resulting charge and current operators are relevant in the Kubo formalism (see Sec.~\ref{subsubsec_Kubo})
and for the derivation of the spin magnetization (see Sec.~\ref{subsubsec_spinmagn}).
We conclude that the current of the Pauli field is composed of two parts: (i) the usual gauge-covariant current
represented by the first two contributions \eqref{eq_gaugeinvCurr1}--\eqref{eq_gaugeinvCurr2},
and (ii) a divergence-free spinorial current \eqref{eq_spincurr} which can be expressed in terms of the spin density (see Eq.~\eqref{eq_spincurrent}). 

We will now give a short derivation of the above important result. 
To do this, we start directly from the classical Pauli equation in the Schr\"{o}dinger form (see \cite[Eq.~(XX.187)]{Messiah} or \cite[Eq.~(1.32)]{Bjorken}),
\begin{equation} \label{pauli_eq}
 \i\hbar \, \partial_t \Psi(\vec x, t) = (\hat H \h \Psi)(\vec x, t) \,,
\end{equation}
where the Pauli Hamiltonian for the spinor field $\Psi(\vec x,t)$ in the presence of the external four-potential $(\varphi/c, \h \vec A)^{\rm T}$ is given by
\begin{equation}
 \hat H = \frac{1}{2m} \h \big(\vec \sigma \cdot (\hat{\vec p} + e \hh \vec A(\vec x,t))\big)^{\mh 2}  - e \h \varphi(\vec x,t) \,, \label{pauli} \smallskip
\end{equation}
with the momentum operator
\begin{equation}
\hat{\vec p}=\frac{\hbar}{\rm i} \h \nabla \,.
\end{equation}
By expanding the product in Eq.~\eqref{pauli} and using the Coulomb gauge,
\begin{equation}
 \nabla \cdot \vec A = 0 \,,
\end{equation}
the Hamiltonian can be written as a sum of two terms,
\begin{equation}
 \hat H = \hat H_0 + \hat H_{\rm int} \,,
\end{equation}
where $\hat H_0$ is the Hamiltonian in the absence of the vector potential,
\begin{equation} \label{freeHam}
 \hat H_0 = \frac{|\hat{\vec p}|^2}{2m} \,,
\end{equation}
and $\hat H_{\rm int}$ is the interaction Hamiltonian given explicitly by
\begin{equation}
 \hat H_{\rm int} = \frac{e}{m} \h \vec A \cdot \hat{\vec p} + \frac{e\hbar}{2m} \, \vec \sigma \cdot (\nabla \times \vec A) + \frac{e^2}{2m} \h  |\vec A|^2 - e \h \varphi \,.
\end{equation}
Here, we have used the well-known matrix identity
\begin{equation}
 (\vec \sigma \cdot \vec A) (\vec \sigma \cdot \vec B) = (\vec A \cdot \vec B) \h \mathbbm 1_{2 \times 2} + \i \h \vec \sigma \cdot (\vec A \times \vec B) \,,
\end{equation}
where $\mathbbm 1_{2 \times 2}$ denotes the $(2 \times 2)$ identity matrix while $\vec A$ and $\vec B$ are ordinary vectors. Now, by taking the expectation value of the Hamiltonian,
\begin{align}
 \langle \Psi \mid \hat H \mid \Psi \rangle = \langle \Psi \mid \hat H_0 \mid \Psi \rangle+\langle \Psi \mid \hat H_{\rm int} \mid \Psi \rangle\,,
 \end{align}
we find $H_{\rm int}(t) \equiv \langle\Psi \mid \hat H_{\rm int}(t)\mid \Psi \rangle$ to be given by
\begin{align}
 H_{\rm int}(t) & = \int\! \de^3\vec x\,\h\Psi^\dagger(\vec x,t) \, \bigg(\frac{e\hbar}{m{\rm i}} \h \vec A(\vec x,t) \cdot \nabla \h + \h \frac{e\hbar}{2m} \, \vec \sigma \cdot (\nabla \times \vec A(\vec x,t)) \nonumber \\[3pt] 
 & \hspace{3.5cm} + \h \frac{e^2}{2m} \h  |\vec A(\vec x, t)|^2 -e \h \varphi(\vec x, t) \bigg) \hh \Psi(\vec x,t)\,. \label{int_term_prev}
\end{align}
After partial integration, this reverts to
\begin{equation}
\begin{aligned}
H_{\rm int}(t)& =\int\! \de^3\vec x\,\h \vec A(\vec x,t) \cdot \frac{e\hbar}{2m{\rm i}}\left(\Psi^\dagger(\vec x,t)\h\nabla\Psi(\vec x,t)-(\nabla\Psi^\dagger(\vec x,t))\h\Psi(\vec x,t) \right) \hspace{-1cm} \\[3pt]
&\quad \, +\int\!\de^3\vec x\,\h\vec A(\vec x,t)\cdot \frac{e\hbar}{2m} \h \nabla\times\left(\Psi^\dagger(\vec x,t) \, \boldsymbol\sigma\,\Psi(\vec x,t)\right) \label{int_term} \\[3pt]
&\quad \, +\int\!\de^3\vec x\,\h\vec A(\vec x,t)\cdot \frac{e^2}{2m} \h \Psi^\dagger(\vec x,t) \h \vec A(\vec x,t) \h \Psi(\vec x,t) \,. \\[5pt]
& \quad \, + \int \! \de^3 \vec x \,\h \varphi(\vec x, t) \, (-e) \h \Psi^\dagger(\vec x, t) \h \Psi(\vec x, t) \,.
\end{aligned}
\end{equation}
In particular, this interaction Hamiltonian is non-linear in the vector potential, and hence it is not of the form usually assumed in the general electromagnetic field theory,
\begin{equation}
H_{\rm int}(t)\not =\int\!\de^3\vec x\,\h\big(\h \rho(\vec x,t) \h \varphi\ext(\vec x,t)-\vec j(\vec x,t)\cdot\vec A\ext(\vec x,t)\big)\,.
\end{equation}
Nevertheless, the charge and current densities can be defined from Eq.~\eqref{int_term}: they are given by the functional derivatives (see e.g.~\cite[Eq.~(1.96)]{Bruus} 
and \cite[p.~391, footnote 25]{Altland})
\begin{align}
\rho(\vec x, t) & = \frac{\delta H_{\rm int}(t)}{\delta \varphi_{\rm ext}(\vec x, t)} \,, \label{defchar} \\[5pt]
\vec j(\vec x,t) & = -\frac{\delta H_{\rm int}(t)}{\delta\vec A_{\rm ext}(\vec x,t)}\,. \label{defcurr}
\end{align}

\vspace{3pt} \noindent
These definitions retain their validity even for non-linear couplings. One now verifies easily that Eq.~\eqref{int_term} implies the expressions \eqref{eq_charge} for the charge density and \eqref{eq_gaugeinvCurr1}--\eqref{eq_spincurr} for the electromagnetic current of Pauli spinors.

\subsection{Kubo formula for electromagnetic current} \label{kubo_current}

Finally, we show that the electromagnetic Kubo formula from Sec.~\ref{subsubsec_Kubo} can be derived as a special case of the generalized Kubo formula stated in Sec.~\ref{subsec_genKubo}. We focus on the spatial current response to an external vector potential as given by Eq.~\eqref{kubo_for_curr}, because in this case we encounter precisely these two complications as compared to the ordinary Kubo formalism: the observable under consideration, i.e.~the electromagnetic current, depends itself on the perturbation, i.e.~on the external potential, and the interaction is non-linear (in fact: quadratic) in the vector potential.

We start from the free Hamiltonian
\begin{equation}
 \hat H_0 = -\frac{\hbar^2}{2m} \int \! \de^3 \vec x \, \h \hat \Psi^\dagger(\vec x) \h (\Delta \hat \Psi)(\vec x) \,, \label{pauli_1}
\end{equation}
which is the second-quantized form of Eq.~\eqref{freeHam}. The interaction Hamiltonian is given by the operator analogon of Eq.~\eqref{int_term_prev}, i.e.,
\begin{equation} \label{pauli_2}
\begin{aligned}
 \hat H_{\rm int} & = \int\! \de^3\vec x\,\h\hat \Psi^\dagger(\vec x) \, \bigg(\frac{e\hbar}{m{\rm i}} \h \vec A(\vec x) \cdot \nabla \h + \h \frac{e\hbar}{2m} \, \vec \sigma \cdot (\nabla \times \vec A(\vec x)) \\[3pt]
 & \hspace{3.2cm} + \h \frac{e^2}{2m} \h  |\vec A(\vec x)|^2 -e \h \varphi(\vec x) \bigg) \hh \hat \Psi(\vec x)\,.
\end{aligned}
\end{equation}
In order to apply the generalized Kubo formula \eqref{eq_GenKuboForm} from Sec.~\ref{subsec_genKubo},
we identify the parameters $X_j$ of the external perturbation with the external vector potential $\vec A(\vec x')$. Furthermore, the observable $\hat O$ corresponds 
to the Pauli current density $\vec j(\vec x)$, which is given by the operator analogon of Eqs.~\eqref{eq_gaugeinvCurr1}--\eqref{eq_spincurr}, i.e.,
\begin{equation}
\begin{aligned}
\,\hat{\!\vec j}(\vec x)[\vec A]=&-\frac{e}{2m} \, \hat \Psi^\dagger(\vec x) \h \bigg(\h\frac{\hbar}{\i} \h \nabla + e \h \vec A(\vec x)\mh\bigg)\hat\Psi(\vec x) \\[4pt]
&-\frac{e}{2m} \, \bigg(\bigg({-\frac{\hbar}{\i} \h \nabla} + e \h \vec A(\vec x)\mh\bigg) \h \hat \Psi^\dagger(\vec x)\bigg) \h \hat \Psi(\vec x)\\[5pt]
&-\frac{e \hbar}{2m} \h \nabla \times \big( \hat \Psi^\dagger(\vec x) \, \boldsymbol\sigma \, \hat \Psi(\vec x) \big) \,.
\end{aligned} \smallskip
\end{equation}
The current density can be split into two contributions,
\begin{equation}
 \,\hat{\!\vec j}(\vec x) = \,\hat{\!\vec j}^{\rm p}(\vec x) + \,\hat{\!\vec j}^{\rm d}(\vec x) \,,
\end{equation}
which are sometimes called the {\itshape paramgnetic current density} and the {\itshape diamagnetic current density} (see e.g.~\cite[App.~2]{Giuliani}). The first contribution does not depend on the vector potential,
\begin{equation}
 \,\hat{\!\vec j}^{\rm p}(\vec x) = \,\hat{\!\vec j}(\vec x) \hh [ \hh \vec A \equiv \vec 0 \h ] \,, \smallskip
\end{equation}
whereas the second contribution is proportional to the vector potential and can be expressed in terms of the charge density operator as
\begin{equation}
 \,\hat{\!\vec j}^{\rm d}(\vec x)[\vec A] = -\frac{e^2}{m} \h \vec A(\vec x) \h \hat \Psi^\dagger(\vec x) \h \hat \Psi(\vec x) = \frac{e}{m} \h \vec A(\vec x) \h \hat \rho(\vec x) \,.
\end{equation}
In particular, the current operator at point $\vec x$ depends only on the vector potential at the same point, such that
\begin{equation}
 \frac{\delta \,\hat{\!j}_k(\vec x)}{\delta A_\ell(\vec x')} \h \bigg|_{\vec A \h \equiv\h \vec 0} = \h \frac{e}{m} \, \delta_{k\ell} \, \delta^3(\vec x - \vec x') \h \hat \rho(\vec x) \,.
\end{equation}
The derivative of the interaction Hamiltonian with respect to the external potential $A_\ell(\vec x')$ yields again the current (with the opposite the sign) by its very definition \eqref{defcurr}.
Thus, the generalized Kubo formula \eqref{eq_GenKuboForm} yields
\begin{equation} \label{kubo_el}
\begin{aligned}
 \frac{\delta j_k(\vec x, t)}{\delta A_\ell(\vec x', t')} \h \bigg|_{\vec A \h \equiv \h \vec 0} &  = \delta(c \h t - c\h t')\, \frac{e}{m} \, \delta_{k\ell} \, \delta^3(\vec x - \vec x') \, \big\langle \Psi_0 \hh \big| \hh \hat \rho(\vec x) \hh \big| \Psi_0 \big\rangle \\[2pt]
 & \quad \, +\frac{\i}{c \h \hbar} \, \varTheta(t - t') \, \big \langle \Psi_0 \hh \big| \big[ \h \,\hat{\!j}^{\hh\rm p}_k(\vec x, t) , \h \,\hat{\!j}_\ell^{\hh\rm p}(\vec x', t') \h \big] \big | \Psi_0 \big\rangle \,,
\end{aligned}
\end{equation}
where on the right hand side the time dependence of the current operator is given in the interaction picture by
\begin{equation}
 \,\hat{\!j}_k^{\hh\rm p}(\vec x, t) = \e^{\i t \hat H_0 / \hbar } \, \,\hat{\!j}_k^{\hh\rm p}(\vec x) \, \e^{-\i t \hat H_0 / \hbar } \,.
\end{equation}
The result \eqref{kubo_el} agrees precisely with Eq.~\eqref{kubo_for_curr} in the main text (where we have suppressed the index ``$\mathrm p$'' for the paramagnetic current and implicitly understood that the time evolution is in the interaction picture). The remaining formulae \eqref{kubo_for_1}--\eqref{kubo_for_3} can be shown analogously.

\end{appendices}

\bibliographystyle{model1-num-names}
\bibliography{masterbib}

\end{document}